\pdfoutput=1

\documentclass[twocolumn]{aastex631}
\usepackage{epstopdf}
\usepackage{xcolor}
\usepackage{xspace}
\usepackage{graphicx}
\usepackage{amssymb}
\usepackage{amsmath}
\usepackage{natbib}
\usepackage{float}
\usepackage{hyperref}
\shorttitle{Radio Emission of Early-type Galaxies at Low Luminosity Range}

\shortauthors{W\'ojtowicz et al.}

\begin{document}

\title{Radio Emission of Nearby Early-type Galaxies at Low and Very-Low Radio Luminosity Range}

\correspondingauthor{A.~W\'ojtowicz}
\email{awojtowicz@oa.uj.edu.pl}

\author[0000-0001-5666-8665]{Anna~W\'ojtowicz}
\affiliation{Astronomical Observatory of the Jagiellonian University, Orla 171, 30-244 Krak\'{o}w, Poland}
\affiliation{Department of Theoretical Physics and Astrophysics, Faculty of Science, Masaryk University, Kotl\'a\v{r}sk\'a 2, Brno, 611 37, Czech Republic}

\author[0000-0001-8294-9479]{{\L}ukasz~Stawarz}
\affiliation{Astronomical Observatory of the Jagiellonian University, Orla 171, 30-244 Krak\'{o}w, Poland}

\author[0000-0002-4377-0174]{C.~C. Cheung}
\affiliation{Naval Research Laboratory, Washington, DC 20375, USA}

\author[0000-0003-0392-0120]{Norbert Werner}
\affiliation{Department of Theoretical Physics and Astrophysics, Faculty of Science, Masaryk University, Kotl\'a\v{r}sk\'a 2, Brno, 611 37, Czech Republic}

\author{Dominik~Rudka}
\affiliation{Astronomical Observatory of the Jagiellonian University, Orla 171, 30-244 Krak\'{o}w, Poland}

\begin{abstract} 

We analyze radio continuum emission of early-type galaxies with dynamical measurements of central super-massive black hole (SMBH) masses, and well-characterized large-scale environments, but regardless on the exact level of the nuclear activity. The 1.4\,GHz radio fluxes collected with $\sim$arcmin resolution for 62 nearby targets (distances $\lesssim 153$ Mpc), correspond to low and very low monochromatic luminosities $L_{\rm r} \sim 10^{35} - 10^{41}$\,erg\,s$^{-1}$. We quantify possible correlations between the radio properties with the main parameters of supermassive black holes, host galaxies, and hot gaseous halos, finding a general bimodality in the radio luminosity distribution, with the borderline between ``radio-bright'' and ``radio-dim'' populations $\log L_{\rm r} / L_{\rm Edd} \simeq -8.5$. We analyze the far-infrared data for the targets, finding that all radio-bright sources, and over a half of radio-dim ones, are over-luminous in radio with respect to the far-infrared--radio correlation. High-resolution radio maps reveal that the overwhelming majority of radio-dim sources are unresolved on arcsecond scale, while the bulk of radio-bright sources display extended jets and lobes of low- and intermediate-power radio galaxies; those jets dominate radio emission of radio-bright objects. Regarding the origin of the radio emission of radio-dim sources, we discuss the two main  possibilities. One is the ADAF model, in which the radio and the nuclear X-ray radiative outputs at very low accretion rates, are both dominated by unresolved jets. The other possibility is that the radio-dim sources, unlike the radio-bright ones, are characterized by low values of SMBH spins, so that their radio emission is not related to the jets, but instead is due to a combination of starforming processes and past nuclear outbursts.

\end{abstract}

\section{Introduction}
\label{sec:intro}

Radio continuum emission of astrophysical sources is produced predominantly through the synchrotron process, whenever relativistic electrons with sufficiently high energies are present, and are accelerated in a strong magnetic field of a system. On galactic scales, such conditions are characteristic of either starformation regions in the interstellar medium (ISM), or of accretion disks and nuclear outflows in Active Galactic Nuclei (AGN). In the latter case, depending on the exact physical conditions and parameters of the central engine --- such as the accretion rate, supermassive black hole (SMBH) spin, or magnetization of the accreting matter -- the outflows may range from sub-relativistic and uncollimated massive winds terminating within host galaxies, to highly relativistic and light but powerful jets reaching extragalactic scales \citep[see, e.g.,][for the recent compendium]{meier12}. 

Relativistic jets, in particular, formed via the extraction of the SMBH rotational energy by the magnetic field supported by the accreting matter within the ergosphere \citep{Blandford77}, are known to be very efficient emitters of radio synchrotron photons, so much that the characteristic of a given AGN to be  ``radio-loud'' became synonymous with being ``jetted''. Yet all AGN are sources of radio synchrotron photons, at least at some level, and while it is widely agreed on, and strongly supported by observations, that relativistic jets provide indeed the bulk of the observed radiative output at radio wavelengths in high-luminosity sources such radio quasars and radio galaxies, there is an ongoing debate on the dominant source of the radio emission in other types of AGN with low and very low radio luminosities. Such ``radio-quiet'' AGN may shine in radio through the emission of their accretion disks, disk coronae, and disk winds, for example, in addition to the starforming regions \citep[for the recent review see][and references therein]{Panessa19}.

The question in this context is the exact distribution of the radio-loudness parameter in the general population of AGN, and to what extent this distribution reflects the efficiency of the jet production by the central engine. In the case of quasar sources, for instance, it was argued that the distribution of the radio-loudness parameter, defined as the ratio of the monochromatic radio (5\,GHz) to optical (B-band) luminosity densities $\mathcal{R} \equiv L_{\rm 5\,GHz,\,\nu}/L_{\rm B,\,\nu}$ \citep{Kellermann89}, is bimodal, with only $\sim 10\%$ of quasars characterized by the largest $\mathcal{R}$ values being jetted \citep[e.g.,][]{Ivezic02}. In the framework of the \citet{Blandford77} model for the jet production, where the resulting jet power scales with the square of the angular velocity of the black hole horizon, this could imply the ``bottom-heavy'' spin distribution of SMBHs in quasars' host galaxies \citep{Wilson95,Moderski98,Hughes03}. On the other hand, the distribution of the radio-loudness for quasars may not be intrinsically bimodal, but instead evolving with redshift \citep[see][]{Singal11,Singal13}.

Quasars constitute however a peculiar class of luminous AGN, with SMBHs accreting at high rates, for which the observed optical emission is dominated by the blackbody-like radiation of optically-thick accretion disks \citep[for the review on the accretion disks physics, see][]{Abramowicz13}; as a result, the radio-loudness parameter $\mathcal{R}$ is a well-defined proxy for the jet production efficiency in quasar sources indeed. The problem arise when applying the same definition of the radio loudness to other types of AGN, for example those with SMBHs accreting at lower rates  ($\ll 1\%$ Eddington), for which the radiative output of accretion disks --- in terms of both the spectral energy distribution, and also the overall radiative efficiency --- is very different  than that of standard optically-thick disks \citep[for a review see][]{Yuan14}. 

Still, with the extra care regarding the $\mathcal{R}$ definition, optical and radio fluxes considered, as well as particular classes and samples of AGN studied, another type of bimodality in the radio-loudness distribution was noted, namely that the values of the $\mathcal{R}$ parameters characterizing AGN hosted by late-type galaxies, are always orders of magnitude lower than those characterizing AGN hosted by early-type galaxies, at \emph{any} accretion rate, including also very sub-Eddington values \citep{Xu99,Terashima03,Sikora07}. In the framework of the spin paradigm for the AGN jet production, this could imply that the spins of SMBHs in spheroids are on average much larger than those of SMBH at the centres of disk galaxies, due to different formation histories of the hosts (convolved with the formation histories of the central black holes).

In this paper, we TC{investigate} the radio emission of nearby early-type galaxies, for which neither an exact level of the nuclear activity, nor did a presence of relativistic radio-emitting jets, played any role in the selection procedure. In the studied sample of 62 sources, we observe a bimodal distribution in the Eddington-normalized radio power $\log L_{\rm r} / L_{\rm Edd}$, where $L_{\rm r}$ is the observed monochromatic 1.4\,GHz luminosity, and
\begin{equation}
L_{\rm Edd} = \frac{4 \pi G M_{\rm BH} m_p c}{\sigma_{\rm T}} \approx 10^{46} \left(\frac{M_{\rm BH}}{10^8 M_{\odot}}\right) \, {\rm erg\,s^{-1}} \, ,
\end{equation}
is the Eddington luminosity corresponding to the black hole mass $M_{\rm BH}$. The borderline between the modes is located at $\log L_{\rm r} / L_{\rm Edd} \simeq -8.5$, and the sample extends down to $\log L_{\rm r} / L_{\rm Edd} \simeq -12$; such a low level of the radio continuum emission of early-type systems, has barely been probed before in a systematic manner.

In general, nearby bright elliptical and lenticular galaxies were long known to be radio emitters, with about one-third (or more) of a sample being typically detected at GHz frequencies in mJy-flux-limit surveys with arcsec/arcmin resolution \citep{Sadler89,Wrobel91}, and even a larger fraction displaying milliarcsec-scale radio cores \citep{Slee94}. More recently, it was also established that the most massive ($> 10^{11} M_{\odot}$) or luminous (absolute K-band magnitudes $< -25.5$) spheroids are in fact \emph{always} detected at radio frequencies \citep{brown2011,Sabater19,Grossova22,Capetti22}, even though they often lack prominent large-scale ($> 4$\,kpc) radio structures \citep{Capetti22}. 

Early-type galaxies with ``core-type'' nuclear profiles of the optical surface brightness (following the Nuker scheme), were in particular argued to always host ``miniature radio galaxies'', with GHz radio luminosities in the range $\sim 10^{36} - 10^{38}$\,erg\,s$^{-1}$ \citep{Capetti06,Balmaverde06,Baldi09,Grandi21}. We note that, at higher radio luminosities $\sim 10^{38} - 10^{40}$\,erg\,s$^{-1}$, early-type galaxies host radio jets with the newly characterized Fanaroff--Riley (FR) type 0 morphology \citep{Baldi15}, followed by classic Fanaroff--Riley type I and II radio galaxies \citep{FR74}, with monochromatic radio powers $> 10^{39}$\,erg\,s$^{-1}$. Several targets from the list of sources analyzed in this paper (see the next Section\,\ref{sec:sample} and Appendix\,\ref{A:data}), do overlap with the samples of ``miniature radio galaxies'', and also with the catalogs of FR\,0, I, and II radio galaxies \citep[see][respectively]{Baldi18,Capetti17a,Capetti17b}.

\section{Source Sample and Data Acquisition}
\label{sec:sample}

In this work we utilized the source list from \citet{gaspari19}, which represents the collection of high-quality optical and X-ray data for 85 galaxies with various morphological types, for which the masses of central SMBHs, $M_{\rm BH}$, have been measured with high precision via direct dynamical methods, i.e. by resolving the stellar or gas kinematics within the SMBH influence regions \citep[the innermost galactic regions dominated by the SMBH gravitational potential; see][]{kormendy2013,bosch2016}. Note that this selection criterium effectively limits the sample to nearby objects with redshifts $z<0.04$. For such, in addition to the main galaxy parameters including the stellar velocity dispersion, $\sigma_{\rm vel}$, the bulge mass, $M_{\rm bulge}$, and the total galaxy $K$-band luminosity, $L_K$, \citeauthor{gaspari19} collected also the basic information regarding the intragroup/intracluster halo parameters, resulting from deep observations with the high-resolution X-ray instruments {\it Chandra} and XMM-{\it Newton}, in particular the halo X-ray luminosity $L_{\rm X}$, and the hot gas temperature $T_{\rm X}$. In this paper we use this extensive dataset (as summarized in Tables\,2 and 3 therein).\footnote{For the large-scale X-ray luminosity and temperature, here we take the $0.3-7$\,keV luminosity of the hot halo within the core radius, and the corresponding core temperature, denoted in \citet{gaspari19} as $L_{\rm x,c}$ and  $T_{\rm x,c}$, respectively.}

The overwhelming majority of galaxies in the sample --- 76 systems out of 85 total --- are early-type objects, i.e. galaxies with Hubble morphological types encompassing ellipticals (E0--E6) and lenticulars (S0--SAB0), at distances between 3.6 and 152.4\,Mpc. The large-scale environments range from the centres of rich clusters, to poor galaxy groups, fields, and even isolated systems.
Although by no means complete or homogeneous, the sample represent the most comprehensive list of nearby galaxies with precisely characterized central black hole, stellar, and hot halo parameters. The main objective of \citet{gaspari19} was to research the link between SMBHs and their large-scale environment, by means of a comprehensive Bayesian correlation analysis between all the gathered system parameters (and their combinations) within the sample, free from the {\it ``nonindependent correlations with unreliable conversion uncertainty''}; such nonindependent correlations may emerge whenever the black hole masses do not result from the direct dynamical measurements, but are instead obtained from the other established correlations and scaling relations such as the $M_{\rm BH} - \sigma_{\rm vel}$ correlation, or the ``AGN fundamental plane''. The goal of this work, generally speaking, is to extend the analysis to the radio range, i.e. to investigate the radio emission of the uniquely gathered sample of galaxies.

\begin{figure}[th!]
    \centering
\includegraphics[width=0.45\textwidth]{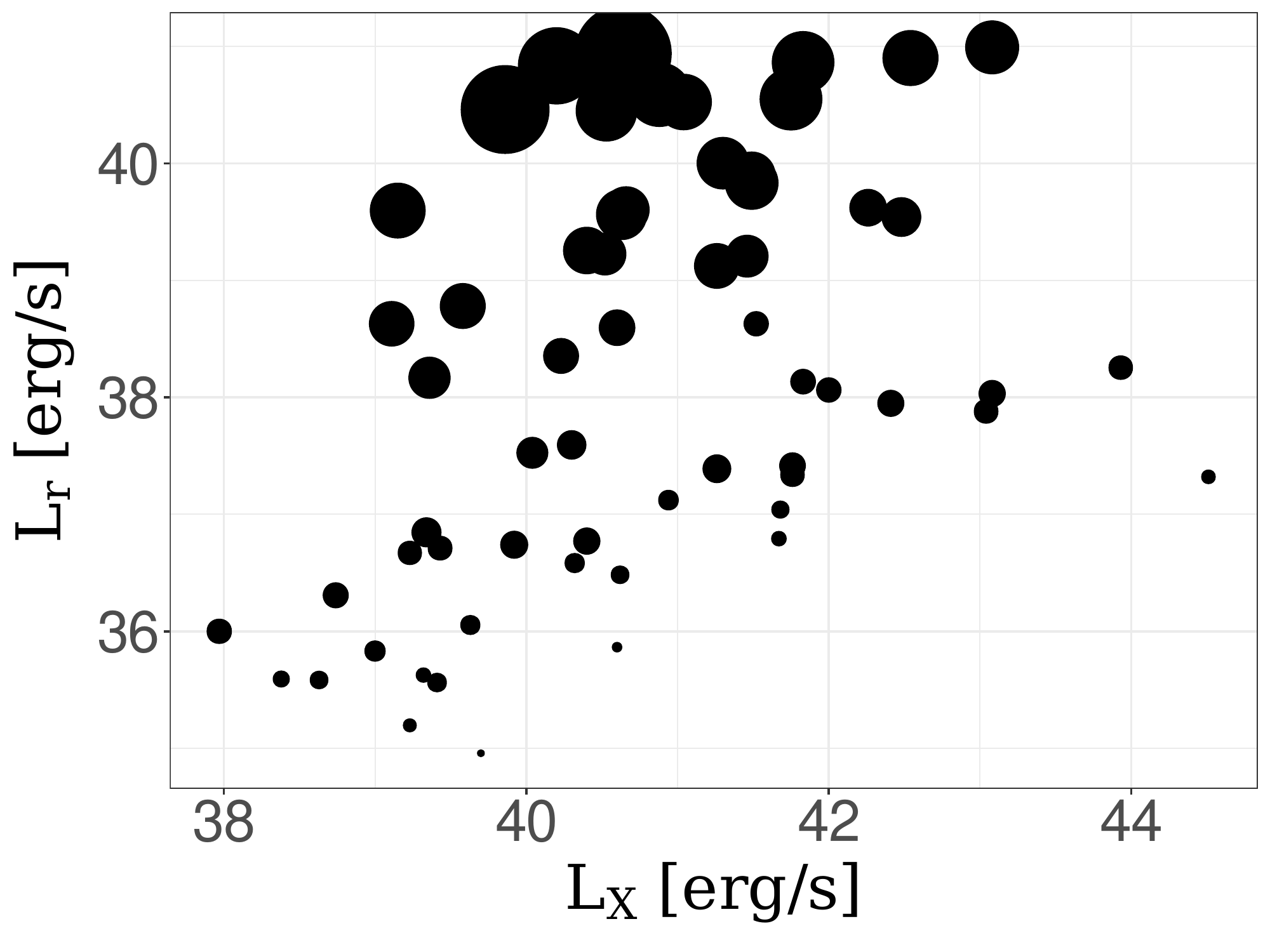} 
\includegraphics[width=0.45\textwidth]{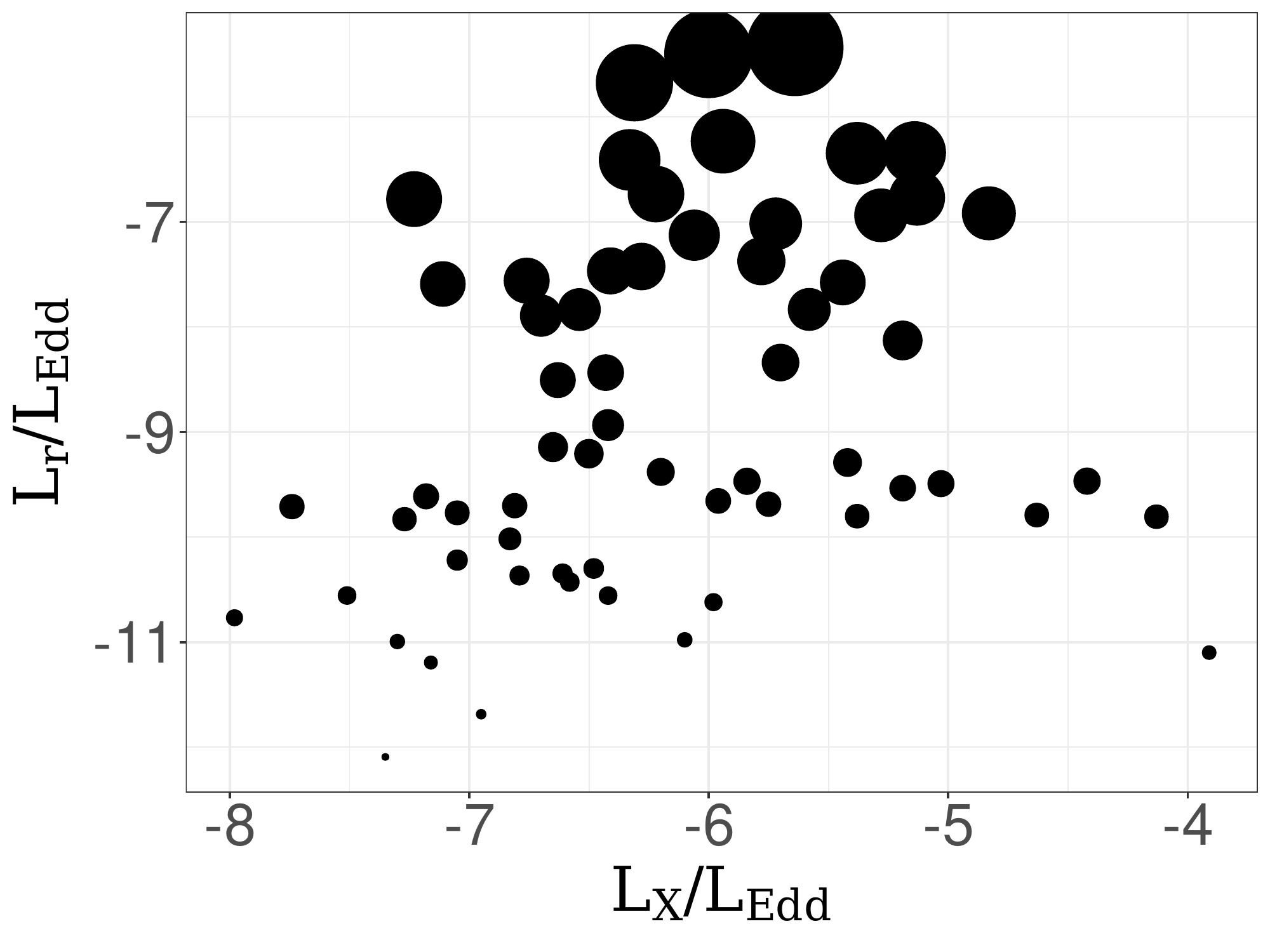}
\includegraphics[width=0.45\textwidth]{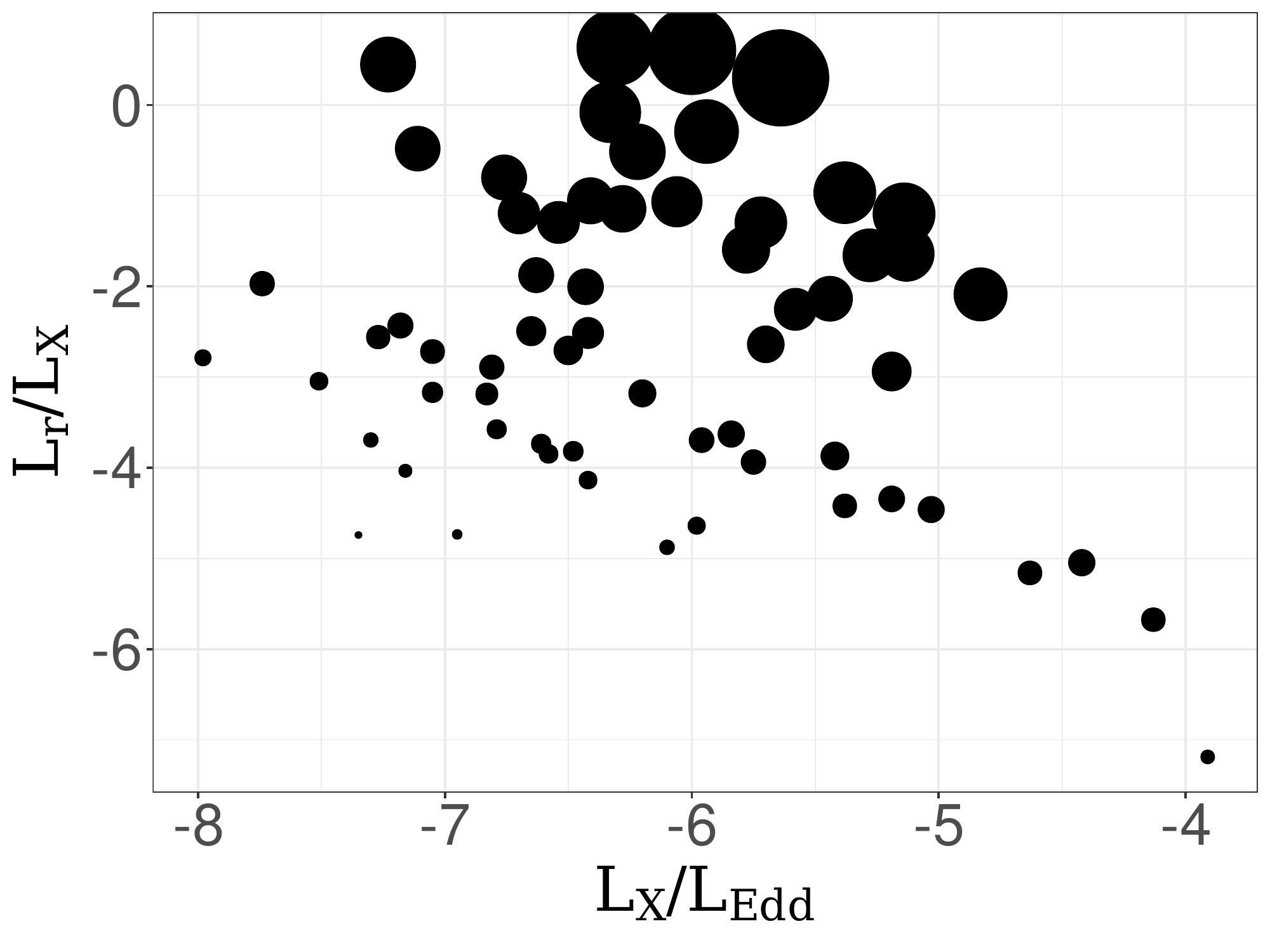}
\caption{{\it Upper panel:} Scaling of the monochromatic radio luminosities of early-type galaxies, $L_{\rm r}$, with the X-ray luminosities of their hot gaseous halos, $L_{\rm X}$; point sizes denote here the logarithm of the Eddington-normalized radio luministy, $\log L_{\rm r}/L_{\rm Edd}$. {\it Middle panel:} Same as the upper panel except that here both luminosities are expressed in Eddington units. {\it Lower panel:} Same as the upper panel, except that here the radio luminosities on the $y$-axis are replaced by the corresponding $L_{\rm r}/L_{\rm X}$ ratios.}
\label{fig:3D}
\end{figure}

\begin{figure}[th!]
    \centering
\includegraphics[width=\columnwidth]{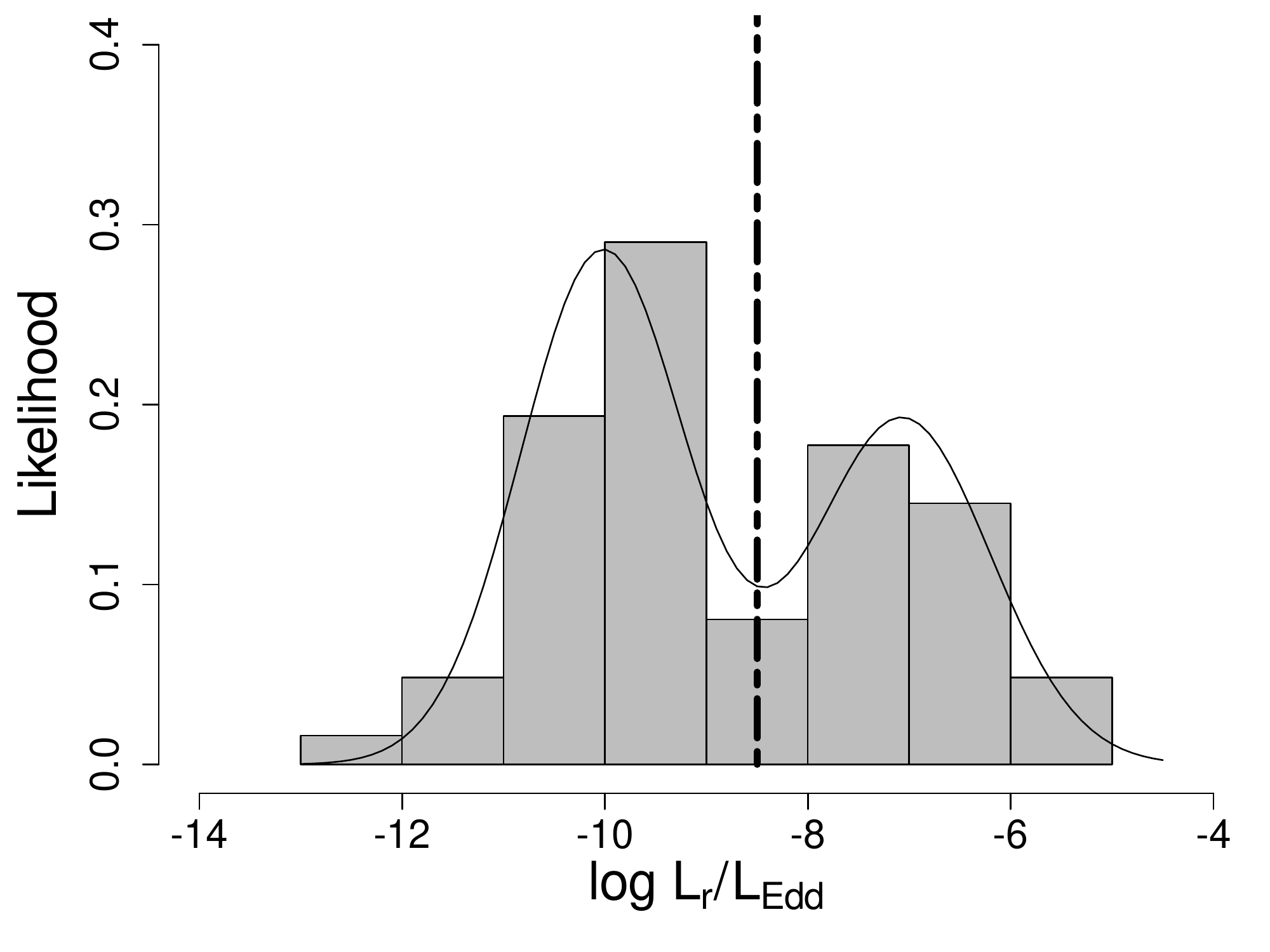} 
\caption{Histogram showing the probability density distribution of the Eddington-normalized monochromatic radio luminosity in the analyzed sample of early-type galaxies. The location of the anti-mode at $\log L_{\rm r}/L_{\rm Edd} \simeq -8.5$ is marked in the plot by vertical dashed line. The Black solid lines represent the fitted probability density function as was described in section \ref{sec:bimodality}.}
 \label{fig:hist}
\end{figure}

It is important to emphasize at this point that, as the sample has been compiled based on \emph{all} the known systems with the direct SMBH masses \emph{and} with the X-ray halo detections, it is also free from the selection bias related to the exact level of the nuclear activity, and/or the source prominence in the radio band. That is to say, the sample includes a mixture of non-active and active galaxies (albeit the latter ones restricted rather to low-luminosity AGN classified as either Low-Ionization Nuclear Emission-line Regions, LINERs, or, at most, Seyferts). Likewise, the galaxies in the sample could be either seemingly radio-silent on the one extreme of the scale, or characterized by a very pronounced jet activity on the other extreme (including the well-known radio galaxies NGC\,5128/Centaurus\,A and NGC\,4486/M\,87).

The vast majority of the galaxies from the \citet{gaspari19} sample were observed by the Very Large Array (VLA) at 1.4\,GHz frequency with $45^{\prime\prime}$ resolution. Most of them were cataloged in  the \citet{brown2011} study of near-infrared-bright (2MASS $K< 9$\,mag) sample of early-type galaxies. They primary utilized the VLA 1.4\,GHz data from \citet{condon1998}, supplementing it with low-resolution ($\approx 12^{\prime}$) single-dish imaging from the 300-ft Green Bank and the 64-m Parkes radio telescopes  for the brightest, extended sources ($S_{\rm 1.4\,GHz}>0.6$\,Jy). The remaining sources were mostly cataloged in \citet{condon1998,condon2002}; complementary radio information for several southern sources could be found in \citet{velzen2012}, and the flux of NGC\,5128 which structure spans $\sim 10$\,deg in the sky was given in \citet{cooper1965}. For nine ellipticals/lenticulars from the list, only upper limits at the level of 0.5\,mJy were available from the VLA observations; five early-type systems are missing any arcmin-scale data at GHz frequencies. The gathered radio data are summarized in Appendix\,\ref{A:data}. In Appendix\,\ref{A:all} we present the scaling of the resulting 1.4\,GHz monochromatic luminosities $L_{\rm r}$ with the main galaxy and halo parameters $M_{\rm BH}$, $M_{\rm bulge}$, $\sigma_{\rm vel}$, $L_K$, $L_{\rm X}$, and $T_{\rm X}$, for all the galaxies from the list of \citet{gaspari19}, including also radio non-detections, as well as nine spirals for a comparison. However, for the main analysis discussed in the following sections, and focusing on early-type systems, we utilize exclusively the sample of 62 \emph{``radio-detected''} ellipticals/lenticulars. 

All the analyzed galaxies with the exception of the southernmost NGC\,5128, have recently been also targeted by the Karl G. Jansky Very Large Array Sky Survey (VLASS) at 3\,GHz, with the resolution of $2.5^{\prime\prime}$ \citep[see][]{lacy20}. We inspected the available VLASS quicklook (epoch 2) maps from the CIRADA database\footnote{See:\url{https://science.nrao.edu/vlass/data-access/vlass-epoch-1-quick-look-users-guide}} \citep{Gordon2020} for the central $2^{\prime}$ regions of the targets to access their radio morphologies. We find clear structures --- consisting of either unresolved cores coinciding with the galactic centres, extended jets and jet-like features, or even amorphous halos in a few cases ---  in 49 out of 61 maps, all presented in Appendix\,\ref{A:VLASS}; the remaining 12 maps display only featureless noise.

The bulk of our targets also have available archival far-infrared (FIR) data from the Infrared Astronomical Satellite \citep[IRAS;][]{Beichman88}, where the angular resolution of $\lesssim 2^{\prime}$ matches roughly the resolution of the utilized VLA data. For our analysis we utilized the 60\,$\mu$m and 100\,$\mu$m IRAS fluxes provided in the corrected data file from \citet{Knapp89}, unless explicitly stated otherwise (see Appendix\,\ref{A:data}). In addition, we investigate the higher-resolution ($\gtrsim 6^{\prime\prime}$) mid-infrared (MIR) data provided by the Wide-field Infrared Survey Explorer (WISE), in three bands W1, W2, and W3 centred at 3.4, 4.6, and 12\,$\mu$m, respectively \citep{Wright10}. To avoid possible contamination issues, we exclusively use magnitudes measured by fitting the extended source profiles, as provided in \citet{Cutri13};  the resulting W1--W2 and W2--W3 colours are listed in Appendix\,\ref{A:data}. 

Finally, we augment the source sample information by the {\it Chandra} data for the central regions of the targets. In particular, we follow \citet{She17}, who re-analyzed the archival {\it Chandra} Advanced CCD Imaging Spectrometer data for 719 galaxies within 50\,Mpc, in order to {\it ``quantify the incidence of AGN activity as a function of Hubble type''}. For the purpose of our analysis, we utilize the nuclear 2--10\,keV luminosities of the overlapping sources, corresponding to the flux extraction regions of about a few arcseconds, depending on the target. Due to a limited number of counts, \citeauthor{She17} could perform only the basic spectral fitting in the majority of the cases, and so the provided nuclear luminosities may still be contributed by, or even dominated by, non-AGN components, such as the innermost regions of the ISM.

\section{Radio Emission of Early-type Galaxies}
\label{sec:radio}

\subsection{Radio---X-ray Correlations}
\label{sec:radioXray}

As discussed in Appendix\,\ref{A:all}, in the \emph{entire} sample of all the early-type galaxies studied here, we do not see any correlation between radio luminosities and black hole masses, while the overall correlations between radio luminosities and the remaining optical and X-ray parameters of the galaxies and their halos, even though modest or even strong (the Pearson's product-moment correlation coefficients $0.40 \leq \rho_{\rm E+L} \leq 0.54$; see Appendix\,\ref{A:all}), are characterized by very large spreads. In particular, a rather loose correlation between $L_{\rm r}$ and $L_{\rm X}$ (with $ \rho_{\rm E+L} = 0.43$), could be surprising to some extent, keeping in mind that the X-ray luminosity of the large-scale hot gaseous halo is expected to be linked to the supply of the matter accreting onto SMBHs, through the cooling rate of the X-ray--emitting plasma $\dot{M}_{\rm cool} \propto L_{\rm X}/T_{\rm X}$ \citep[see the discussion in][and references therein]{gaspari19}. 

\begin{figure*}[th!]
    \centering
\includegraphics[width=0.45\textwidth]{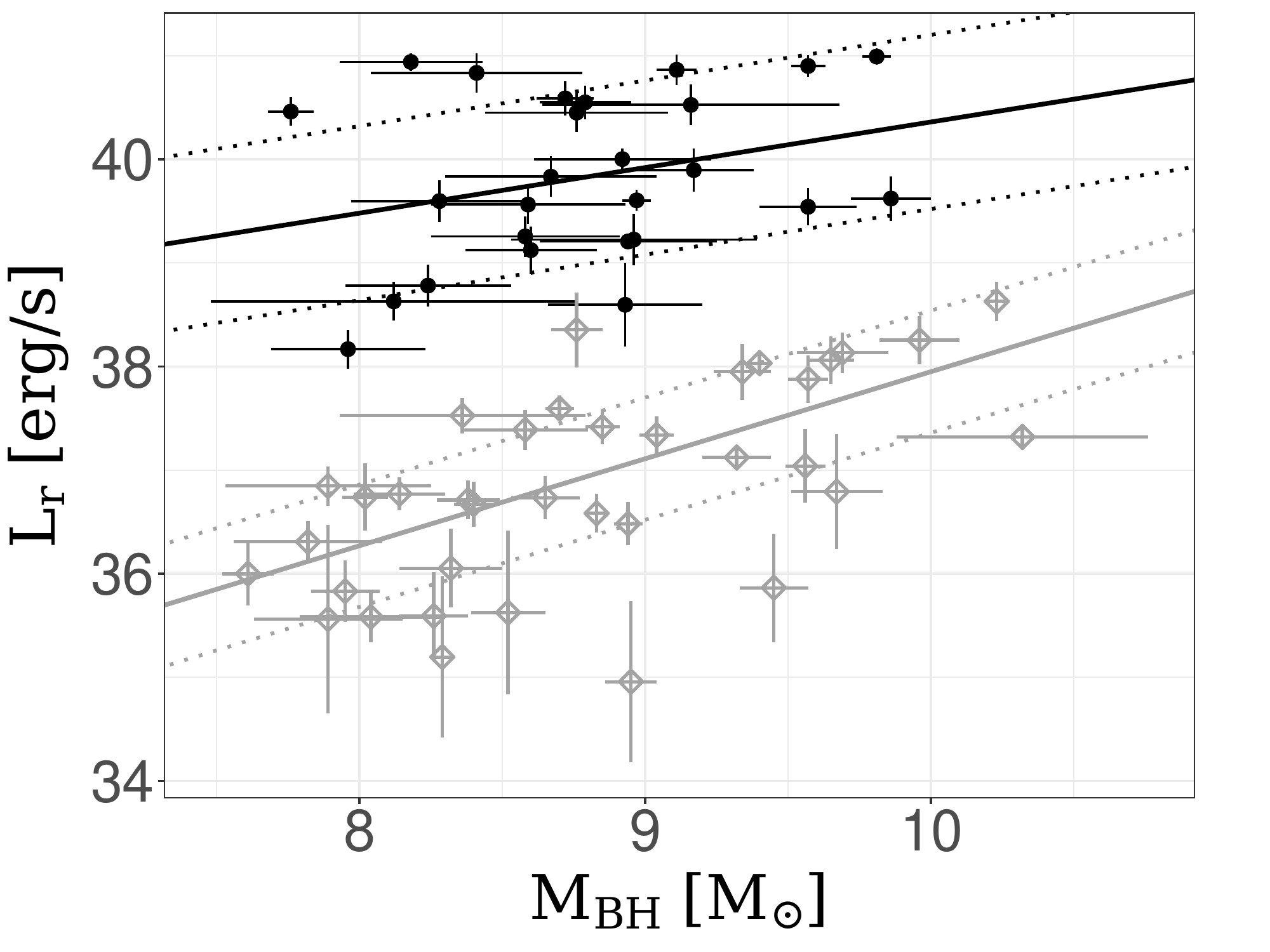} 
\includegraphics[width=0.45\textwidth]{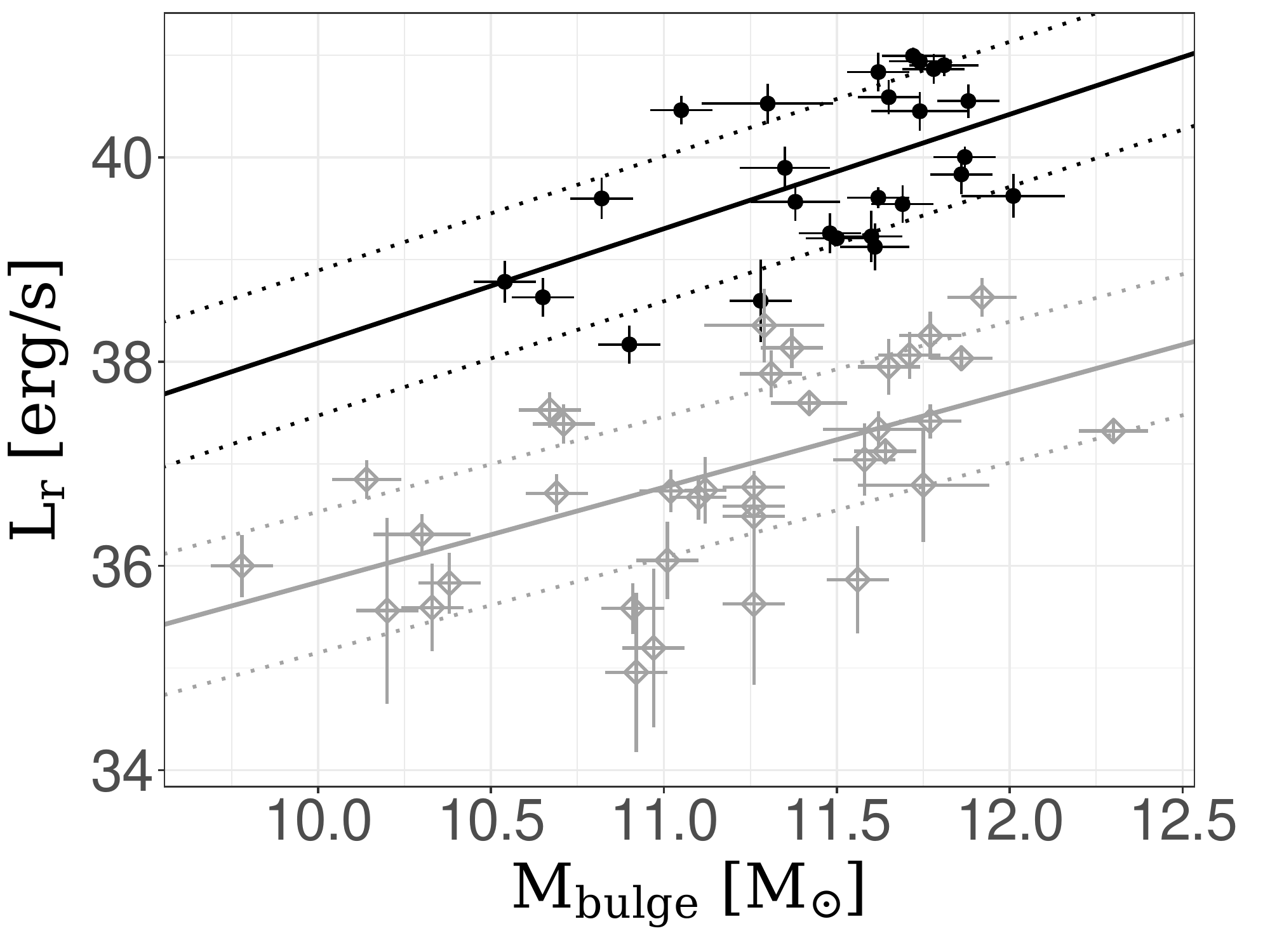}
\includegraphics[width=0.45\textwidth]{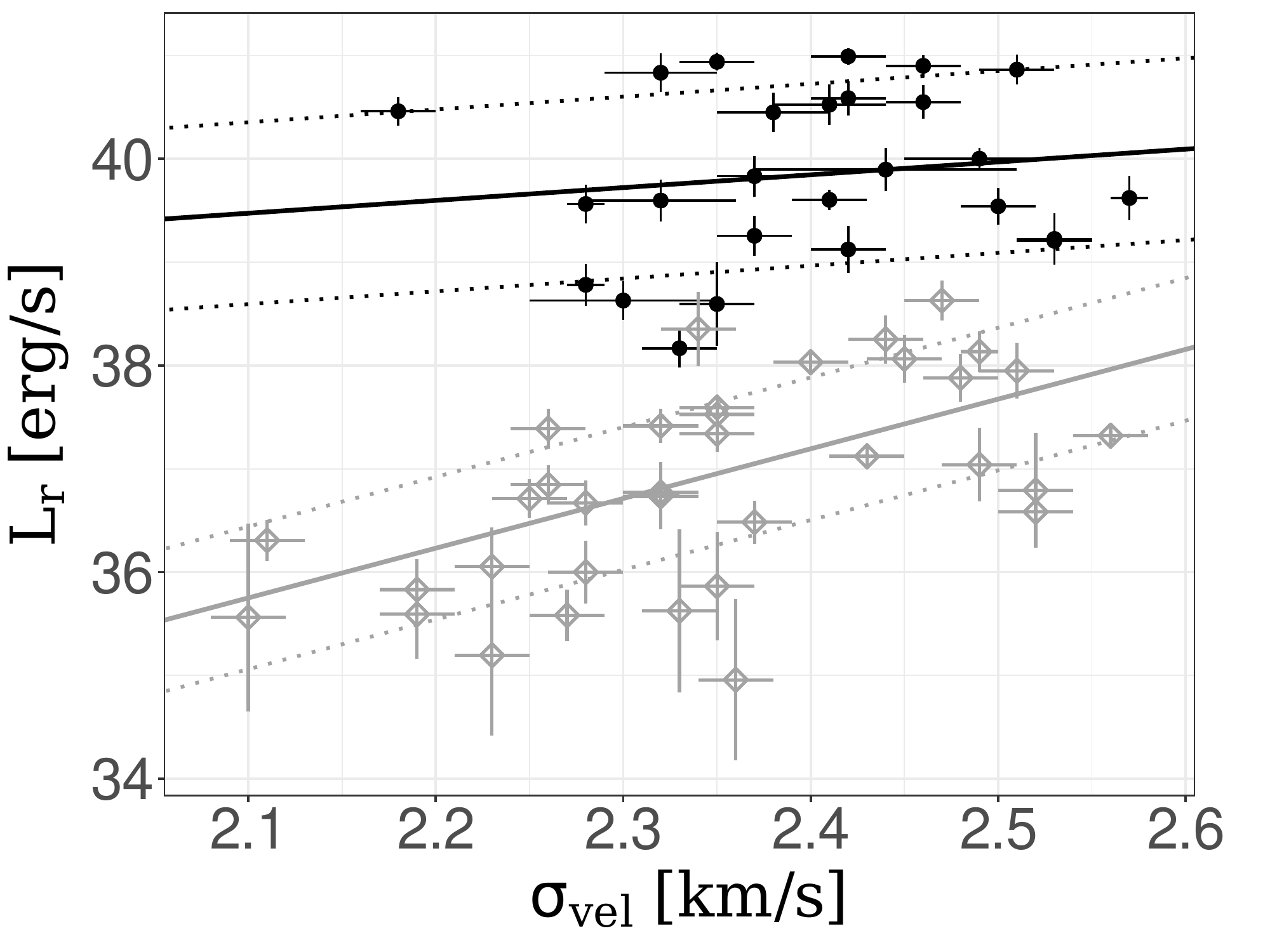}
\includegraphics[width=0.45\textwidth]{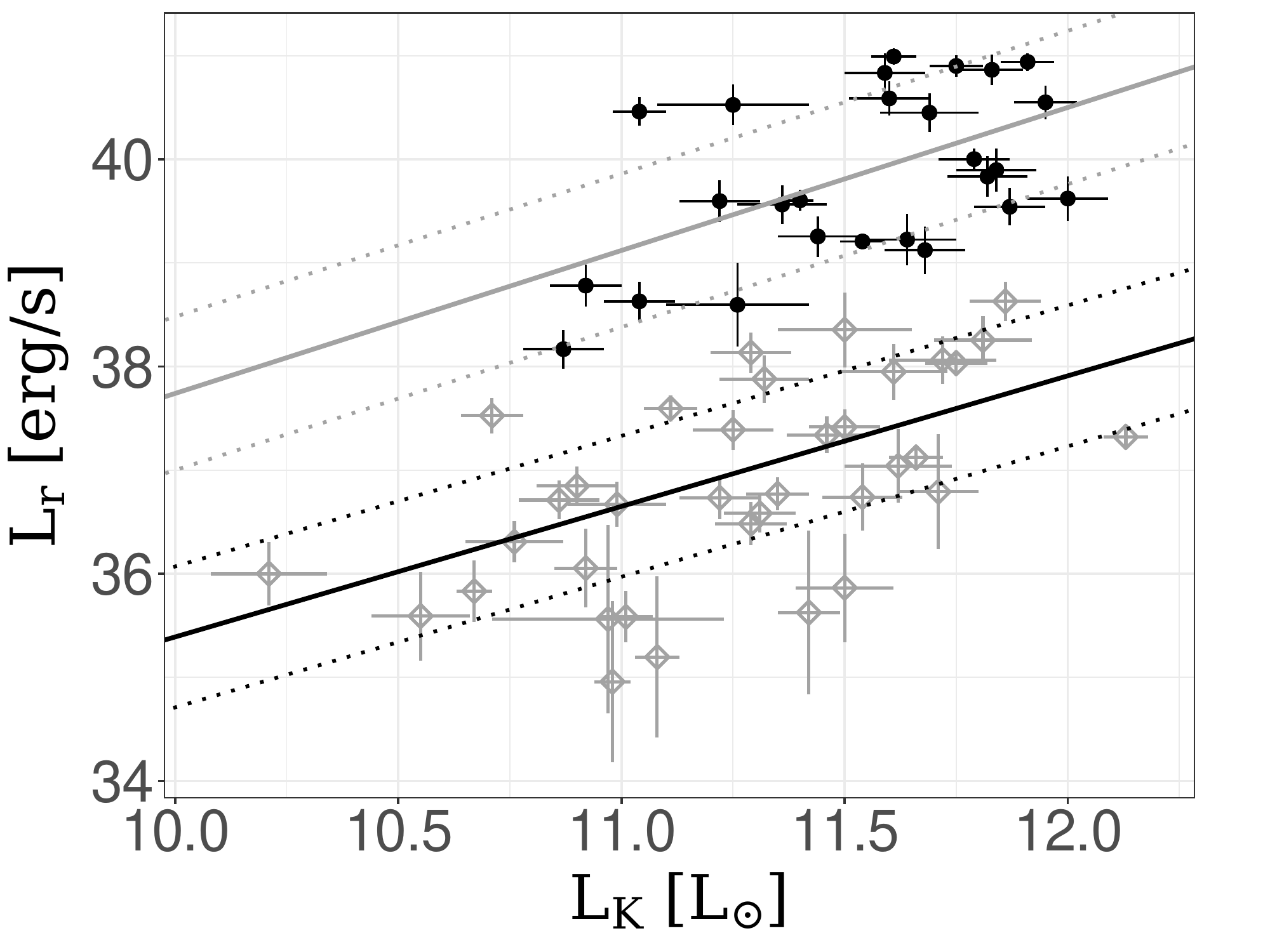}
\includegraphics[width=0.45\textwidth]{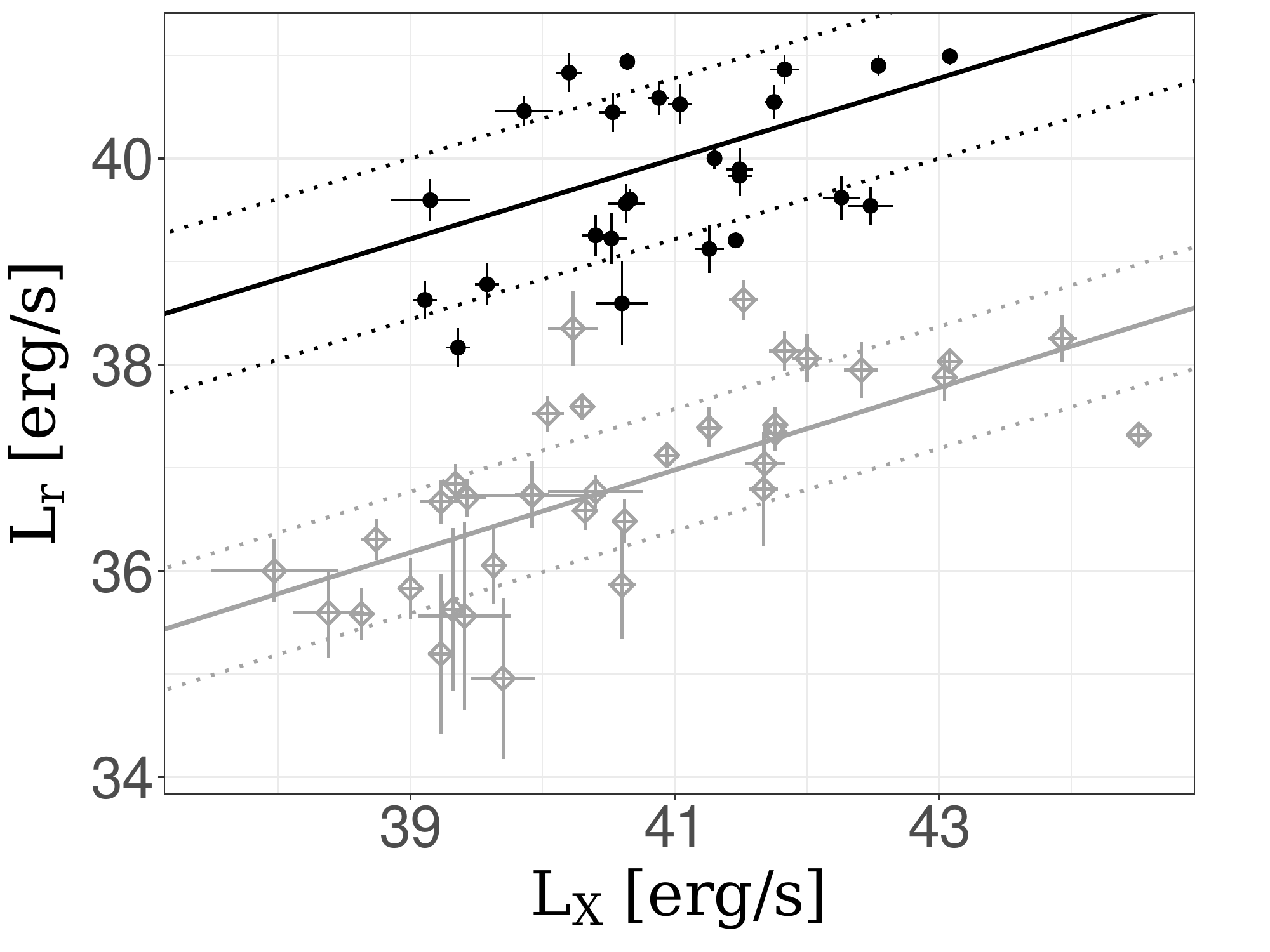}
\includegraphics[width=0.45\textwidth]{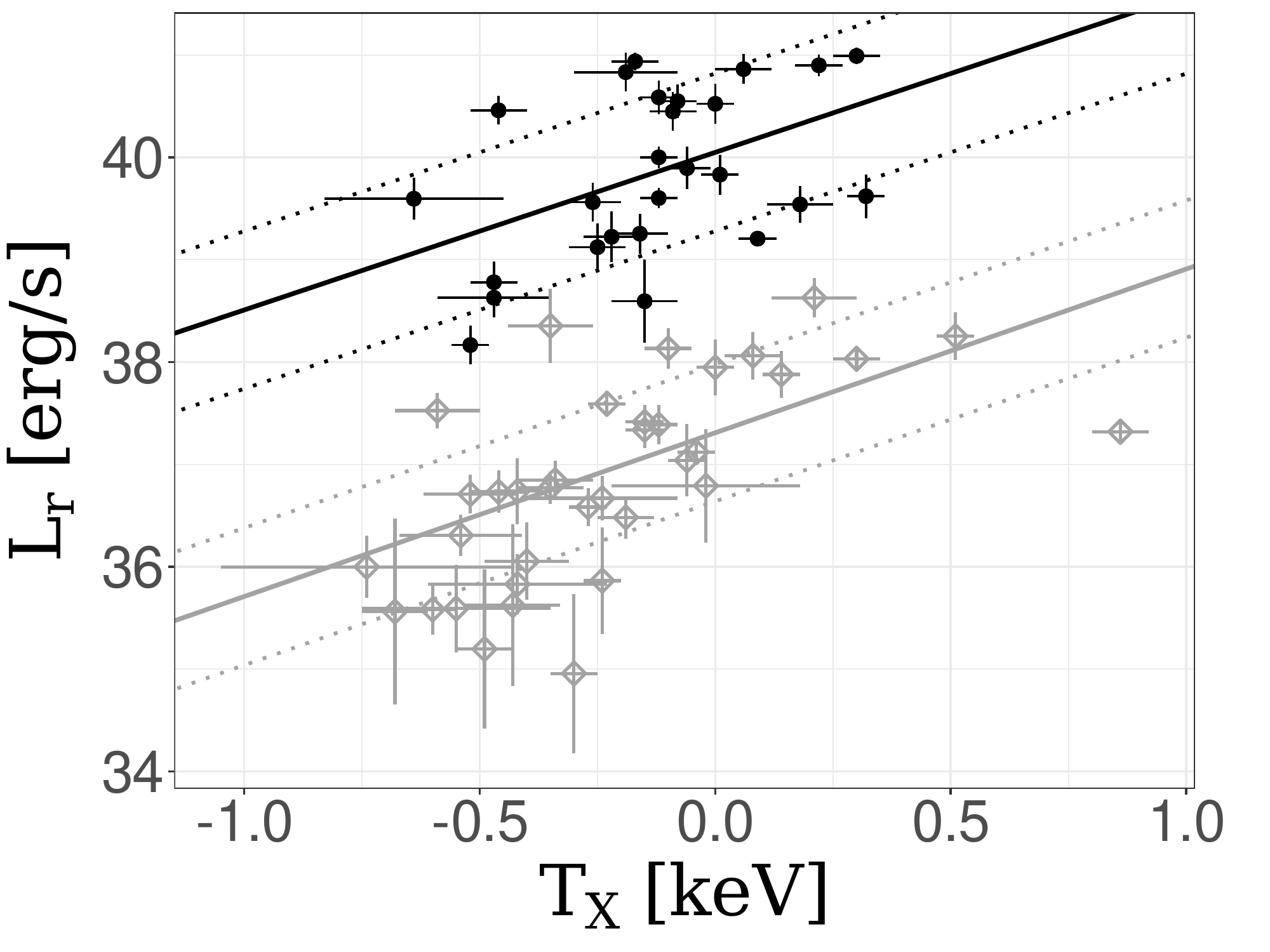}
\caption{Results of the regression analysis performed separately for the radio-dim and radio-bright sub-samples (denoted in all the panels by grey open diamonds and black filled circles, respectively) . Solid lines represent the mean trend in the sub-samples, while the $\pm  \tilde{\sigma}$ deviations are marked by dotted lines. See Table\,\ref{tab:bayes} for the median values $\tilde{\theta} =\{\tilde{a},\tilde{b};\tilde{\sigma}_{\rm int}\}$.}
\label{fig:bayes}
\end{figure*}

In Figure\,\ref{fig:3D} we present the scaling of the monochromatic radio luminosities of our early-type galaxies with the X-ray luminosities of their hot gaseous halos, coding the data point sizes to denote the logarithm of the Eddington-normalized radio luminosity, $\log L_{\rm r}/L_{\rm Edd}$, which ranges in the sample from --12.09 to --5.34. As evident from the figure, the data cluster in two distinct correlation branches on the plot, the upper and lower branch populated by sources with large and small values of the $\log L_{\rm r}/L_{\rm Edd}$ ratio, respectively. The two branches are preserved in all the representations $L_{\rm r} - L_{\rm X}$, $L_{\rm r}/L_{\rm Edd} - L_{\rm X}/L_{\rm Edd}$, and $L_{\rm r}/L_{\rm X} - L_{\rm X}/L_{\rm Edd}$, implying a well-defined borderline between the two data clusters at a certain value of $\log L_{\rm r}/L_{\rm Edd}$. Each sequence separately shows seemingly better correlation with much narrower spread when compared to the entire sample of all early-type galaxies, what is particularly evident in the bottom panel of Figure\,\ref{fig:3D}. 

Note that if the $L_{\rm X}/L_{\rm Edd}$ ratio could be considered as a proxy for the available supply of the accreting matter in the Eddington units, the $L_{\rm r}/L_{\rm X}$ parameter should correspond to the efficiency of the production of the radio emission for a given accretion reservoir. And this production efficiency seems bimodial for every $L_{\rm X}/L_{\rm Edd}$ value captured in the sample.

\subsection{Radio Luminosity Distribution}
\label{sec:bimodality}

The overall distribution of the $\log L_{\rm r}/L_{\rm Edd}$ parameter is shown in Figure\,\ref{fig:hist}. We investigate the apparent bimodality using the function \texttt{modetest} implemented in the CRAN \texttt{multimode\footnote{\url{https://www.jstatsoft.org/article/view/v097i09}}} package within the R statistical software environment \citep{Ameijeiras-Alonso2018}. This function tests the number of modes in the observed data set, for a particular specified method; we performed the analysis using the latest excess-mass method by \citet{Ameijeiras-Alonso2016, Ameijeiras-Alonso2019}. The two null hypothesis were tested, namely that the true number of modes is equal to either $mod_0=1$ or $mod_0=2$ (the alternative hypotheses being the true number of modes is greater than that specified in the null hypothesis). According to this excess mass test, we cannot rule out the null hypothesis that true number of modes in our data set is equal to $mod_0=1$, as the corresponding $p-$value\,$=0.16$; the probability of observing the distribution shown in Figure\,\ref{fig:hist} further increases up to $p-$value\,$=0.44$ when the null hypothesis is that the number of mode is equal to $mod_0=2$.

Still, it should be cautioned that our sample consisting of only 62 objects, may be too small to robustly detect the multi-modal behavior in the data. We thus followed an alternative approach, locating first the anti-mode under the assumption of the presence of two modes in our dataset. In particular, using the \texttt{locmode} function from CRAN \texttt{multimode} package, we found the location of the anti-mode at $\log L_{\rm r}/L_{\rm Edd} \simeq -8.5$ (see Figure\,\ref{fig:hist}). Based on this location, we divide our dataset into two sub-samples: 36 sources with $\log L_{\rm r}/L_{\rm Edd} \leq -8.5$, denoted hereafter as ``radio-dim'' objects, and 26 sources with $\log L_{\rm r}/L_{\rm Edd} > -8.5$, called the ``radio-bright'' ones.
First, we confirm with a simple Shapiro-Wilk normality test implemented as \texttt{shapiro.test} in the R statistical software environment, that the distributions of these sub-samples can both be approximated by Gaussian distributions. This is unlike the case of the entire sample, for which the $\log L_{\rm r}/L_{\rm Edd}$ distribution deviates from normal at the 95\% confidence level.
We used the \texttt{Mclust} function implemented in \texttt{mclust} package in R to identified the true number of modes in our dataset. We fit the Gaussian mixture models with number of components ranging from one to nine and unequal variances. Based on this analysis and according to Bayesian Information Criterion (BIC) we found that two mode Gaussian mixture model provides the best description of our data.
Next, in order to further test the bimodal hypothesis, we use a maximum likelihood method, implemented in the \texttt{mle2} function of \texttt{bbmle} package in R to fit
%, to compare two models fitted to the $\log L_{\rm r}/L_{\rm Edd}$ distribution in the whole sample, namely (A) a model consisting of a single normal distribution, and (B)
a mixture of two Gaussians $f(x) = p\times f_1 + (1-p)\times f_2$, where $f_1$ and $f_2$ denote normal distributions, and $p$ is the mixture weight. 
%We observe that the two-Gaussian mixture model (B) provides a better description of our data according to the Akaike Information Criterion (function \texttt{AIC} in the \texttt{stats4} package). 
The parameters of the mixture distribution $f(x)$ obtained through the maximum likelihood method are the means $\mu_1=-10.02\pm0.17$ and $\mu_2=-7.07\pm0.22$, and the standard deviations $\sigma_1=0.81\pm0.13$ and $\sigma_2=0.87\pm0.17$, with the mixture weight of $p=0.58\pm0.07$. Based on those values, we use the criterium of \citet{Schilling2002}, which states that {\it ``The Mixture Density $f(x) = p\times f_1 + ( 1-p)\times f_2$ Is Bimodal If and Only If $|\mu_2-\mu_1|$ Exceeds $(\sigma_1 + \sigma_2)$ Times the Value Indicated"}, where the ``Value Indicated'' depends on the standard deviation ratio and the weight (see Table\,2 therein). In our case, $\sigma_1/\sigma_2\approx0.95$ and $p\approx0.6$, implying according to \citet{Schilling2002} that our mixture density is bimodal if and only if $|\mu_2-\mu_1|>1.24\times (\sigma_1 + \sigma_2)$. This condition is satisfied for the values of the model parameters obtained through the maximum likelihood method (within $1\sigma$ uncertainty ranges) as provided above. This supports the presence of the two modes in the analyzed dataset.

\subsection{Regression Analysis}
\label{sec:bayes}

In real astronomical observations of a given sample of sources, besides obvious dependencies of certain measured parameters, there might be additional ones, which are hard to disentangle/identify, but which result in a wider spread in the regression studies. One can however quantitatively incorporate this spread into the analysis, through the Bayesian approach. Unlike traditional methods, such as the least-square fitting, in the Bayesian approach the intrinsic spread is fitted simultaneously with other free parameters of the model.

\begin{deluxetable}{ccccc}[!th]
\tabletypesize{\footnotesize}
%\tabletypesize{\scriptsize}
\tablecaption{Summary of the Bayesian linear regression analysis, and the Pearson's correlation analysis, for the radio-dim and radio-bright sub-samples.}
\label{tab:bayes}
\tablewidth{0pt}
\tablehead{
\colhead{Relation/sub-sample} & \colhead{$\tilde{a}$}  & \colhead{$\tilde{b}$} & \colhead{$\tilde{\sigma}_{\rm int}$}& \colhead{$\rho$}\\
\colhead{(i)} & \colhead{(ii)} & \colhead{(iii)} & \colhead{(iv)} & \colhead{(v)}
}
\startdata
\\
$L_{\rm r} - M_{\rm BH}$ & & & & \\
radio-bright & $35.96^{+3.09}_{-2.54}$ & $0.44^{+0.29}_{-0.35}$ & $0.84^{+0.16}_{-0.12}$ & 0.25 \\
radio-dim & $29.55^{+1.51}_{-1.45}$ & $0.84^{+0.16}_{-0.17}$ & $0.59^{+0.12}_{-0.10}$ & 0.55 \\
\\
$L_{\rm r} - M_{\rm bulge}$& & & &  \\
radio-bright &$ 26.98^{+3.82}_{-4.36}$ &$1.12^{+0.38}_{-0.33}$ & $0.71^{+0.14}_{-0.11}$ & 0.53\\
radio-dim & $26.54^{+2.35}_{-2.63}$& $0.93^{+0.23}_{-0.21}$ & $0.69^{+0.13}_{-0.10} $ & 0.52 \\
\\
$L_{\rm r} - \sigma_{\rm vel}$ & & & & \\
radio-bright & $36.87^{+4.35}_{-4.57}$& $1.24^{+1.89}_{-1.82}$ & $0.88^{+0.16}_{-0.13}$ &0.12\\
radio-dim & $25.65^{+2.89}_{-2.57}$ & $4.81^{+1.09}_{-1.22}$ & $0.69^{+0.12}_{-0.10}$ & 0.54 \\
\\
$L_{\rm r} - L_{\rm K}$ & & & &  \\
radio-bright & $23.94^{+6.34}_{-5.58}$ & $1.38^{+0.48}_{-0.55}$ & $0.74^{+0.14}_{-0.11}$ & 0.50\\
radio-dim & $22.79^{+3.25}_{-3.40}$ & $1.26^{+0.30}_{-0.29}$ & $0.68^{+0.12}_{-0.10}$ & 0.52\\
\\
$L_{\rm r} - L_{\rm X}$ &   &   &   &   \\
radio-bright &$24.01^{+5.22}_{-6.09}$& $0.39^{+0.15}_{-0.13}$&$0.78^{+0.15}_{-0.11} $& 0.47\\
radio-dim &$20.58^{+2.78}_{-2.61}$ & $0.40\pm 0.06$ &$0.59^{+0.12}_{-0.09}$ & 0.66\\
\\
$L_{\rm r} - T_{\rm X}$ &   &   &   &   \\
radio-bright &$40.05 \pm  0.18$& $1.54^{+0.68}_{-0.66}$&$0.77^{+0.15}_{-0.11} $& 0.43\\
radio-dim &$37.31\pm  0.15$ & $1.60^{+0.38}_{-0.36}$ &$0.67^{+0.13}_{-0.10}$ & 0.56 \\
\\
\enddata
\tablecomments{Col.~(ii)-(iv) --- The median values of the parameter PDFs, $\tilde{\theta} =\{\tilde{a},\tilde{b};\tilde{\sigma}_{\rm int}\}$, with the associated marginalized $68.3\%$ uncertainties. Col.~(v) --- Pearson's product-moment correlation coefficient $\rho$, calculated with the measurement uncertainties taken into account, as described in Appendix\,\ref{A:all}.}
\end{deluxetable}

In the framework of this approach, we investigate the univariate linear regression using the APEMoST\footnote{Automated Parameter Estimation and Model Selection Toolkit; \url{ http://apemost.sourceforge.net/}, 2011 February.} algorithm \citep{Gruberbauer2009}, for the response variable $Y = L_{\rm r}$, and the predictor variables either $X=M_{\rm BH}$, $M_{\rm bulge}$, $\sigma_{\rm vel}$, $L_K$, $L_{\rm X}$, or $T_{\rm X}$. We assume the linear trend $Y = a+b \, X+\epsilon$, with a normally distributed noise $\epsilon \sim \mathcal{N}(0,\sigma)$. The standard deviation of the noise distribution, $\sigma$, can be expressed as the combined uncertainty
\begin{equation}
\sigma=\sqrt{\sum\limits_{i=1}^{N} \sigma_{\rm int}^2+\sigma_{Y_i}^2+(b\times \sigma_{X_i})^2 }
\end{equation}
in the sample of $N=62$ objects, where $\sigma_{\rm int}$ is the intrinsic spread, while $\sigma_{X_i}$ and $\sigma_{Y_i}$ denote the uncertainties associated the respective measurements, $X_i$ and $Y_i$. We seek to determine multidimensional probability density function (PDF) of the parameter set $\theta= \{a,b, \sigma_{\rm int}\}$. 

In the regression analysis, we following closely \citet{Ostorero17} and \citet{wojtowicz21}, performing $2\times 10^6$ MCMC iterations and twenty chains to ensure a sufficiently complete sampling of the parameter space. We assume flat uninformative priors for the $a$ and $b$ parameters, with the parameter space boundaries set to  $[-100,100]$. The prior of the intrinsic spread $\sigma_{\rm int}$, which is always a positive parameter, is given by the PDF that describes a variate with mean $r/\mu$ and variance $r/\mu^2$, namely,
\begin{equation}
P\!\left(\sigma_{\rm int}|\mathcal{M}\right)=\frac{\mu^r}{\Gamma(r)} \, x^{r-1} \, \exp(-\mu x) \, ,
\end{equation}
where $x= 1/\sigma_{\rm int}$, and $\Gamma(r)$ is the Euler Gamma function; in our calculations, we set $r=\mu=10^{-5}$ and the variability interval boundaries $[0.01,1000]$. The random number generator was set with bash command \texttt{GSL\_RNG\_TYPE="taus"} and the initial seed of the random number generator was set with \texttt{GSL\_RNG\_SEED=\$RANDOM}. 

We perform our regression analysis separately for the radio-dim and radio-bright sub-samples. For these, we calculate also the standard Pearson's product-moment correlation coefficients, $\rho$,  taking into account the measurement uncertainties (see Appendix\,\ref{A:all}). The resulting values are summarized in Table\,\ref{tab:bayes}, where we also list the medians of the parameter PDFs following from the regression analysis, $\tilde{\theta} =\{\tilde{a},\tilde{b};\tilde{\sigma}_{\rm int}\}$, with the associated uncertainties. The regression analysis results are also visualized in Figure\,\ref{fig:bayes}

The main conclusion following from the correlation and regression analyses presented above, is that the radio-dim sub-sample correlates in general better with the central black hole mass, or with the other optical and X-ray parameters of the galaxies and their halos, than the radio-bright sub-sample. The difference is the most pronounced in the two particular relations $L_{\rm r} - M_{\rm BH}$ and $L_{\rm r} - \sigma_{\rm vel}$: while for the radio-dim sub-sample we find strong ($\rho \simeq 0.55$) correlations $L_{\rm r} \propto M_{\rm BH}^{0.84\pm  0.2}$ and $L_{\rm r} \propto \sigma_{\rm vel}^{4.81\pm  1.2}$, for the radio-bright sample the analogous correlations appear much looser and, overall, weaker ($\rho < 0.25$). This suggests that, while in the case of radio-dim sources the observed radio emission follows general scalings with the other parameters of the systems, parameters which are tightly linked through co-evolution of SMBHs with their galactic hosts \citep[with the $M_{\rm BH} - \sigma_{\rm vel}$ relation being the most fundamental one; see][and references therein]{kormendy2013,bosch2016}, in the case of radio-bright objects the observed radio emission uncouples, at least to some extent, from those parameters and scaling relations.
\begin{figure}[th!]
    \centering
\includegraphics[width=\columnwidth]{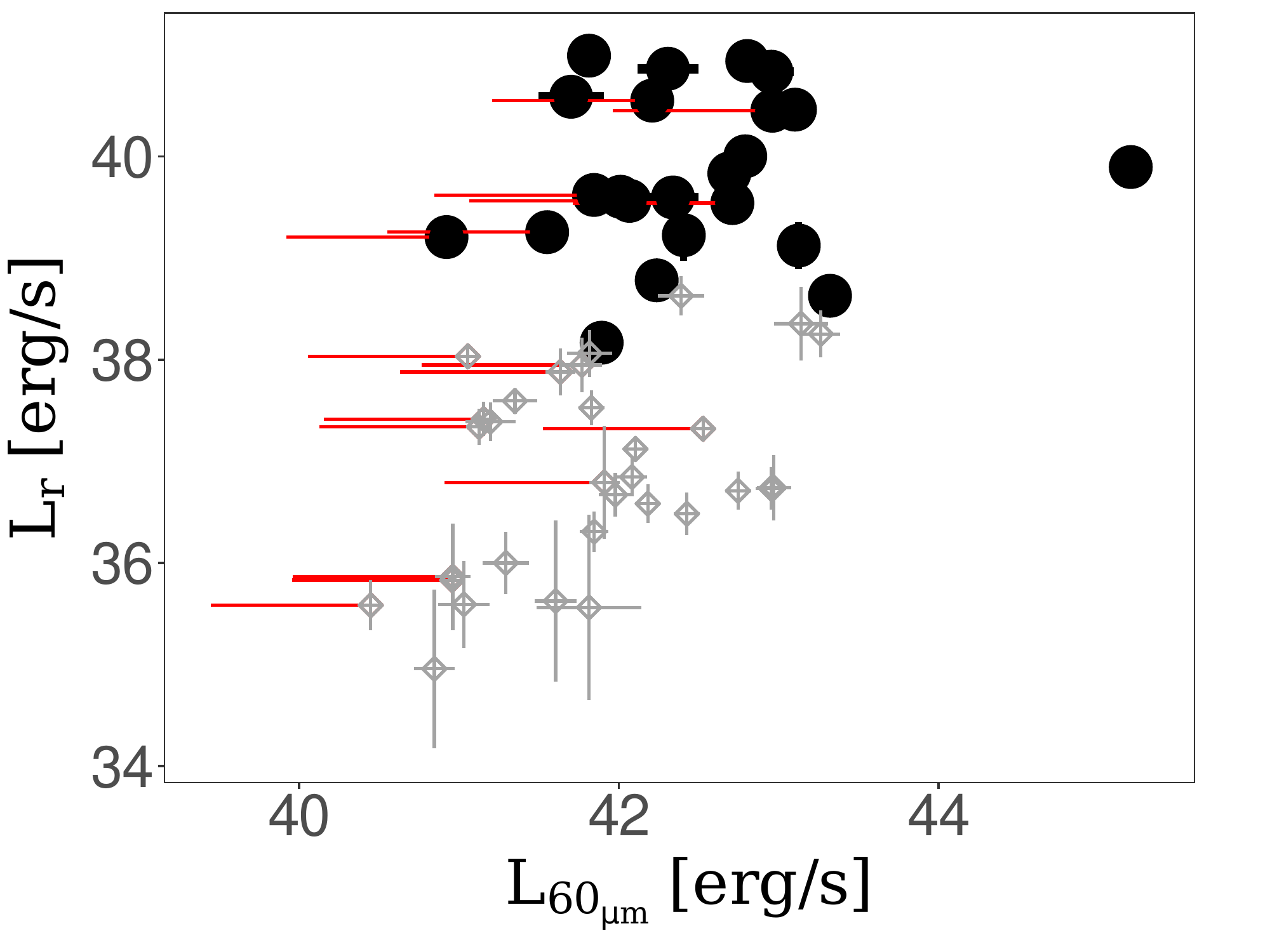} 
\includegraphics[width=\columnwidth]{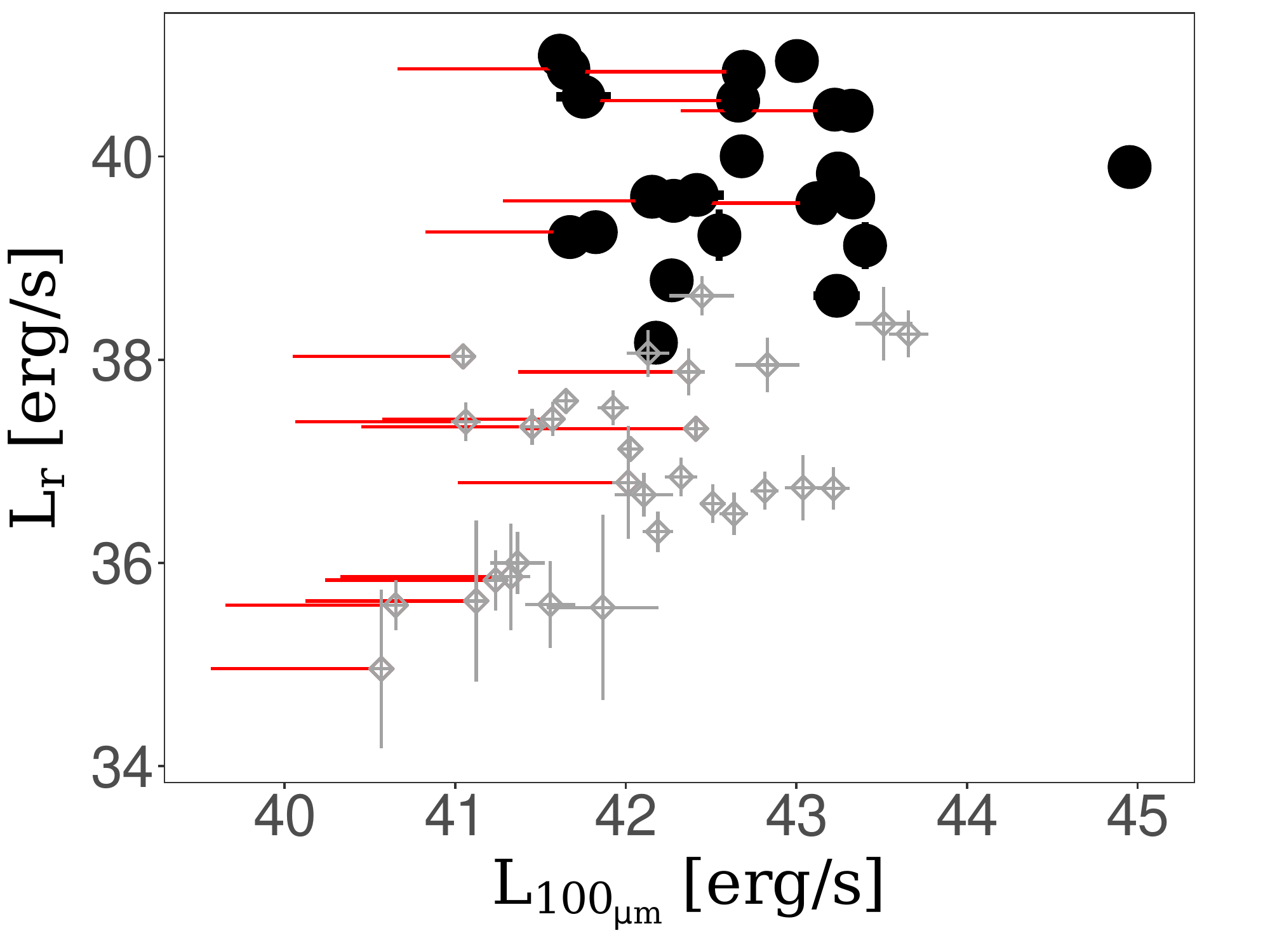} 
\caption{Scaling of the VLA 1.4\,GHz luminosities with the IRAS 60\,$\mu$m and 100\,$\mu$m luminosities for the analyzed early-type galaxies (upper and lower panels, respectively). Upper limits are indicated by red lines. Radio-dim sources are denoted by grey open diamonds, and radio-bright objects by black filled circles.}
 \label{fig:IRAS}
\end{figure}

\subsection{Arcsecond Radio Morphologies}
\label{sec:VLASS}

In Appendix\,\ref{A:VLASS}, we present the $2.5^{\prime\prime}$-resolution VLASS images at 3\,GHz for 25 radio-dim galaxies, noting that for the remaining 11 objects from this sub-sample, the VLASS maps display only featureless noise. Out of those 25 targets, 19 solely consist of unresolved radio cores coinciding with the galactic centres. Six radio-dim sources show some extension in the VLASS data, although the emerging structures are all rather compact and weak; moreover, only one of those, NGC\,3665, display a clear double-jet morphology. Interestingly, NGC\,3665 is characterized by the Eddington-normalized monochromatic radio luminosity $\log L_{\rm r}/L_{\rm Edd}  = -8.51 \pm  0.37$, exactly at the borderline between radio-dim and radio-bright objects (see Section\,\ref{sec:bimodality} above). 

The other five radio-dim galaxies that display some extension in the VLASS maps are: (i) NGC\,1407 --- with the jet-like feature embedded within a large-scale diffuse radio halo, proposed to be a relic from the previous cycle of the source activity \citep{Giacintucci12}; (ii) NGC\,1600 --- with the double lobe-like structure reminiscent of FR-II radio galaxies, but without an identified radio core \citep[see also][]{Grossova22}; (iii) NGC\,3842 --- with the compact elongated structure possibly reminiscent of Wide-Angle Tail (WAT) radio galaxies, and a very faint arcsecond radio core \citep[see][]{Liuzzo10}; (iv) NGC\,4636 --- with the compact and weak jet-like structure we see on the VLASS map, which is also present at lower radio frequencies \citep{Giacintucci11}; (v) NGC\,4472 --- with the elongated lobe-like structure followed by low surface-brightness tails extending up to $\sim$\,arcmin scale \citep{Grossova22}. 

All in all, we conclude that the overwhelming majority of radio-dim sources in our sample of early-type galaxies is either undetected (11/36) or unresolved (19/36) at GHz frequencies in the VLASS survey with arcsec-scale resolution. Only several (6/36) display some elongated features on the VLASS maps, and only one of those is characterized by a clear morphology consisting of a prominent radio core with two-sided symmetric jets; this one outlier, NGC\,3665, is in fact located at the borderline between radio-dim and radio-bright populations.

We also note that, for a few other sources from the radio-dim sub-sample, some weak  extension/elongation could be noted in the deep VLA data presented in the literature, namely for NGC\,4649 and NGC\,4636 \citep{Stanger86}; likewise, some structure consisting of multiple compact radio components could be spotted on milli-arcsec scale in NGC\,4552 \citep{Nagar02} and NGC\,5846 \citep{Filho04}. These structures, even though consistent with some kind of a nuclear radio activity, do not provide however clear cases for the presence of relativistic jets in the systems. 

In Appendix\,\ref{A:VLASS}, we also present the VLASS images for 24 radio-bright galaxies. We note that the southernmost radio-bright object NGC\,5128/Centaurus\,A, which is the famous radio galaxy with complex multi-scale jet morphology \citep[e.g.,][]{Morganti99}, is not covered by the survey. In addition, the radio-bright source NGC\,6086 is not detected in the VLASS maps; this brightest cluster galaxy in Abell\,2162, displays a double-lobed morphology on the NVSS image, but with no radio core \citep[see][]{Liuzzo10}, and so its non-detection in the VLASS map may is not surprising. Out of the remaining 24 objects shown in the Appendix\,\ref{A:VLASS}, 16 are clearly extended jet sources, with radio morphologies representing various FR types; among them are the famous FR\,I radio galaxies NGC\,4486/M\,87 or NGC\,315.\footnote{We note that the VLASS map for the giant radio galaxy NGC\,6251 includes some artefact features, resulting in a confusing/messy appearance of the target.} In addition to those, two radio-bright sources are clearly extended on the VLASS maps, but display unclear/amorphous morphologies. These are NGC\,5419, a probable radio relic associated with a poor cluster \citep{Subrahmanyan03}, and NGC\,6240S, a peculiar post-merger ULIRG with double nucleus \citep[see][]{Komossa03,Colbert1994}.

There are also six radio-bright sources which appears unresolved on the presented VLASS maps. Among them, there is the radio galaxy at the center of the Abell\,1836 cluster, PKS\,B1358--113, with clear FR\,II large-scale morphology \citep[see][]{Stawarz14}. Three sources in this group display jet structures as well, but on milli-arcsecond scales, possibly accompanied by slightly resolved features on the scale of a few arcseconds, namely NGC\,1052 \citep{Nakahara20}, NGC\,2110 \citep{Mundell00}, and NGC\,4278 \citep{Bondi04}. Finally, NGC\,5077 seems truly unresolved on both arcsec and milli-arcsecond scales, while NGC\,1459 is missing any higher-resolution radio imaging data.

To sum up, we conclude that the vast majority (16/26) of radio-bright sources in our sample of early-type galaxies display complex large-scale morphologies at GHz frequencies in the VLA survey with arcsec-scale resolution, characteristic for radio galaxies of various FR types, undoubtedly indicating ongoing jet activity. Several from the remaining radio-bright objects (6/26) are extended on either larger scales, or smaller (milli-arcsec) scales, again consistent with the presence of radio jets. There are only a few unresolved sources in this group, and the two extended ones for which the arcmin-scale radio structure may not be related to the jet activity (NGC\,5419 and NGC\,6240S). This is in stark contrast to radio-dim population.

\begin{figure}[th!]
    \centering
\includegraphics[width=0.98\columnwidth]{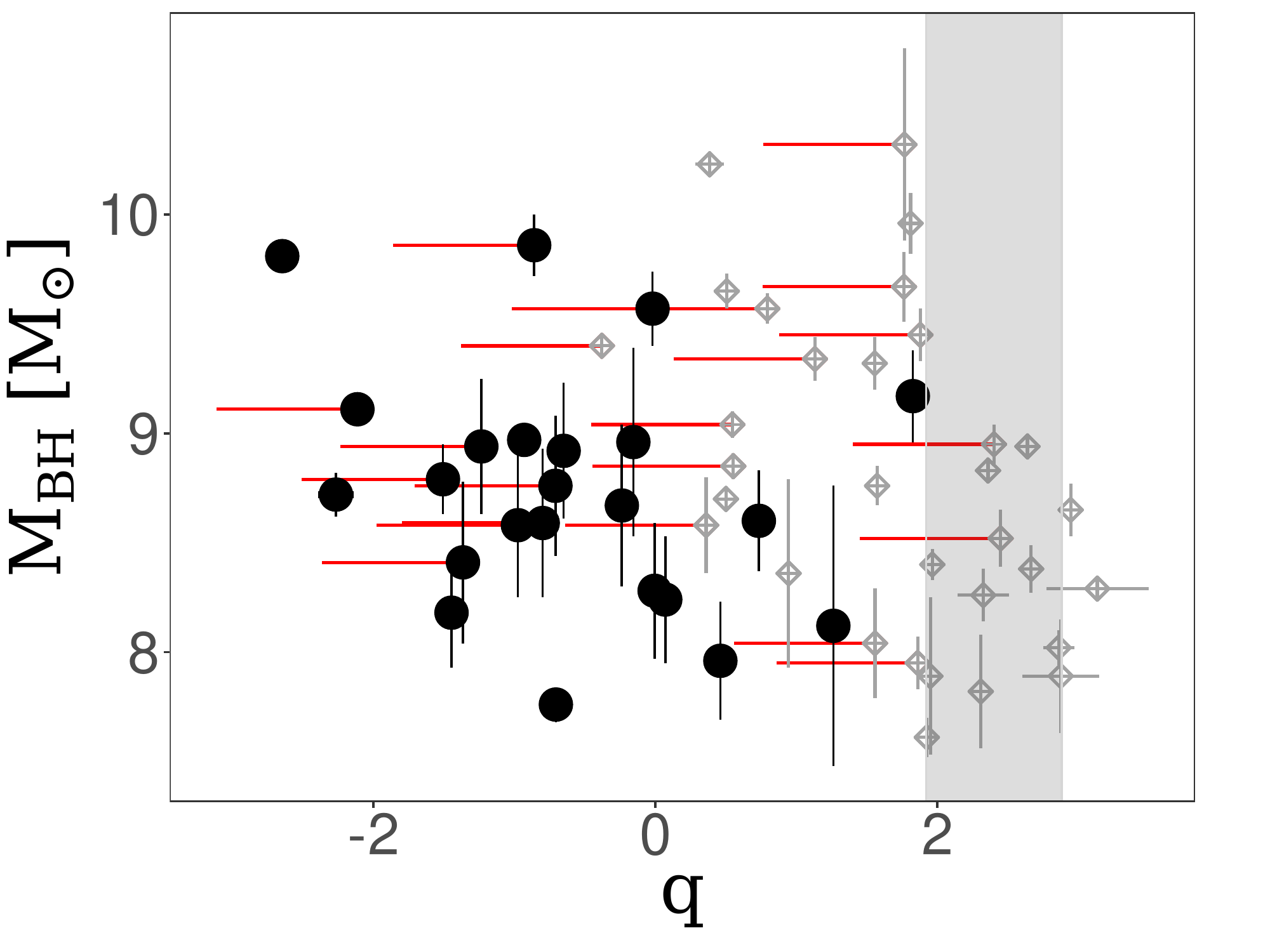} 
\includegraphics[width=0.98\columnwidth]{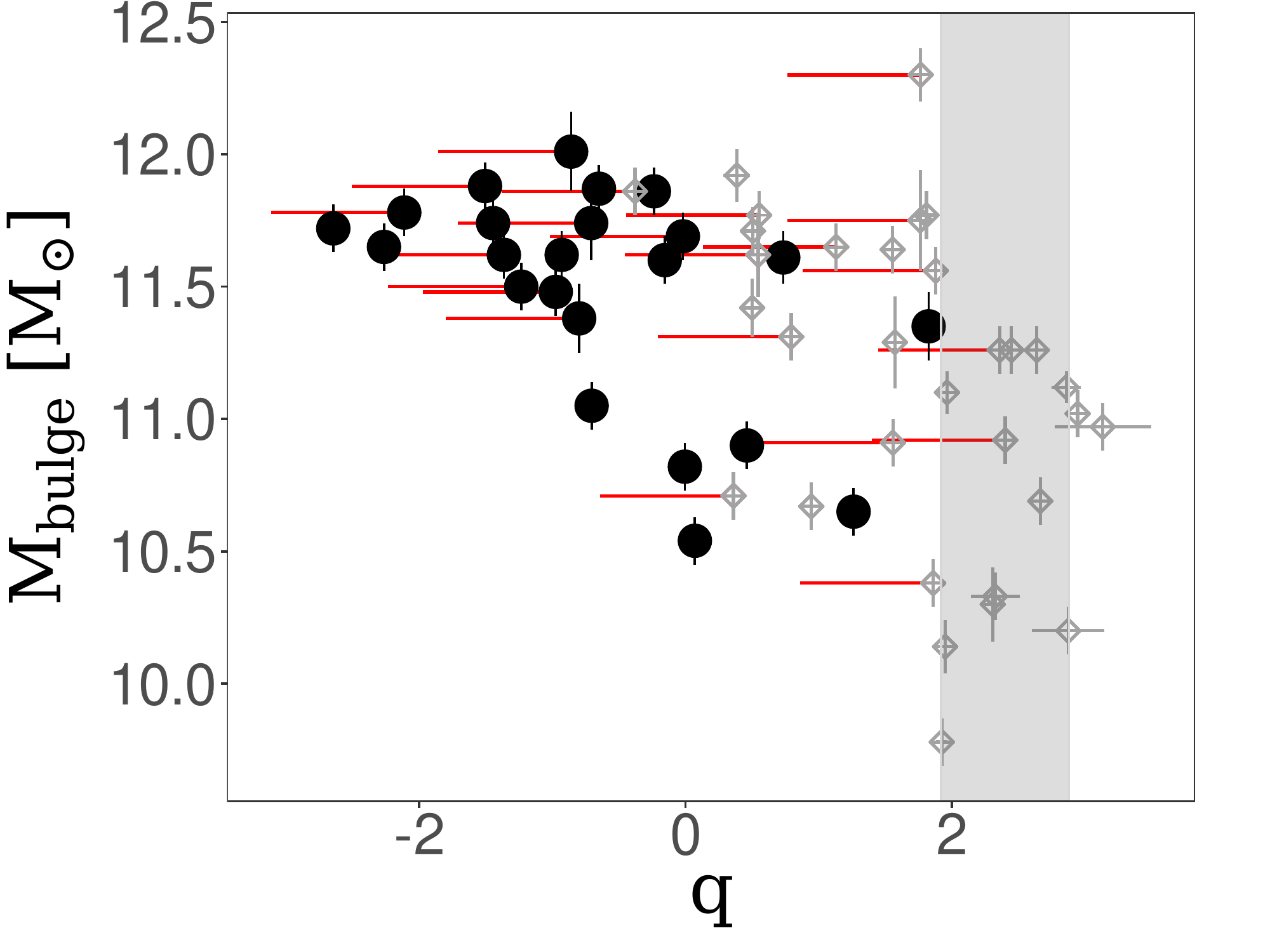} 
\includegraphics[width=0.98\columnwidth]{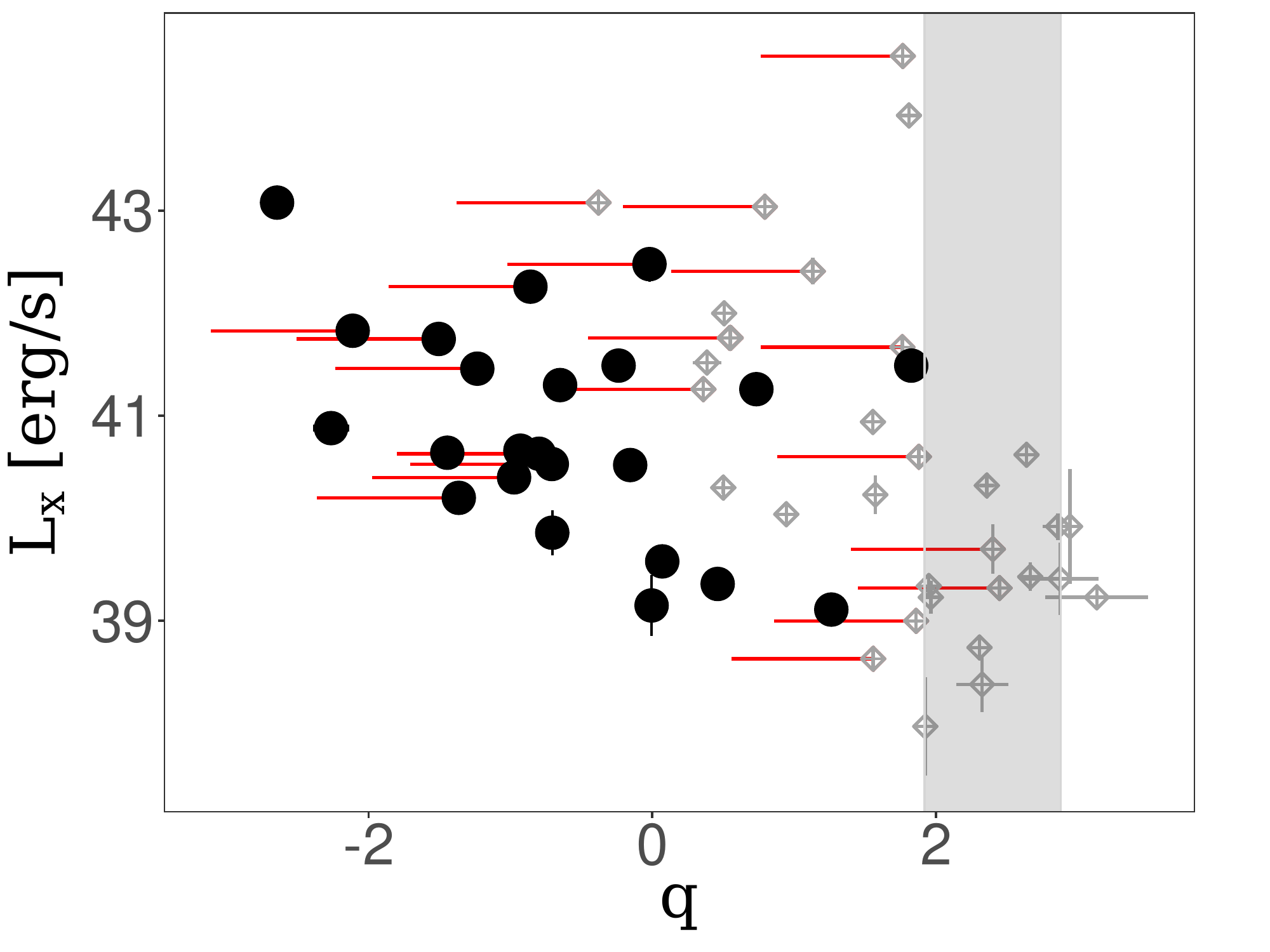} 
\caption{Dependence of the FIR-radio energy flux ratio $q$ on the black hole mass $M_{\rm BH}$, the bulge mass $M_{\rm bulge}$, and the large-scale X-ray luminosity $L_{\rm X}$, for the early-type galaxies analyzed in this paper (top, middle, and bottom panels, respectively). The shaded areas in all the panels denote the range corresponding $q  = 2.40 \pm  2 \sigma_q$ with $\sigma_q = 0.24$ \citep{Ivison10}. Upper limits are indicated by red lines. Radio-dim sources are denoted by grey open diamonds, and radio-bright objects by black filled circles.}
 \label{fig:q}
\end{figure}

\begin{figure}[th!]
    \centering
\includegraphics[width=\columnwidth]{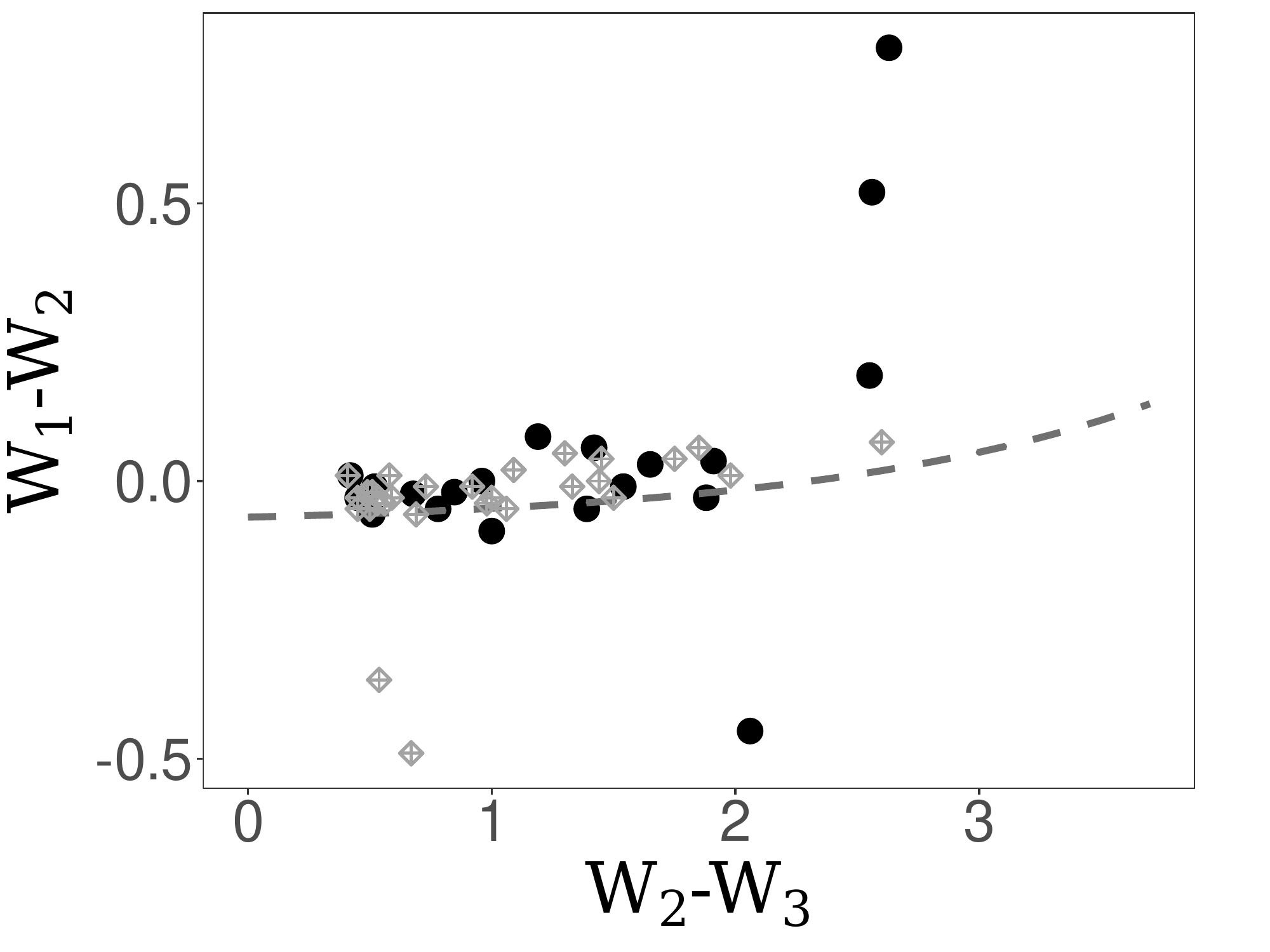} 
\caption{WISE colours W2--W3 vs. W1--W2 for the analyzed early-type galaxies. Dashed curve corresponds to the ``star formation sequence'' by \citet{Jarrett19} (see Equation\,\ref{eq:WISE} for the functional form of the sequence). Radio-dim sources are denoted by grey open diamonds, and radio-bright objects by black filled circles.}
 \label{fig:WISE}
\end{figure}

\section{Infrared Diagnostics}
\label{sec:IR}

In Figure\,\ref{fig:IRAS}, we present the scaling of the VLA 1.4\,GHz luminosities for the analyzed sources, with the IRAS 60\,$\mu$m and 100\,$\mu$m luminosities (or upper limits), whenever available. For the IRAS-detected sources, the radio and FIR luminosities do not display any correlation. We note that the outlier in the plot with the extreme values of the FIR luminosities at the level of $\sim 10^{45}$\,erg\,s$^{-1}$, is the one Ultraluminous Infrared Galaxy (ULIRG) in the sample with the LINER-type nucleus, NGC\,6240S \citep[see][]{Veilleux09}.

In the case the observed radio emission is dominated by the starforming regions within the ISM, one expect to see a clear FIR-radio correlation, corresponding to a well-defined and universal value of the logarithm of the energy flux ratio
\begin{equation}
q \equiv \log \frac{F_{\rm FIR}/3.75\times 10^{12}\,{\rm W\,m^{-2}}}{S_{\rm 1.4\,GHz}/{\rm W\,m^{-2}\,Hz^{-1}}}
\label{eq:q}
\end{equation}
as introduced by \citet{Helou85}, where $F_{\rm FIR}/{\rm W\,m^{-2}} = 1.26 \times 10^{-14} \, [2.58 \times S_{\rm 60\,\mu m}/{\rm Jy} + S_{\rm 100\,\mu m}/{\rm Jy}]$. We calculate the $q$ values for the early-type galaxies analyzed here, and in Figure\,\ref{fig:q} we present the resulting scaling of this parameter with the black hole and galactic bulge masses, as well as with the large-scale X-ray luminosities. The shaded areas in all the panels denote the range corresponding to the median $q  = 2.40$ with $\pm  2 \sigma_q$ dispersion, where $\sigma_q = 0.24$, as established by \citet{Ivison10} for the galaxies in the GOODS-North field, based on the {\it Herschel} and VLA observations \citep[see also in this context][]{Magnelli15,Delhaize17,Giulietti22}. As follows, all the radio-bright sources, and over half of the radio-dim ones, are over-luminous in radio with respect to the FIR-radio correlation. That is, the bulk of the radiative output of these systems at radio frequencies cannot be ascribed to starforming ISM. Moreover, as can be noted in the figure, the objects with larger values of $M_{\rm bulge}$, and possibly also $L_{\rm X}$, tend to be ``radio-excess'', in the sense $q < 1.92$, although one cannot claim here any statistically significant correlation.
 
We note that the radio-dim sources which seem to follow relatively closely the FIR-radio correlation, are predominantly (though not exclusively) those targets which are not detected in the VLASS, but have measured VLA radio fluxes at $45^{\prime\prime}$ resolution (see Section\,\ref{sec:VLASS} above).

\begin{figure}[th!]
    \centering
\includegraphics[width=\columnwidth]{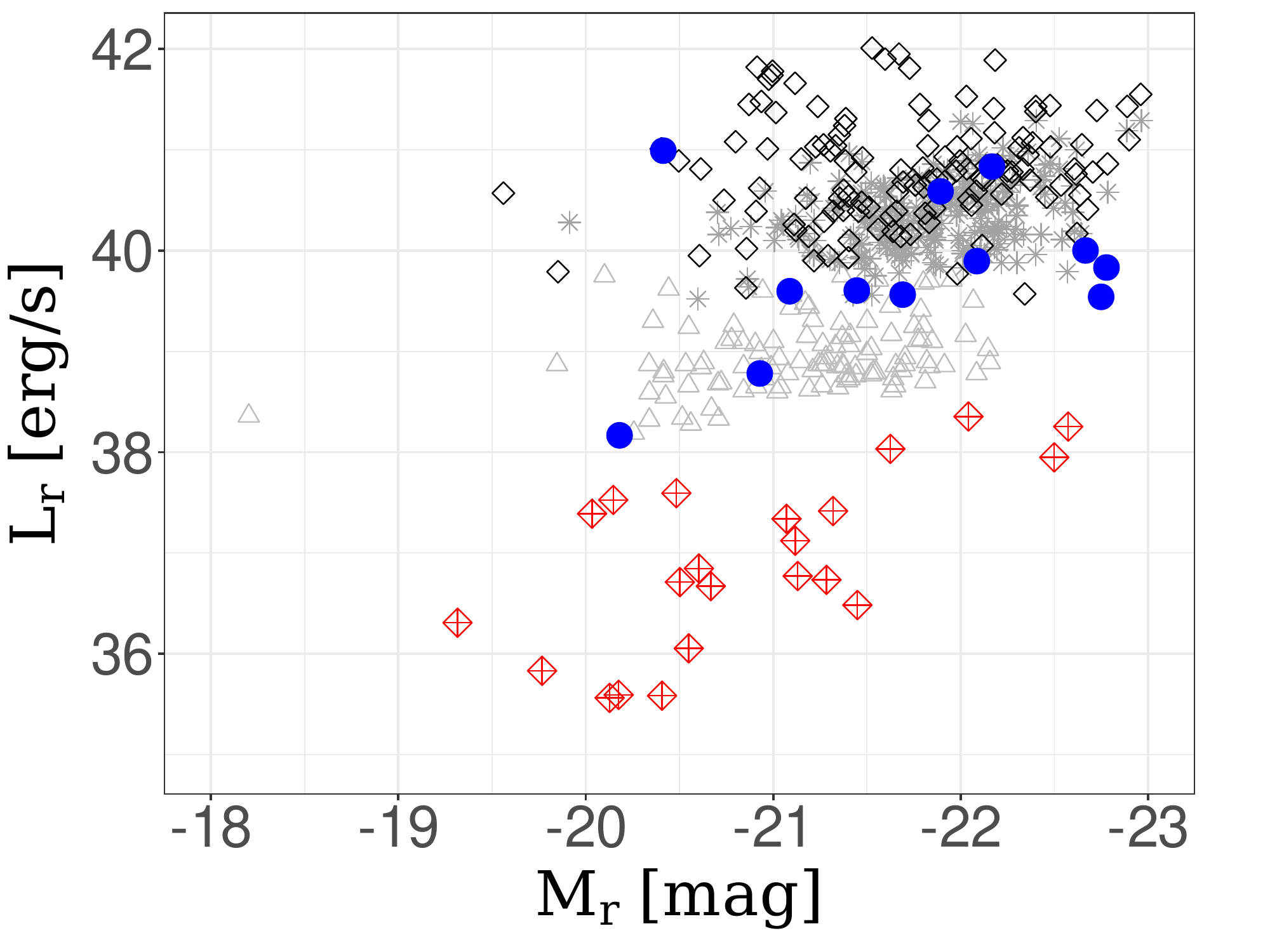} 
\caption{Absolute r-band magnitude $M_{\rm r}$ vs. 1.4\,GHz luminosity $L_{\rm r}$ for radio-detected early-type galaxies. The radio-dim and radio-bright sources analyzed in this paper are denoted by red open diamonds and blue circles, respectively, FR\,0s \citep{Baldi18} are shown as open grey traingles, FR\,Is \citep{Capetti17a} as grey stars, and FR\,IIs \citep{Capetti17b} as open black squares.}
 \label{fig:FR}
\end{figure}

Finally, in Figure\,\ref{fig:WISE} we present the WISE colours for the analyzed early-type galaxies, W2--W3 vs. W1--W2. As shown, our targets follow closely the tight ``star formation sequence'' approximated by the functional form
\begin{equation}
({\rm W1-W2}) = 0.015 \times e^{({\rm W2-W3})/1.38} - 0.08
\label{eq:WISE}
\end{equation}
as introduced by \citet{Jarrett19}, and denoted in the figure by dashed curve, implying low starformation rates characteristic of spheroids (W2--W3\,$<$\,1.5) and intermediate disks (1.5\,$<$\,W2--W3\,$<$\,3.0). The two outliers here are the low-excitation Seyfert galaxy NGC\,5128/Centaurus\,A \citep[see][]{Marconi00}, and Seyfert type 2 galaxy NGC\,2110 \citep{Evans07}, which both approach the region of the WISE colour-colour diagram that is occupied by AGN \citep[see, e.g.,][]{Jarrett11,Stern12,Mateos12}, i.e., more precisely, by the sources with the central MIR emission dominated by the AGN component (in particular, by circumnuclear hot dusty torus, re-processing to infrared the UV/X-ray emission of the matter accreting onto SMBHs). All the other targets analyzed here are therefore ISM-dominated in the infrared range, and so the FIR diagnostics discussed above regarding the origin of the observed radio emission, are robust, in the sense that they robustly identify the ``radio-excess'' objects.

\section{Discussion}
\label{sec:discu}

Figure\,\ref{fig:FR} presents the distribution of various types of radio-detected early-type galaxies on the plane defined by the absolute r-band magnitude of the host, $M_{\rm r}$, and the monochromatic 1.4\,GHz luminosity $L_{\rm r}$, including the radio-dim and radio-bright sources analyzed in this paper, as well as FR\,0 radio galaxies from \citet{Baldi18}, FR\,I radio galaxies from \citet{Capetti17a}, and finally FR\,II radio galaxies from \citet{Capetti17b}, for a comparison. As follows, our radio-bright sources, with radio luminosities within the range $\sim 10^{38} - 10^{41}$\,erg\,s$^{-1}$, overlap with the FR radio galaxies in the diagram, as expected (see Section\,\ref{sec:VLASS} above). Our radio-dim sources, on the other hand, populate lower radio luminosities within the range $\sim 10^{35} - 10^{38}$\,erg\,s$^{-1}$, hardly explored in a systematic manner before. We note in this context that, for example, 9 out of 14 ``miniature radio galaxies'' of \citet{Baldi09}, are included in our radio-dim sub-sample, and only one (NGC\,4278) in our radio-bright sub-sample.

Figure\,\ref{fig:FR} reveal also that the division between various types of radio-detected early-type galaxies, including the radio-dim/radio-bright division discussed in this paper, may be a function of the optical luminosity of the host galaxy, the trend which is well established for the FR\,I/FR\,II division \citep{Ledlow96}. This is related to distinct $L_{\rm r} - L_{\rm K}$, and $L_{\rm r} - M_{\rm buldge}$ correlations followed by the radio-dim and radio-bright populations (see Section\,\ref{sec:bayes} above).

The observed radio emission of radio-bright sub-sample, just like in FR radio galaxies, originates therefore in relativistic jets, launched by the central engines with different efficiencies and in a wider range of kinetic powers, but nonetheless undoubtedly present in all the targets. The question emerges, however, what is the origin of the radio emission in radio-dim population of early-type galaxies. 

For some, in particular those obeying the FIR--radio correlation (see Section\,\ref{sec:IR}), radio synchrotron photons can, in principle, be produced in starforming regions, corresponding to the observed monochromatic radio luminosities up to even $\sim 10^{37}$\,erg\,s$^{-1}$ \citep{Wrobel88,condon2002}. And in fact, while for the general population of spheroids, starformation rates are expected to be low, meaning SFR\,$<0.1 M_{\odot}$\,yr$^{-1}$, for giant ellipticals they may reach even $\sim 1\,M_{\odot}$\,yr$^{-1}$ \citep{Kokusho17,Capetti22}. In our sub-sample of radio-dim sources, the starformation rates estimated for some of the targets by \citet{OSullivan18}, range from SFR\,$\simeq 0.004 M_{\odot}$\,yr$^{-1}$ for NGC\,3923, up to $\simeq 0.2 M_{\odot}$\,yr$^{-1}$ for NGC\,4636, NGC\,5813 , NGC\,5846, and NGC\,7619. These, along with the radio--SFR correlation discussed in \citet{Yun01} and \citet[see equation\,2 therein]{Ackermann12}, would correspond to the 1.4\,GHz luminosities ranging from $\lesssim 10^{35}$\,erg\,s$^{-1}$ up to $\sim 4 \times 10^{36}$\,erg\,s$^{-1}$ indeed.

The other possibility for the origin of the observed radio emission of radio-dim sources, are the accretion disks and their winds. Keeping in mind that the bulk of the sources in the sample analyzed here are low-luminosity AGN, or even seemingly non-active galaxies, the nuclear accretion disks are expected to be of the ``hot, radiatively inefficient'' types \citep{Yuan14}. For such, the fundamental property is the unambiguous presence of the jets. \citet{Yuan05} showed that, assuming those jets dominate the radio continuum emission of the systems, while accretion disks account for the bulk of the nuclear X-ray photons, the ``Advection-Dominated Accretion Flow'' (ADAF) model for radiatively inefficient disks, can explain the radio--X-ray correlation observed in Galactic X-ray binaries (XRBs), of the form $(L_{\rm r}/L_{\rm Edd}) \simeq 4.6 \times 10^{-8} \, (L_{\rm X,\,nuc}/L_{\rm Edd})^{0.7}$. As discussed further by \citet{YuanCui05}, such a correlation should hold, however, only for nuclear X-ray luminosities above the critical value
\begin{equation}
\log \frac{L_{\rm X,\,cr}}{L_{\rm Edd}} = -5.36 - 0.17 \, \log  \frac{M_{\rm BH}}{M_{\odot}} \, ,
\end{equation}
below which the jet is expected to dominate in the X-ray band as well, steeping the radio--X-ray correlation to $(L_{\rm r}/L_{\rm Edd}) \propto (L_{\rm X,\,nuc}/L_{\rm Edd})^{1.23}$.

\begin{figure}[th!]
    \centering
\includegraphics[width=\columnwidth]{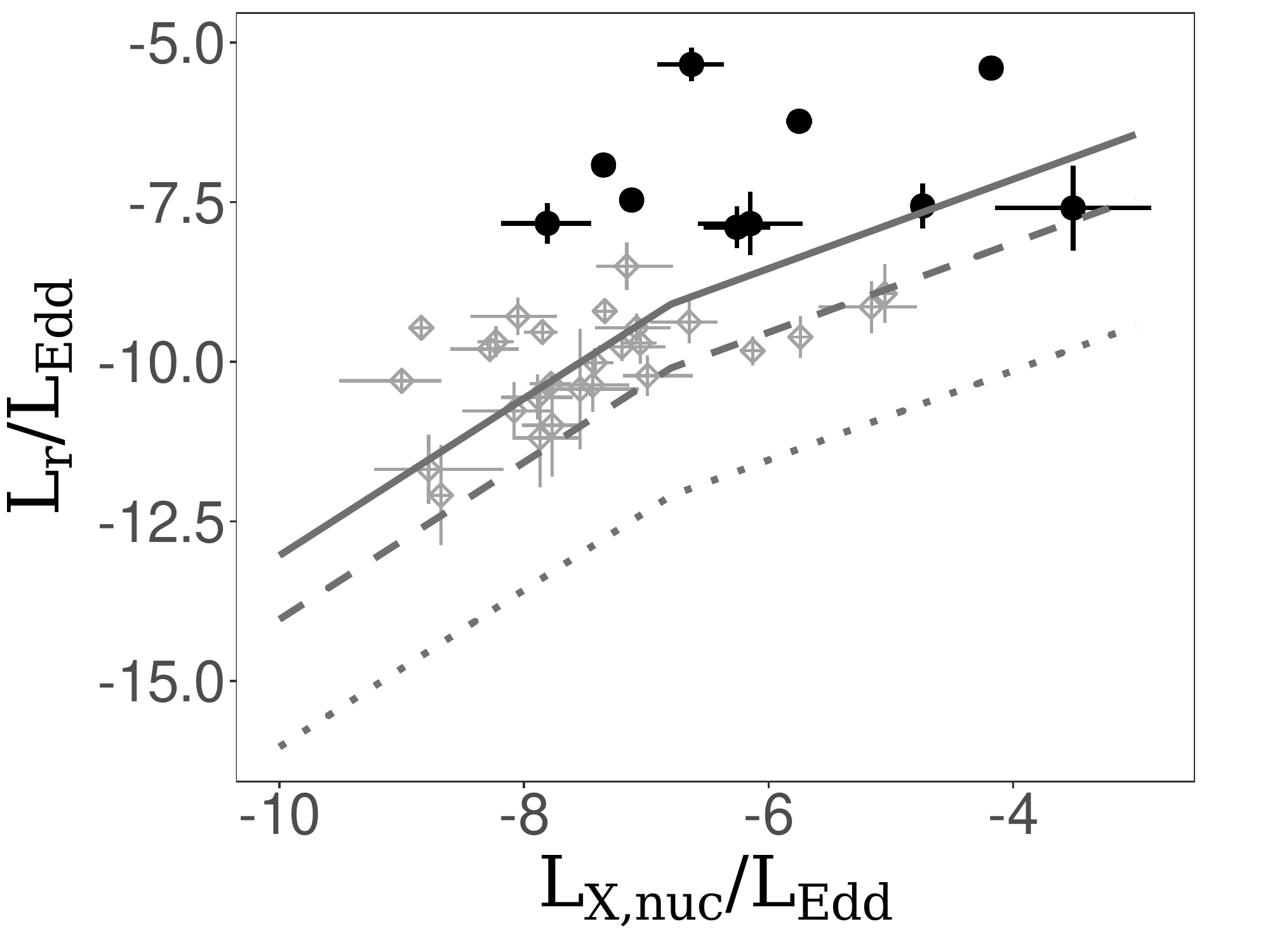} 
\caption{Distribution of the radio-dim and radio-bright sources (open grey diamonds and black filled circles, respectively) on the $(L_{\rm X,\,nuc}/L_{\rm Edd}) - (L_{\rm r}/L_{\rm Edd})$ plane, where the nuclear X-ray fluxes were taken from \citet{She17}. The ADAF scaling relation of \citet{YuanCui05} are given by grey dotted, dashed, and solid curves, as explained in the text (see Section\,\ref{sec:discu}).}
 \label{fig:nuclear}
\end{figure}

In Figure\,\ref{fig:nuclear}, we present the distribution of the studied sources on the $(L_{\rm X,\,nuc}/L_{\rm Edd}) - (L_{\rm r}/L_{\rm Edd})$ plane, where the nuclear X-ray fluxes were taken from \citet[see Section\,\ref{sec:sample}]{She17}. In addition, in the Figure we also plot the original \citet{YuanCui05} scaling as discussed above (grey dotted curve), where for the critical X-ray luminosity we assumed the average value $ \log M_{\rm BH}/M_{\odot} = 8.5$, resulting in $\log L_{\rm X,\,cr}/L_{\rm Edd} \simeq -7$. As shown, the observed radio luminosities are under-predicted by the ADAF model by orders of magnitudes, although the general dependence seems to be accounted for intriguingly well. And since the normalization in the original \citet{YuanCui05} relation follows exclusively from matching the observed outburst-state of XTE\,J1118+480, in Figure\,\ref{fig:nuclear} we include in addition the same relation but with the normalization elevated 100 and 1,000 times (dashed and solid curves, respectively). 

The above exercise demonstrates that, it is plausible to explain the properties of the analyzed sample in the framework of the ADAF model for the nuclear activity of the radio-detected early-type galaxies. In this model, radio-dim \emph{and} radio-bright objects are \emph{all} jetted, and differ only in the accretion rate, so that in radio-bright sources accreting at higher rates, the observed radio emission is dominated by the jets, and the nuclear X-ray emission by the accretion flow, while in radio-dim sources accreting at lower rates, both the radio and the nuclear X-ray radiative outputs, are dominated by compact and typically unresolved jets.

The caveats to the above interpretation are that (i) the jet production efficiency of ADAFs, have to be then at least two--three orders of magnitude higher in AGN than in XRBs, and (ii) the bi-modality in the $L_{\rm r}/L_{\rm Edd}$ distribution for early-type galaxies, remains unexplained. The former caveat may however not be an issue, as we know that the radio emission in the ADAF model strongly depends on the black hole mass: SMBHs have in general much higher radio loudness than stellar-mass black holes, and this can be quantified in the framework of the so-called `Fundamental Plane of the Black Hole Activity' \citep{Merloni03,Falcke04}.

Finally, the other remaining possibility is that radio-dim and radio-bright objects differ in the spins of central SMBHs, and that the the bi-modality in the $L_{\rm r}/L_{\rm Edd}$ distribution we see, reflects closely the distribution of the black hole spins. In the framework of this scenario, the radio emission observed from radio-dim objects --- characterized by low values of the SMBH spins --- is not related to the jets, but instead is due to a combination of starforming processes and past nuclear outbursts as proposed by \citet{Jiang10}. Radio-bright objects, on the other hand, host highly-spinning SMBHs, and therefore produce radio-emitting jets via the \citet{Blandford77} mechanism with high efficiency.

\begin{figure*}[th!]
    \centering
\includegraphics[width=0.75\textwidth]{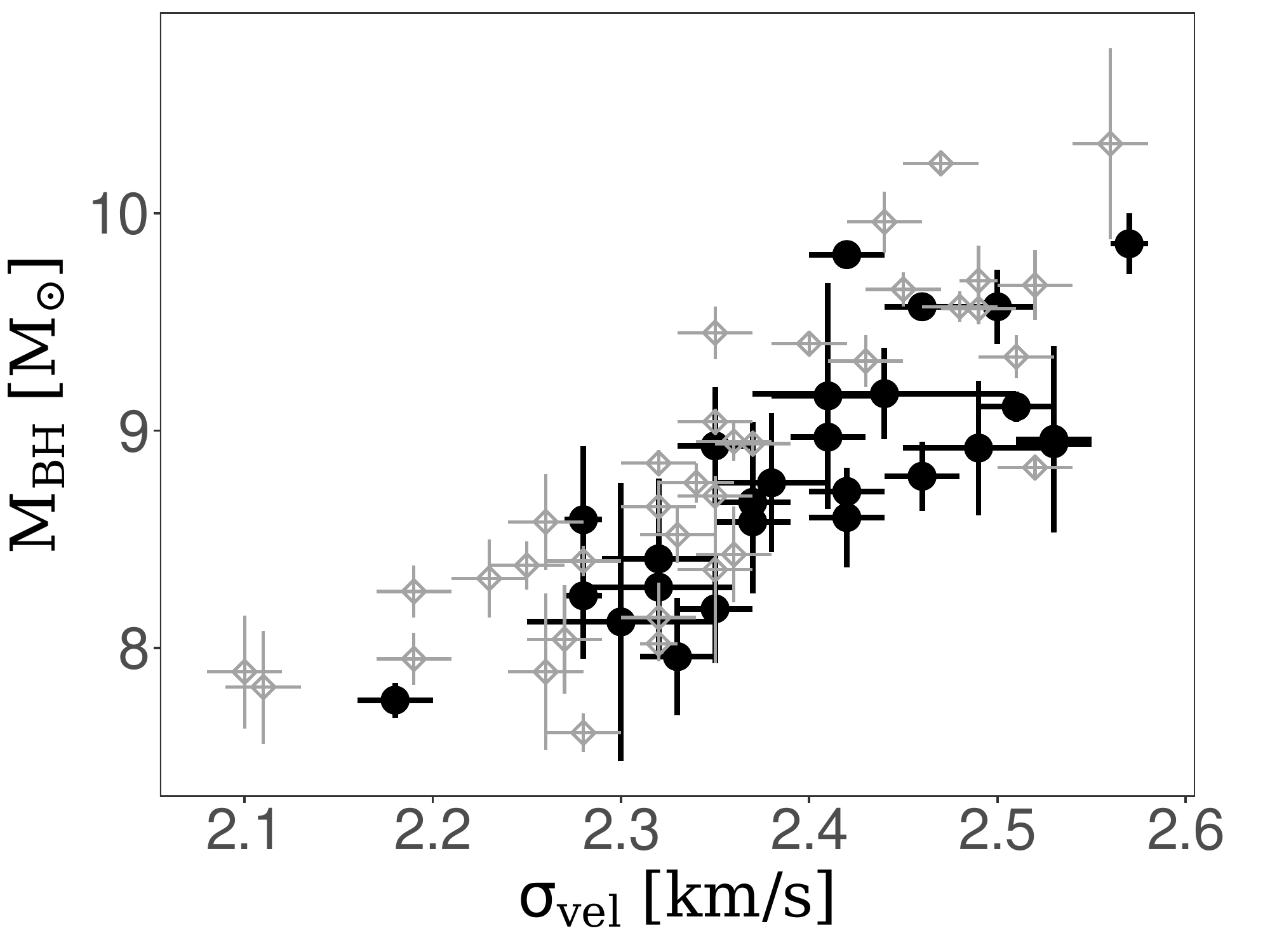} 
\caption{$M_{\rm BH} - \sigma_{\rm vel}$ relation for radio-dim and radio-bright objects in the analyzed sample of early-type galaxies (grey open diamonds and black filled circles, respectively).}
 \label{fig:Msigma}
\end{figure*}

If correct, such an interpretation would imply different evolutionary histories for SMBHs in radio-dim and radio-bright sub-samples, as during the black hole growth across cosmological epochs, changes in the spin value are determined by the frequency and types of black hole mergers and/or accretion events \citep[see in this context, e.g.,][]{Volonteri13,Dubois14,Fiacconi18}. Without much more speculation on this possibility, here we only note that such distinct evolutionary paths, should then be reflected in the SMBH--host galaxy scaling relations observed in the current epoch, in particular in the fundamental $M_{\rm BH} - \sigma_{\rm vel}$ relation. In Figure\,\ref{fig:Msigma}, we present this relation for the objects from our sample, marking separately radio-dim and radio-bright objects. As follows, the distribution of both sub-samples on the diagram seems to be different indeed, with radio-bright sources displaying on average systematically larger stellar velocity dispersion values $\sigma_{\rm vel}$ for a given black hole mass $M_{\rm BH}$, when compared to radio-dim sources. This potentially intriguing but tentative finding, will be analyzed in more detail in the follow-up publication.

\section{Conclusions}
\label{sec:concl}

We present the results of our studies regarding radio emission of a sample of nearby (redshifts $z<0.04$) early-type galaxies from the Gaspari et al. (2019) sample, for which the masses of central black holes are measured via direct dynamical methods, and the group/cluster hot gaseous halos are well characterised through high-resolution X-ray observations. The sample includes 76 nearby systems, for which neither an exact level of the nuclear activity, nor did a presence of relativistic radio-emitting jets, play any role in the selection procedure. The 1.4\,GHz radio fluxes collected with $\sim$\,arcmin resolution for 62 targets from the list, span a wide range from $\sim 1$\,mJy for the dimmest source up to 1330\,Jy for the brightest one; the corresponding 1.4\,GHz monochromatic luminosities are low and very low, ranging from $L_{\rm r} \sim 10^{41}$\,erg\,s$^{-1}$ down to $10^{35}$\,erg\,s$^{-1}$. For those, we study the correlations with the main parameters of central supermassive black holes, host galaxies, and hot gaseous halos, finding a general bimodality in the radio luminosity distribution, with the borderline between ``radio-bright'' and ``radio-dim'' population $\log L_{\rm r} / L_{\rm Edd} \simeq -8.5$. 

We analyze the far-infrared data for the targets with the matching arcmin resolution, finding that all radio-bright sources, and also the majority of radio-dim ones, are over-luminous in radio with respect to the FIR-radio correlation, as quantified by the values of the corresponding $q$-parameter $q<1.9$. As such, the radio emission of the radio-excess sources is not likely to result from starformiation processes within galactic hosts. The additional analysis of the mid-infrared colours support low starformation rates for the analyzed targets, characteristic for spheroids and intermediate-disk systems. Finally, the investigation of the high-resolution radio maps reveal that the overwhelming majority of radio-dim sources are unresolved on arc-second scale, while the bulk of radio-bright sources display complex extended structures typically of a double-lobe, double-jet, or a core-jet type. And indeed, on the absolute r-band magnitude vs. 1.4\,GHz luminosity plane, the targets forming our radio-bright population overlap with radio galaxies of the Faranoff-Riley types 0, I, and even II. Radio emission of those originates therefore in relativistic jets, launched by the central engines with different efficiencies and in a wider range of kinetic powers, depending on the exact parameters of the central engine.

Regarding the origin of the radio continuum emission observed from radio-dim sources, we discuss the two main possibility. One is the ADAF model, in which the radio and the nuclear X-ray radiative outputs at very low accretion rates characterizing the targets, are both dominated by compact and typically unresolved jets. This interpretation would imply that the jet production efficiency of ADAFs, have to be then at least two--three orders of magnitude higher in AGN than in Galactic XRBs, consistently however with the scaling following the Fundamental Plane of the Black Hole Activity. The other possibility is that the general bimodality in the $L_{\rm r}/L_{\rm Edd}$ distribution we observe, reflects the spin distribution of SMBHs in the centres of early-type galaxies, with radio-dim and radio-bright population being characterized by low and large spin values, respectively. In this scenario, the radio emission observed from low-spin radio-dim objects, is not related to the jets, but instead is due to a combination of starforming processes and past nuclear outbursts. In support of this idea, we present a tentative finding that both sub-samples differ in the distribution on the $M_{\rm BH} - \sigma_{\rm vel}$ diagram, in the sense that the radio-bright sources display on average systematically larger stellar velocity dispersion values $\sigma_{\rm vel}$ for a given black hole mass $M_{\rm BH}$.

\begin{acknowledgements}
The authors thank A.~A.~Zdziarski and M.~Tarnopolski for valuable comments on the manuscript.
A.~W. and {\L}.~S. were supported by the Polish NSC grant 2016/22/E/ST9/00061. A.~W. and N.~W. were supported by the GACR grant 21-13491X. Research by C.~C.~C. at the Naval Research Laboratory is supported by NASA DPR S-15633-Y.

\end{acknowledgements}

\appendix

\section{Radio and Infrared Data.} 
\label{A:data}

In Table\,\ref{tab:data}, we summarize the gathered $45^{\prime\prime}$-resolution data at 1.4\,GHz for the 62 radio-detected early-type galaxies from the list of \citet{gaspari19}. We note that, in addition to those 62 sources, nine early-type galaxies from \citeauthor{gaspari19} have only the VLA upper limits at the level of 0.5\,mJy \citep[NGC\,0821, NGC\,1023,  NGC\,1374, NGC\,3377, NGC\,4291, NGC\,4459, NGC\,4473, NGC\,4621, and NGC\,5576; see][]{brown2011}, while five are missing any arcmin-scale data at GHz frequencies (NGC\,1277, NGC\,4342, NGC\,5845, NGC\,6861, and NGC\,7768). The corresponding far-infrared IRAS fluxes, as well as mid-infrared WISE colours, are provided in the Table as well.

\startlongtable
\begin{deluxetable}{ccccccccccc}
\tabletypesize{\footnotesize}
\tabletypesize{\scriptsize}
\tablecaption{Radio and infrared data for radio-detected early-type galaxies from \citet{gaspari19}.}
\label{tab:data}
\tablewidth{0pt}
\tablehead{
\colhead{Name} &\colhead{$D_L$}& \colhead{$\rm M_{BH}$}& \colhead{$S_{\rm 1.4\,GHz}$}  & \colhead{Ref.} &   \colhead{$\log L_{\rm r}/L_{\rm Edd}$} & \colhead{$S_{\rm 100\,\mu m}$} &  \colhead{$S_{\rm 60\,\mu m}$} & \colhead{$q$} &  \colhead{W1--W2} &  \colhead{W2--W3}\\
 \colhead{(i)} & \colhead{(ii)} & \colhead{(iii)} & \colhead{(iv)} & \colhead{(v)} & \colhead{(vi)} & \colhead{(vii)} & \colhead{(viii)} & \colhead{(ix)}&\colhead{(x)}&\colhead{(xi)}
}
\startdata
A1836B&152.4$\pm$8.4& 9.57$\pm$0.06&2044$\pm$39&V12&--6.77$\pm$0.12&---&---&---&--0.07&0.74\\
IC1459&28.9$\pm$3.7& 8.96$\pm$0.43&1200$\pm$100&B11&--7.84$\pm$0.5&1.18$\pm$0.1&0.51$\pm$0.03&--0.16$\pm$0.04&--0.02&0.85\\
IC4296&49.2$\pm$3.6&9.11$\pm$0.07&18000$\pm$1000&B11&--6.35$\pm$0.16&$<$0.05&0.14$\pm$0.06&$<$--2.11&--0.02&0.68\\
M105&10.7$\pm$0.5& 8.04$\pm$0.25&2$\pm$0.5&B11&--10.56$\pm$0.35&$<$0.11&$<$0.04&$<$1.56&--0.36&0.54\\
Mrk1216&94$\pm$9.4& 9.69$\pm$0.16&9.2$\pm$0.6&C98&--9.66$\pm$0.25&---&---&---&--0.06&0.69\\
M49& 17.1$\pm$0.6& 9.40$\pm$0.04&220$\pm$10&B11&--9.47$\pm$0.09&$<$0.11&$<$0.06&$<$--0.38&---&---\\
M60&16.5$\pm$0.6& 9.32$\pm$0.12&29$\pm$1&B11&--10.3$\pm$0.14&1.09$\pm$0.06&0.78$\pm$0.03&1.56$\pm$0.02&--0.49&0.67\\
M84& 18.5$\pm$0.6& 8.97$\pm$0.05&7000$\pm$600&B11&--7.47$\pm$0.11&1.16$\pm$0.12&0.5$\pm$0.03&--0.93$\pm$0.04&---&---\\
M87& 16.7$\pm$0.6& 9.81$\pm$0.05&210000$\pm$10000&B11&--6.92$\pm$0.09&0.41$\pm$0.1&0.39$\pm$0.04&--2.64$\pm$0.05&0.08&1.19\\
M89&15.3$\pm$1.0& 8.70$\pm$0.05&100$\pm$3&B11&--9.21$\pm$0.13&0.53$\pm$0.06&0.16$\pm$0.05&0.5$\pm$0.06&---&---\\
NGC0315& 57.7$\pm$2.8& 8.92$\pm$0.31&1800$\pm$100&B11&--7.02$\pm$0.33&0.4$\pm$0.1&0.31$\pm$0.05&--0.65$\pm$0.06&0.04&1.91\\
NGC0541& 63.7$\pm$6.4& 8.59$\pm$0.34& 538.2$\pm$11.9 & V12 & --7.13 $\pm$0.4&$<$0.13&$<$0.05&$<$--0.80&---&---\\
NGC0741& 65.7$\pm$6.6& 8.67$\pm$0.37&940$\pm$60&B11&--6.94$\pm$0.42&1.13$\pm$0.11&0.19$\pm$0.02&--0.24$\pm$0.04&--0.03&0.45\\
NGC1052& 18.1$\pm$1.8& 8.24$\pm$0.29&1100$\pm$100&B11&--7.56$\pm$0.35&1.58$\pm$0.07&0.88$\pm$0.03&0.07$\pm$0.04&0.19&2.55\\
NGC1316&18.1$\pm$0.6& 8.18$\pm$0.25&150000$\pm$10000&B11&--5.34$\pm$0.26&8.11$\pm$1.9&3.07$\pm$0.03&--1.44$\pm$0.06&--0.09&1.00\\
NGC1332& 22.3$\pm$1.9& 8.83$\pm$0.04&4.6$\pm$0.5&B11&--10.35$\pm$0.19&1.81$\pm$0.06&0.51$\pm$0.03&2.36$\pm$0.05&--0.03&1.00\\
NGC1399& 20.9$\pm$0.7& 8.94$\pm$0.31&2200$\pm$100&B11&--7.83$\pm$0.32&0.3$\pm$0.08&$<$0.03&$<$--1.23&---&---\\
NGC1407& 28.0$\pm$3.4& 9.65$\pm$0.08&88$\pm$4&B11&--9.69$\pm$0.24&0.48$\pm$0.07&0.14$\pm$0.03&0.51$\pm$0.06&--0.03&0.53\\
NGC1550& 51.6$\pm$5.6& 9.57$\pm$0.07&17$\pm$2&B11&--9.79$\pm$0.24&$<$0.24&$<$0.03&$<$0.79&---&---\\
NGC1600& 64.0$\pm$6.4& 10.23$\pm$0.04&62$\pm$3&B11&--9.7$\pm$0.2&0.19$\pm$0.07&0.1$\pm$0.03&0.39$\pm$0.1&--0.03&0.45\\
NGC3091& 51.2$\pm$8.3& 9.56$\pm$0.07&2.5$\pm$0.5&B11&--10.62$\pm$0.36&---&---&---&---&---\\
NGC3585& 20.5$\pm$1.7& 8.52$\pm$0.13&0.6$\pm$0.5&B11&--10.99$\pm$0.8&$<$0.09&0.16$\pm$0.04&$<$2.45&--0.01&0.73\\
NGC3607& 22.6$\pm$1.8& 8.14$\pm$0.16&6.9$\pm$0.4&B11&--9.47$\pm$0.22&---&---&---&0.04&1.75\\
NGC3608& 22.8$\pm$1.5&8.32$\pm$0.18 &1.3$\pm$0.5&B11&--10.37$\pm$0.42&---&---&---&--0.02&0.51\\
NGC3842& 92.1$\pm$10.6& 9.96$\pm$0.14&12.6$\pm$1.26&C98&--9.81$\pm$0.27&1.49$\pm$0.19&0.36$\pm$0.06&1.81$\pm$0.06&--0.05&0.45\\
NGC3862& 84.6$\pm$8.5& 8.41$\pm$0.37&5689$\pm$131&V12&--5.68$\pm$0.42&$<$0.19&0.21$\pm$0.05&$<$--1.36&0.06&1.42\\
NGC3923& 20.9$\pm$2.7& 9.45$\pm$0.12&1$\pm$0.5&B11&--11.69$\pm$0.54&$<$0.14&$<$0.04&$<$1.88&0.01&0.58\\
NGC4261& 32.4$\pm$2.8& 8.72$\pm$0.10&22000$\pm$1000&B11&--6.23$\pm$0.19&0.15$\pm$0.05&0.08$\pm$0.04&--2.26$\pm$0.13&---&---\\
NGC4278& 15.0$\pm$1.5& 7.96$\pm$0.27&390$\pm$10&B11&--7.89$\pm$0.33&1.86$\pm$0.06&0.58$\pm$0.05&0.46$\pm$0.02&0&0.96\\
NGC4636& 13.7$\pm$1.4& 8.58$\pm$0.22&78$\pm$3&B11&--9.29$\pm$0.29&$<$0.17&0.14$\pm$0.04&$<$0.36&--0.03&0.59\\
NGC4697& 12.5$\pm$0.4& 8.29$\pm$0.04&0.6$\pm$0.5&B11&--11.19$\pm$0.78&1.24$\pm$0.08&0.46$\pm$0.02&3.13$\pm$0.36&--0.01&0.92\\
NGC4889& 102.0$\pm$5.2& 10.32 $\pm$0.44&1.2$\pm$0.04&B11&--11.1$\pm$0.45&$<$0.07&$<$0.05&$<$1.77&--0.05&0.50\\
NGC5018& 40.5$\pm$4.9& 8.02$\pm$0.08&2$\pm$0.5&B11&--9.38$\pm$0.33&1.86$\pm$0.09&0.95$\pm$0.04&2.86$\pm$0.11&0.05&1.30\\
NGC5077& 38.7$\pm$8.4& 8.93$\pm$0.27&157$\pm$5&B11&--8.43$\pm$0.49&---&---&---&--0.05&0.78\\
NGC5128& 3.6$\pm$0.2& 7.76$\pm$0.08& 1330000$\pm$133000 & C65 & --5.40$\pm$0.16 & 360.57$\pm$0.53&162.91$\pm$0.12&--0.70$\pm$0.04&0.52&2.56\\
NGC5328& 64.1$\pm$7.0& 9.67$\pm$0.16&0.9$\pm$0.5&B11&--10.98$\pm$0.58&$<$0.07&$<$0.03&$<$1.76&--0.04&0.53\\
NGC5419& 56.2$\pm$6.1& 9.86$\pm$0.14&790$\pm$60&B11&--8.34$\pm$0&0.23$\pm$0.07&$<$0.04&$<$--0.86&0.01&0.42\\
NGC5813& 32.2$\pm$2.7& 8.85$\pm$0.06&15$\pm$1&B11&--9.53$\pm$0.18&$<$0.1&$<$0.02&$<$0.55&0.01&0.41\\
NGC5846& 24.9$\pm$2.3& 9.04$\pm$0.06&21$\pm$1&B11&--9.8$\pm$0.19&$<$0.13&$<$0.04&$<$0.55&---5&---\\
NGC6086& 138.0$\pm$11.5& 9.57$\pm$0.17&109$\pm$10.9&C02&--8.13$\pm$0.25&$<$0.19&$<$0.04&$<$--0.02&--0.06&0.51\\
NGC6251& 108.4$\pm$9.0& 8.79$\pm$0.16&1800$\pm$100&B11&--6.34$\pm$0.23&$<$0.11&$<$0.02&$<$--1.51&--0.45&2.06\\
NGC7052& 70.4$\pm$8.4& 8.60$\pm$0.23&160$\pm$10&B11&--7.58$\pm$0.32&1.42$\pm$0.14&0.45$\pm$0.03&0.73$\pm$0.04&--0.01&1.54\\
NGC7619& 51.5$\pm$7.4& 9.34$\pm$0.10&20$\pm$1&B11&--9.49$\pm$0.29&0.71$\pm$0.23&$<$0.04&$<$1.13&--0.02&0.51\\
NGC7626& 38.1$\pm$3.8& 8.58$\pm$0.33&740$\pm$50&B11&--7.43$\pm$0.38&$<$0.13&$<$0.04&$<$--0.97&--0.01&0.52\\
UGC12064& 72.5 $\pm$ 6.7& 9.16$\pm$0.52&3800$\pm$380&C02&--6.74$\pm$0.56&---&---&---&--0.05&1.39\\
NGC0383& 59.2$\pm$5.9& 8.76$\pm$0.32&4800$\pm$200&B11&--6.41$\pm$0.37&$<$1.68$^{\star}$&$<$0.44$^{\star}$&$<$--0.71&0.03&1.65\\
NGC0524& 24.2$\pm$2.2& 8.94$\pm$0.05&3.1$\pm$0.4&B11&--10.56$\pm$0.21&2.05$\pm$0.13&0.76$\pm$0.03&2.64$\pm$0.06&0&1.44\\
NGC2110& 29.1$\pm$2.9& 8.12$\pm$0.64&30$\pm$10&B11&--7.59$\pm$0.67&5.68$\pm$1.36$^{\dagger}$&4.13$\pm$0.21$^{\dagger}$&1.26$\pm$0.04&0.78&2.63\\
NGC2787& 7.4$\pm$1.2& 7.61$\pm$0.09&10.9$\pm$0.5&B11&--9.71$\pm$0.32&1.18$\pm$0.21&0.6$\pm$0.05&1.92$\pm$0.04&--0.01&1.33\\
NGC3115& 9.5$\pm$0.4& 8.95$\pm$0.09&0.6$\pm$0.5&B11&--12.09$\pm$0.78&$<$0.11&0.13$\pm$0.04&$<$2.4&---&---\\
NGC3245& 21.4$\pm$2.0& 8.38$\pm$0.11 &6.7$\pm$0.5&B11&--9.77$\pm$0.22&3.97$\pm$0.09&2.03$\pm$0.04&2.66$\pm$0.03&0.01&1.98\\
NGC3414& 25.2$\pm$2.7& 8.40$\pm$0.07&4.4$\pm$0.4&B11&--9.83$\pm$0.23&0.56$\pm$0.18&0.25$\pm$0.03&1.96$\pm$0.08&--0.04&0.98\\
NGC3665& 34.7$\pm$6.7& 8.76$\pm$0.09&112$\pm$4&B11&--8.51$\pm$0.37&7.53$\pm$0.18&1.91$\pm$0.04&1.57$\pm$0.02&0.07&2.60\\
NGC3801& 46.3$\pm$4.6& 8.28$\pm$0.31&1100$\pm$100&B11&--6.78$\pm$0.37&2.8$\pm$0.09&0.17$\pm$0.05&0$\pm$0.04&--0.03&1.88\\
NGC3998& 14.3$\pm$1.3& 8.36$\pm$0.43&98$\pm$3&B11&--8.93$\pm$0.46&1.15$\pm$0.12&0.55$\pm$0.03&0.94$\pm$0.03&0.06&1.85\\
NGC4026& 13.4$\pm$1.7& 8.26$\pm$0.12&1.3$\pm$0.5&B11&--10.77$\pm$0.45&0.56$\pm$0.13$^{\diamond}$&0.1$\pm$0.03$^{\diamond}$&2.33$\pm$0.18&--0.02&0.49\\
NGC4036& 19.0$\pm$ 1.9& 7.89$\pm$0.36&11.6$\pm$0.5&B11&--9.14$\pm$0.41&1.63$\pm$0.15&0.56$\pm$0.05&1.95$\pm$0.03&0.02&1.09\\
NGC4203& 14.1$\pm$1.4& 7.82$\pm$0.26&6.1$\pm$0.5&B11&--9.61$\pm$0.33&2.16$\pm$0.08&0.59$\pm$0.03&2.31$\pm$0.04&0.04&1.45\\
NGC4526& 16.4$\pm$1.8& 8.65$\pm$0.12&12$\pm$0.5&B11&--10.02$\pm$0.24&17.1$\pm$0.09&5.56$\pm$0.05&2.94$\pm$0.02&---&---\\
NGC4564& 15.9$\pm$0.5& 7.95$\pm$0.12&1.6$\pm$0.5&B11&--10.22$\pm$0.32&$<$0.19&$<$0.06&$<$1.86&--0.04&0.56\\
NGC4596& 16.5$\pm$6.2& 7.89$\pm$0.26&0.8$\pm$0.5&B11&--10.43$\pm$0.95&0.75$\pm$0.04&0.4$\pm$0.02&2.87$\pm$0.27&--0.03&1.50\\
NGC6240S& 105.0$\pm$10.5& 9.17$\pm$0.21&426$\pm$42.6&C02&--7.37$\pm$0.3&22.7$\pm$3.4$^{\star}$&24.1$\pm$3.62$^{\star}$&1.83$\pm$0.07&---&---\\
\enddata
\tablecomments{{\bf Col~(i)} --- name of the source; {\bf Col~(ii)} --- distance in Mpc (from surface brightness fluctuations or redshifts; \citet{bosch2016, saglia16}; {\bf Col~(iii)} --- logarithm of the black hole mass in $M_\odot$ from \citet{bosch2016}; {\bf Col~(iv)} --- 1.4\,GHz VLA flux density in mJy; {\bf Col~(v)} --- reference to VLA data including C65: \citet{cooper1965}, C98: \citet{condon1998}, C02: \citet{condon2002}, B11: \citet{brown2011}, and V12: \citet{velzen2012}; {\bf Col~(vi)} --- logarithm of the monochromatic 1.4\,GHz radio luminosity in Eddington units, for the distances and black hole masses as given in \citet{gaspari19}; {\bf Col~(vii-viii)} --- 60\,$\mu$m and 100\,$\mu$m IRAS flux densities in Jy from \citet{Knapp89}, unless stated otherwise: $^{\star}$ \citet{Golombek1988}, $^{\dagger}$ \citet{Moshir1990}, $^{\diamond}$ \citet{Rice1988};  {\bf Col~(ix)} --- the FIR-radio correlation parameter $q$; {\bf Col~(x--xi)} --- WISE colours based on \citet{Cutri13} magnitudes.
}
\end{deluxetable}
\section{Radio Luminosity Scalings.} 
\label{A:all}

\begin{figure}[th!]
    \centering
\includegraphics[width=0.45\textwidth]{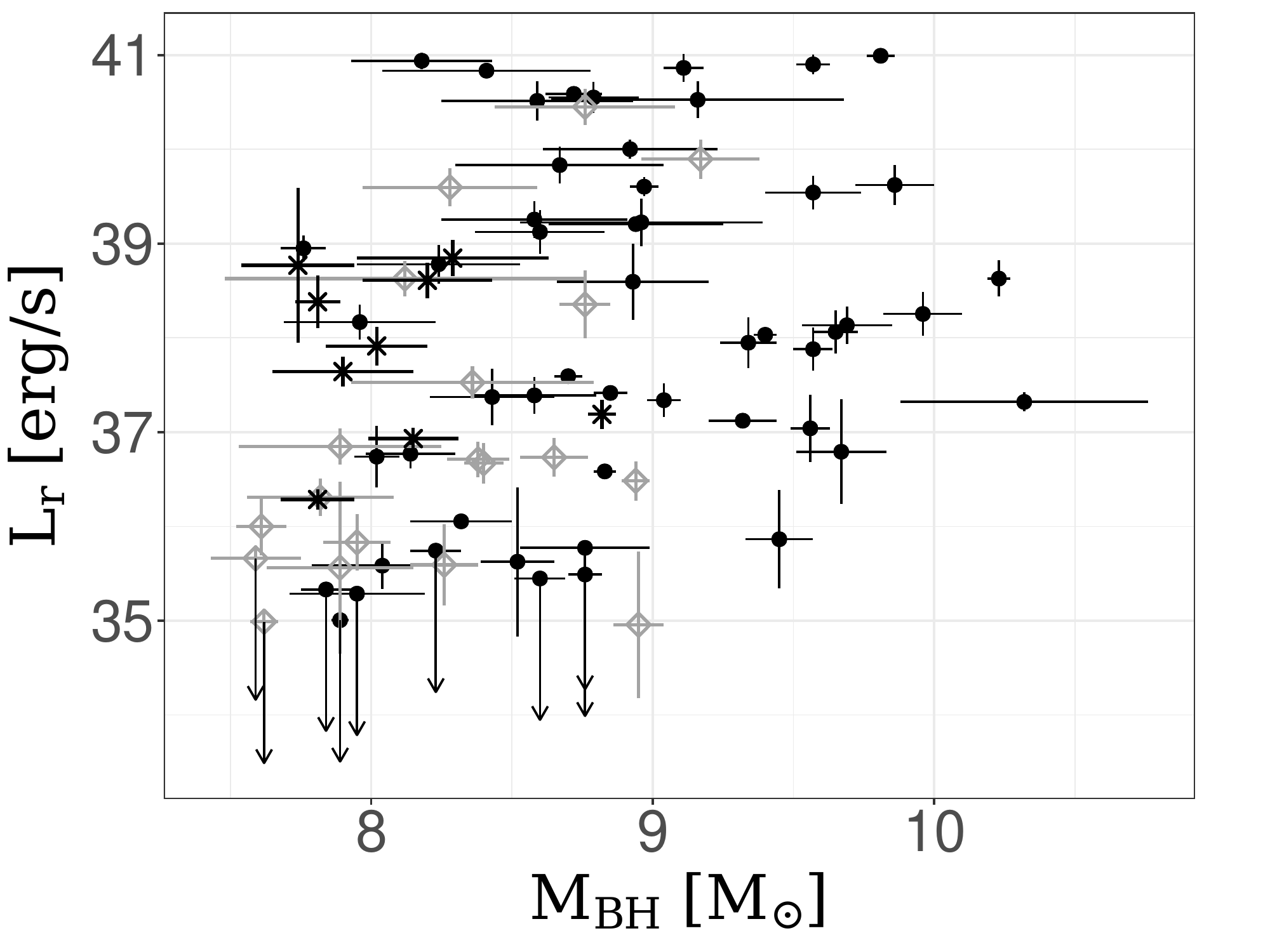} 
\includegraphics[width=0.45\textwidth]{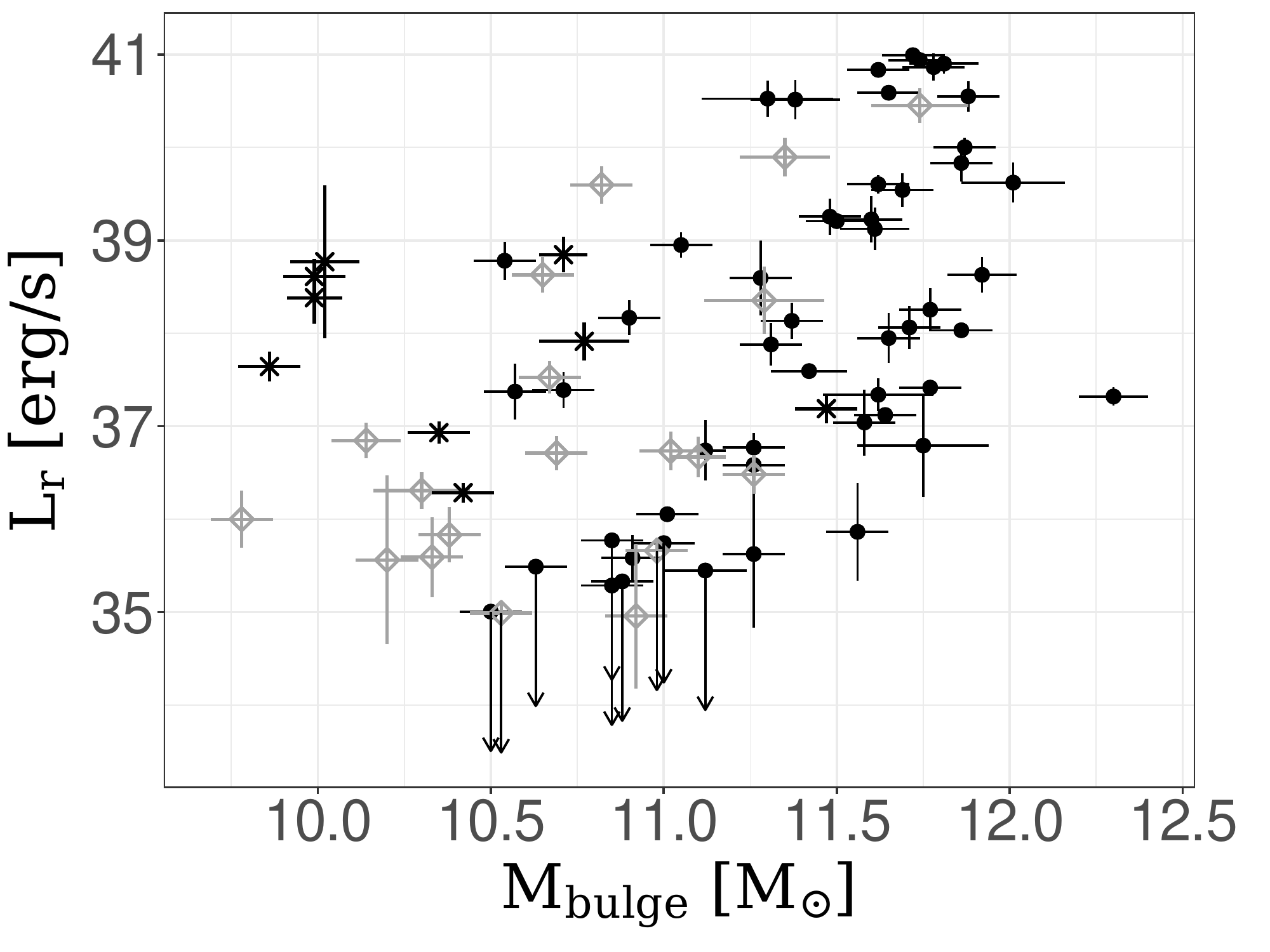}
\includegraphics[width=0.45\textwidth]{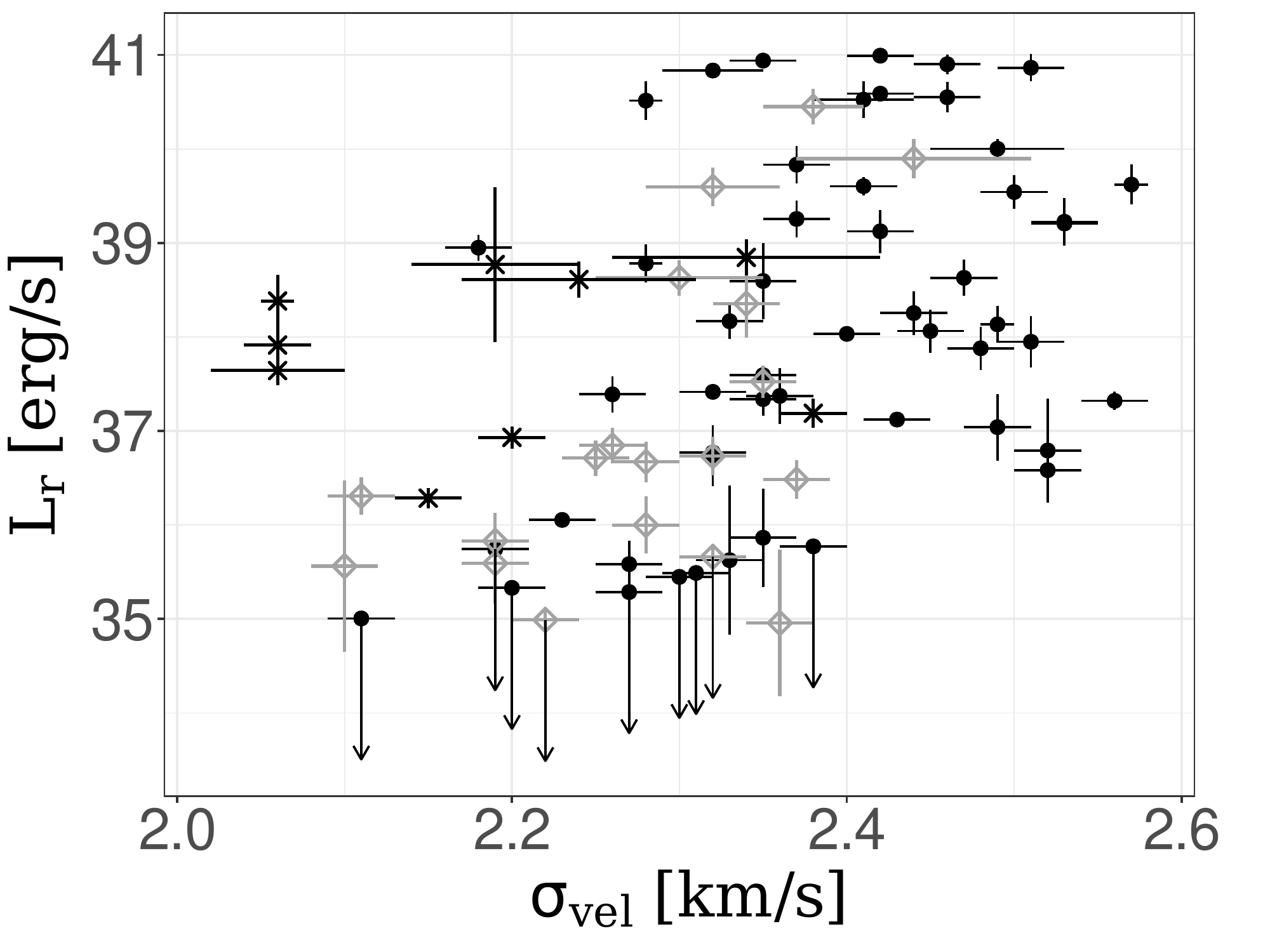}
\includegraphics[width=0.45\textwidth]{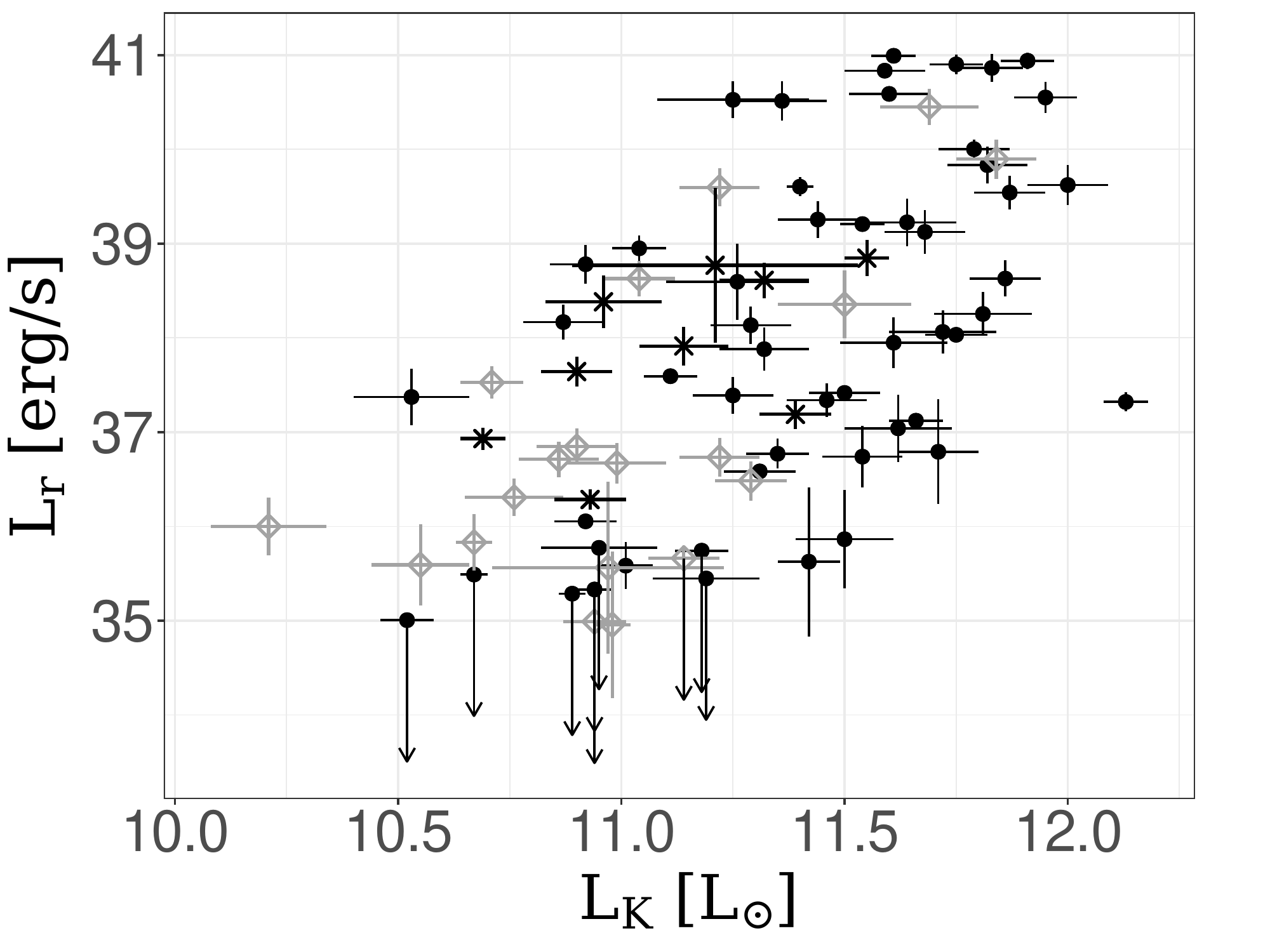}
\includegraphics[width=0.45\textwidth]{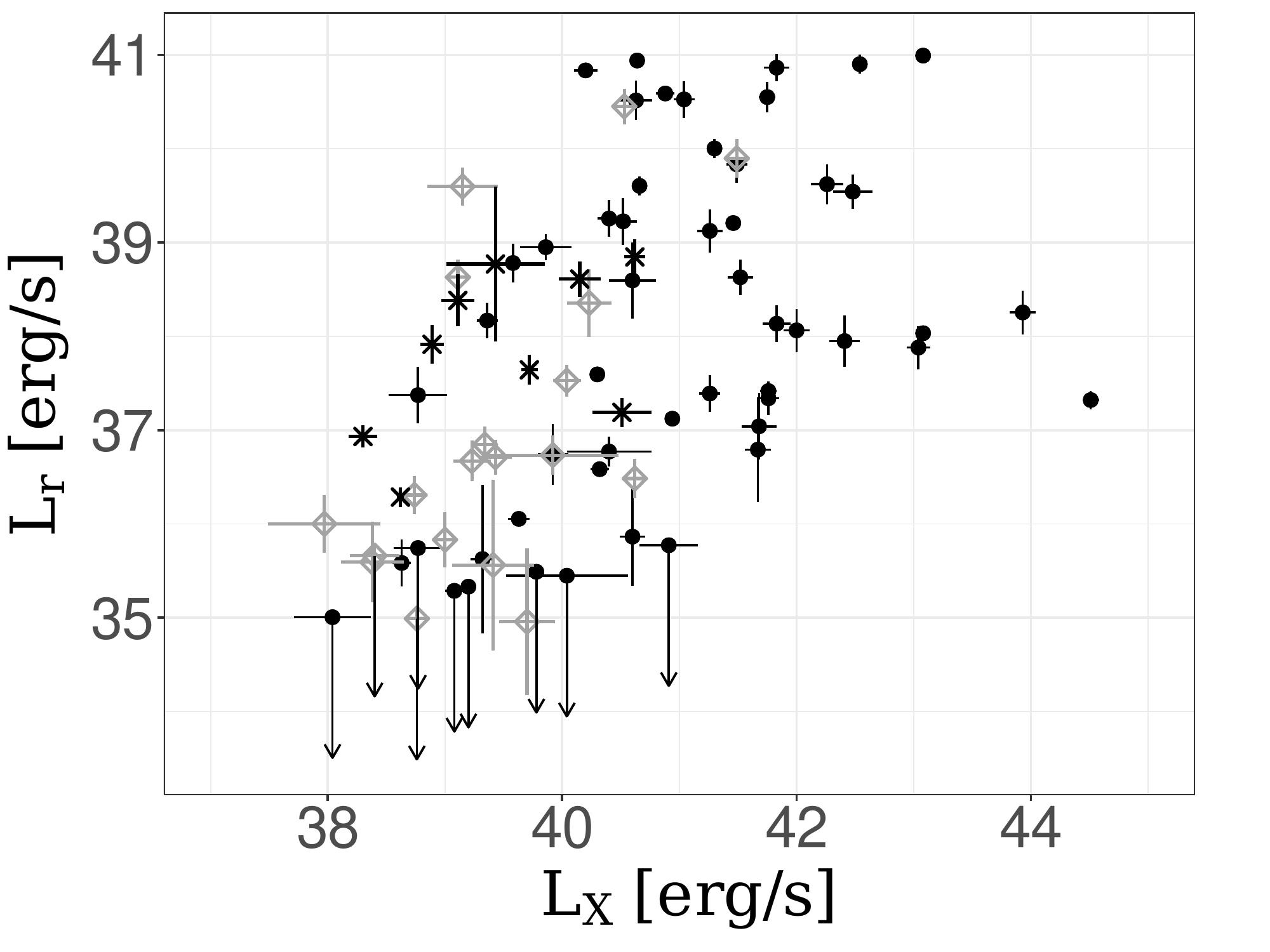}
\includegraphics[width=0.45\textwidth]{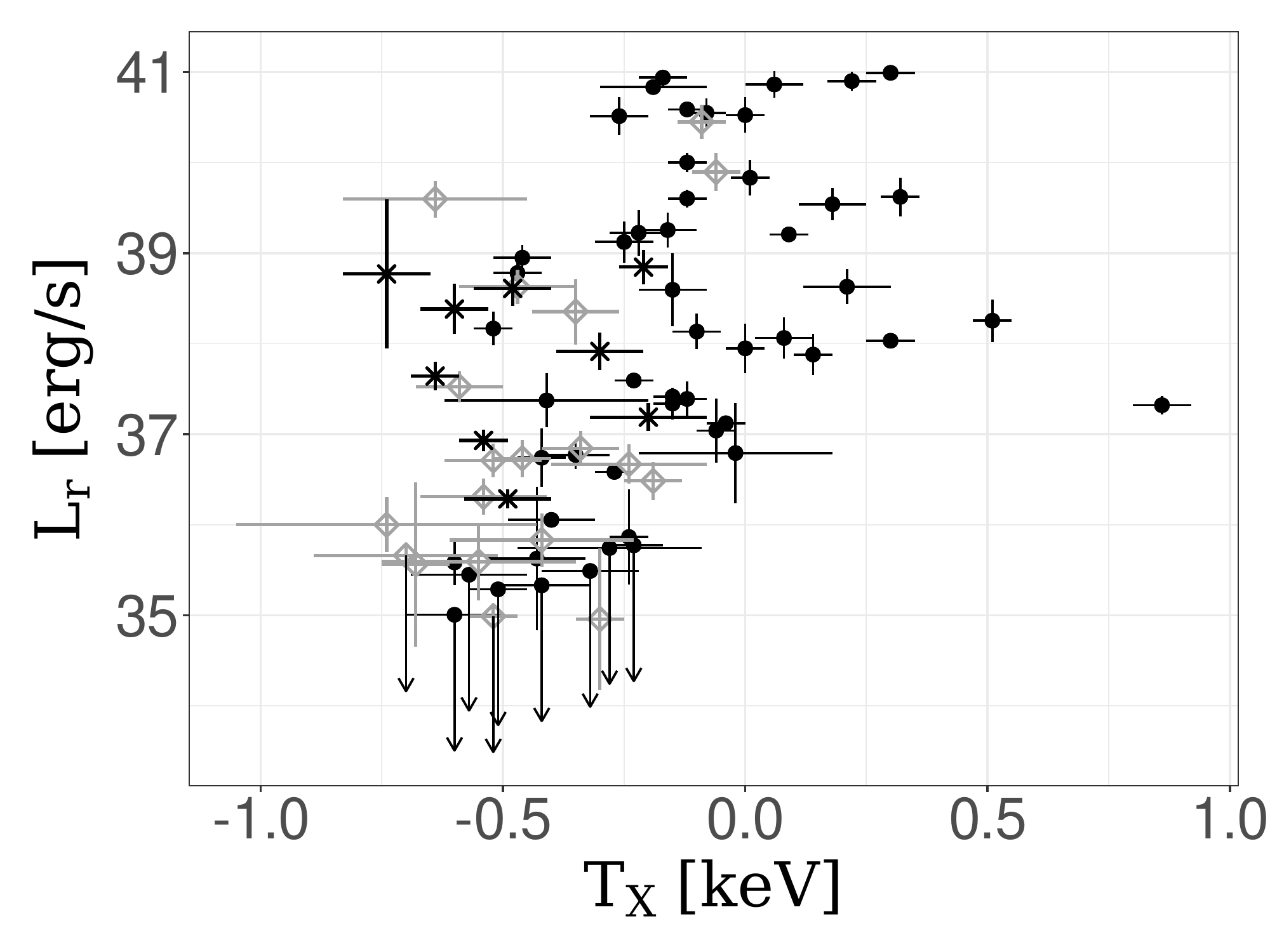}
\caption{Scaling of the 1.4\,GHz monochromatic luminosities $L_{\rm r}$ with the main optical galaxy and X-ray halo parameters $M_{\rm BH}$, $M_{\rm bulge}$, $\sigma_{\rm vel}$, $L_K$, $L_{\rm X}$, and $T_{\rm X}$, for all the galaxies from the list of \citet{gaspari19} with the available VLA radio data. On all the panels, black dots denote elliptical galaxies, grey empty squares denote lenticular galaxies, and black stars denote spiral galaxies. Radio upper limits (corresponding to the flux level of 0.5\,mJy) are indicated by vertical arrows.}
\label{fig:all}
\end{figure}

\begin{figure}[th!]
    \centering
\includegraphics[width=0.45\textwidth]{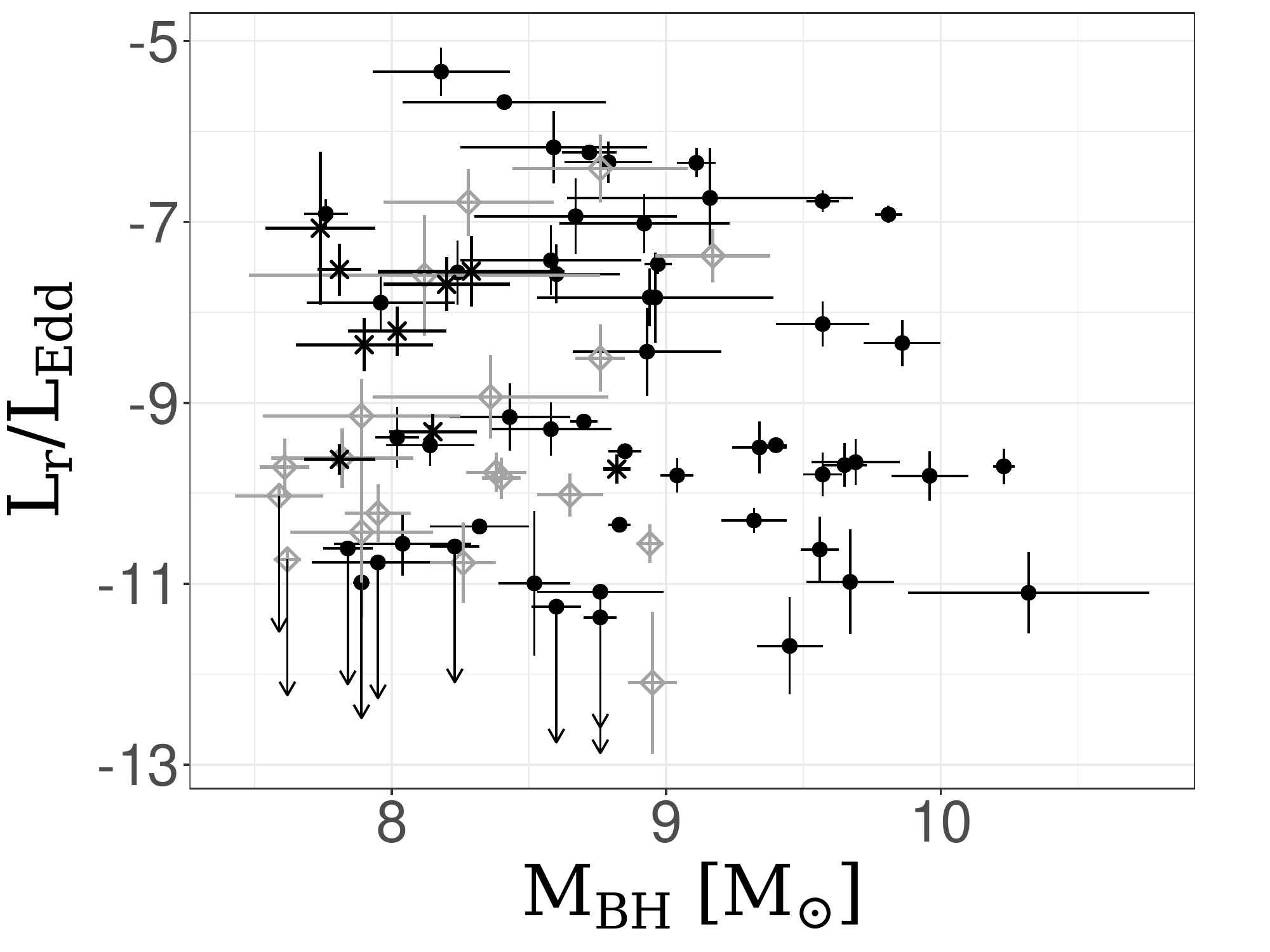} 
\includegraphics[width=0.45\textwidth]{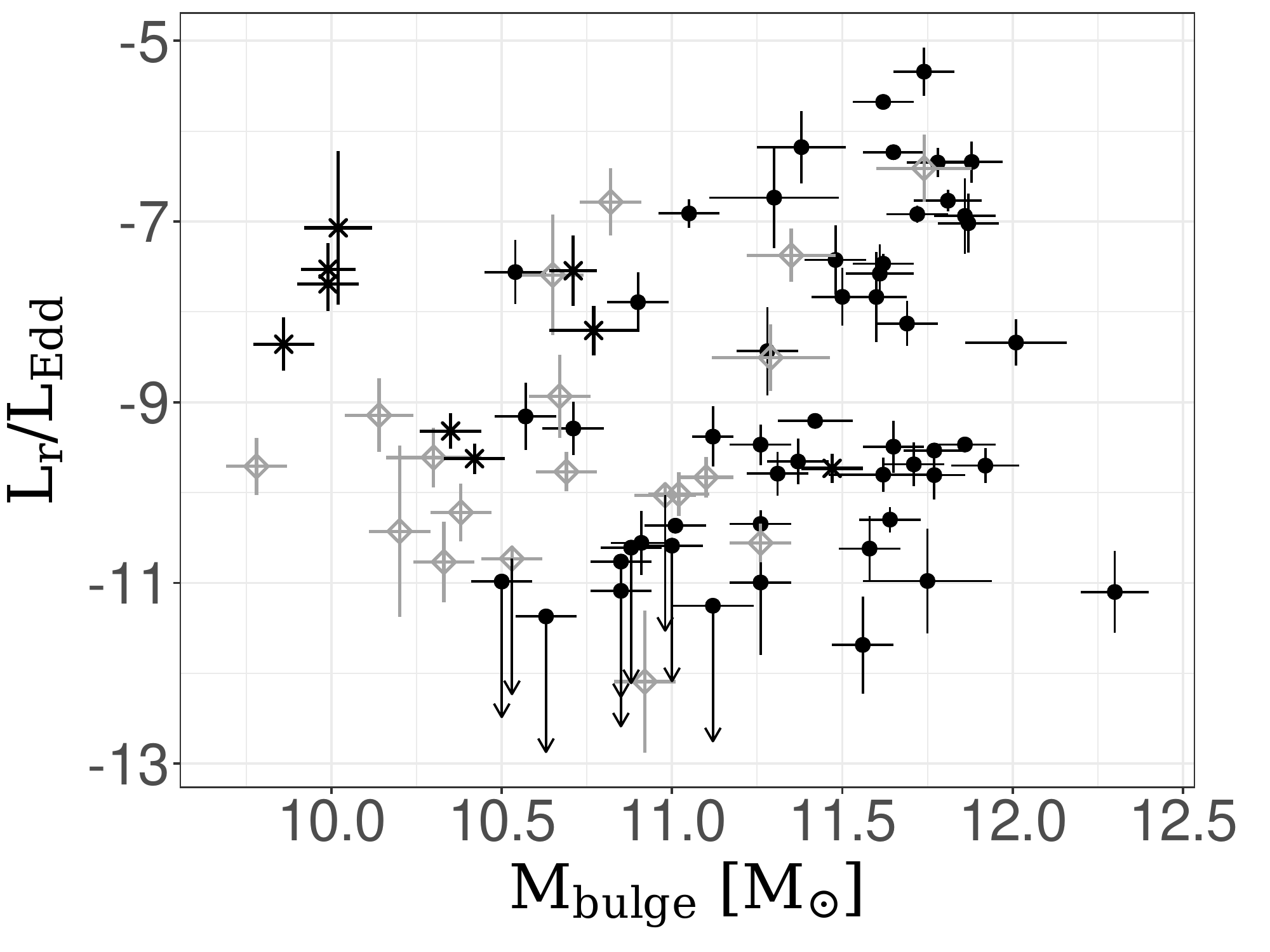}
\includegraphics[width=0.45\textwidth]{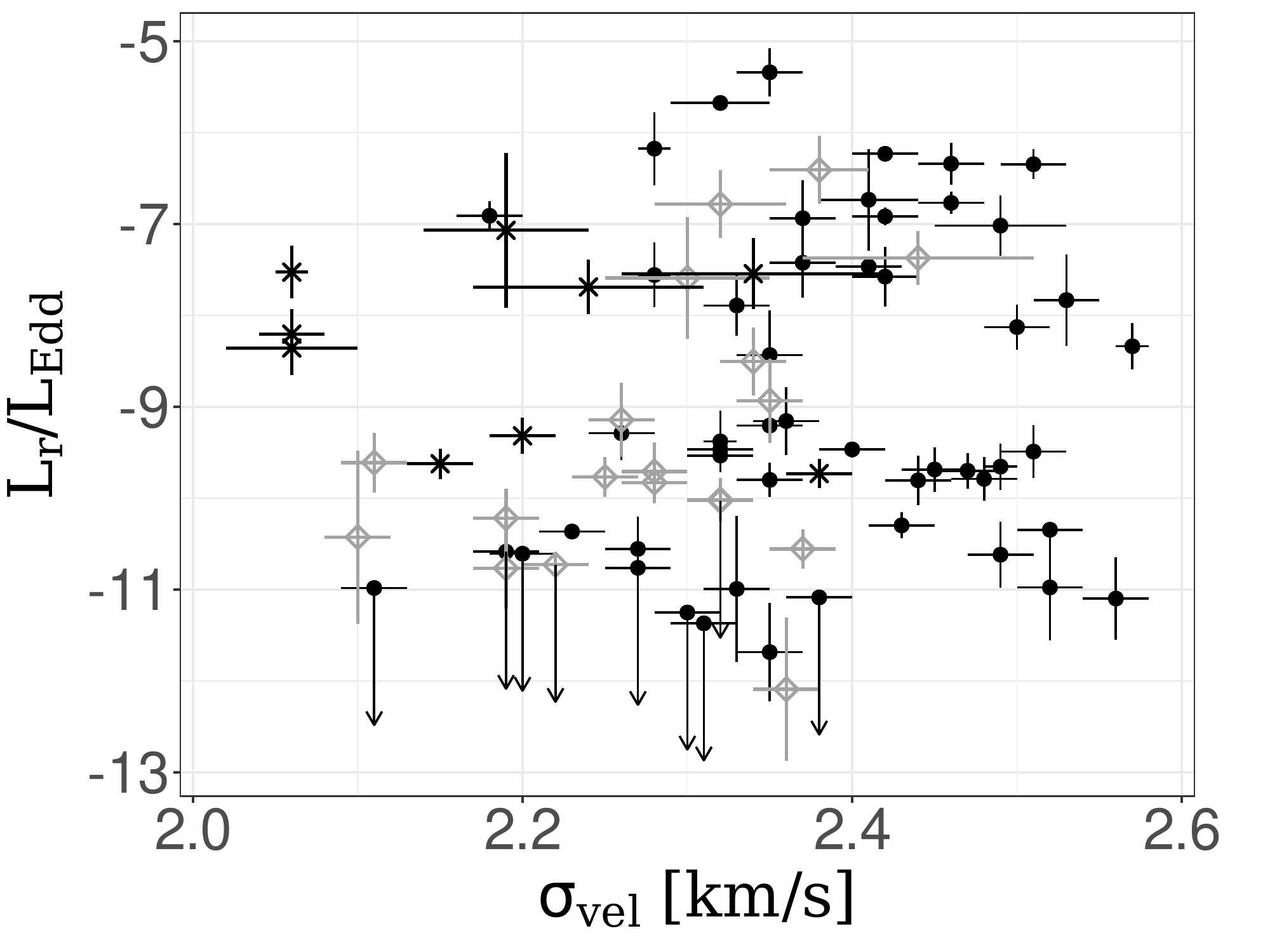}
\includegraphics[width=0.45\textwidth]{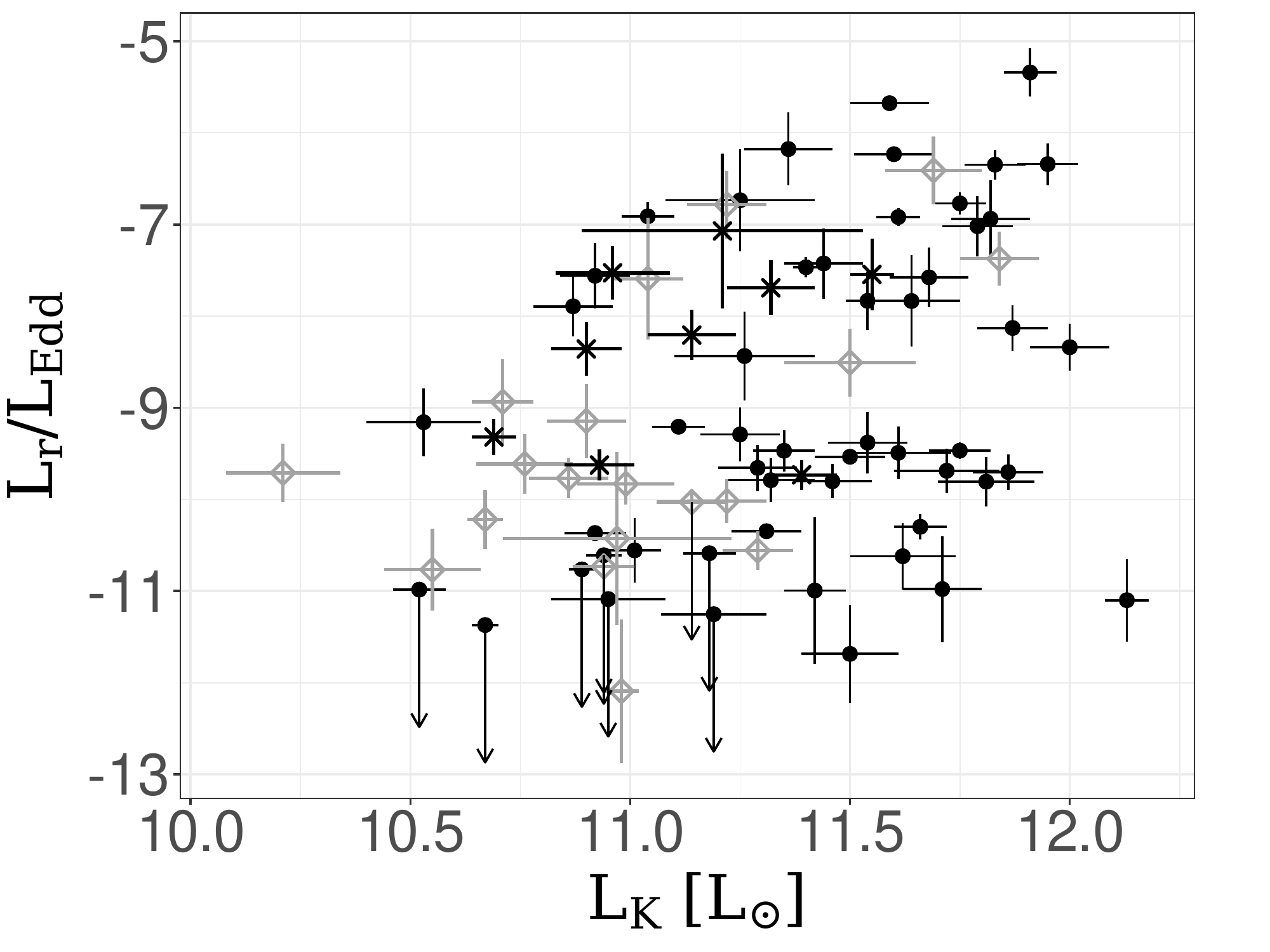}
\includegraphics[width=0.45\textwidth]{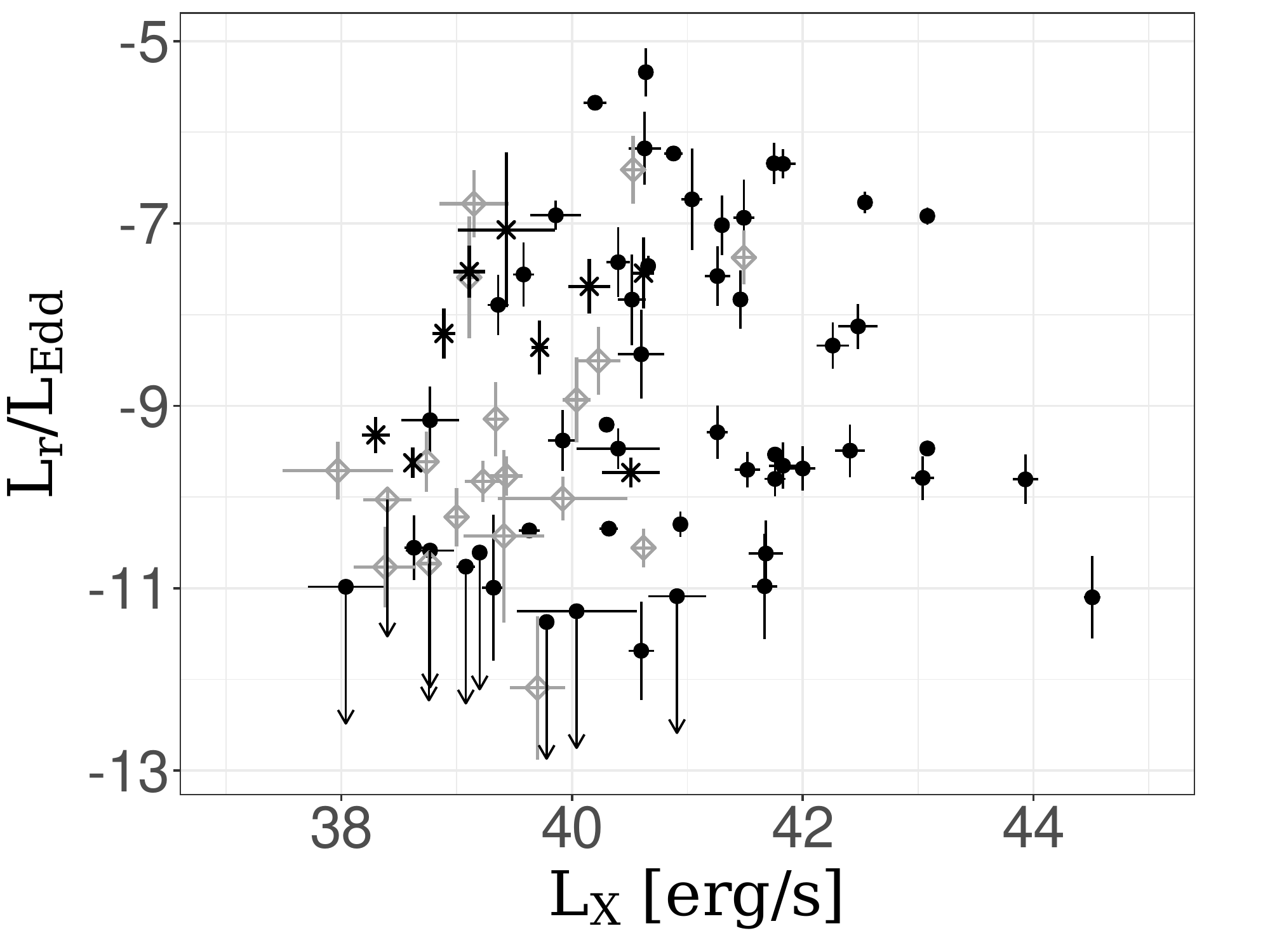}
\includegraphics[width=0.45\textwidth]{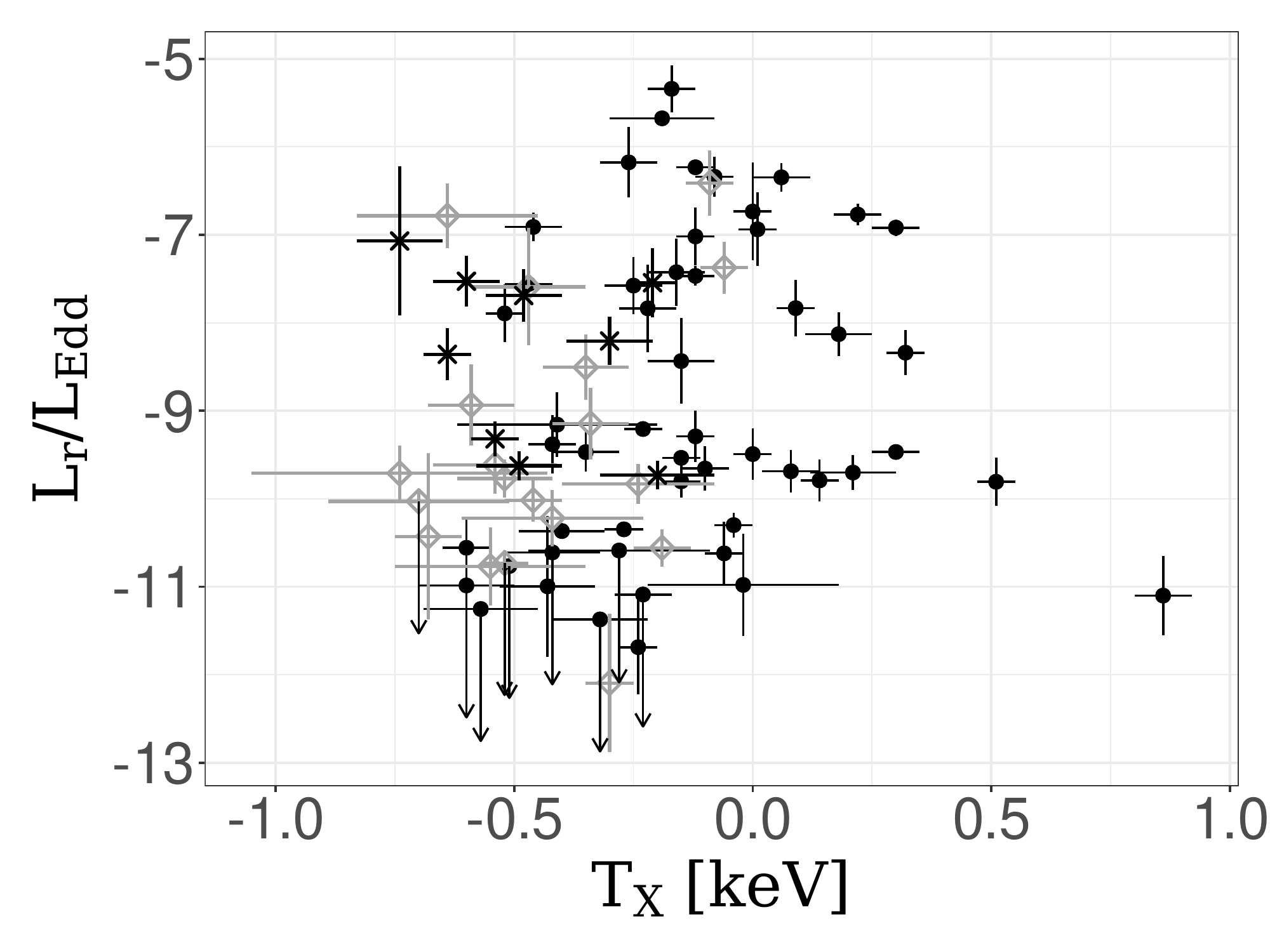}
\caption{Same as Figure\,\ref{fig:all}, except that here the 1.4\,GHz monochromatic luminosities are expressed in Eddington units, $L_{\rm r}/L_{\rm Edd}$.}
\label{fig:allEdd}
\end{figure}

In Figure\,\ref{fig:all} we present the scaling of the 1.4\,GHz monochromatic luminosities, $L_{\rm r}$, with the main galaxy and halo parameters --- the central black hole mass, $M_{\rm BH}$, the galactic bulge mass, $M_{\rm bulge}$, the stellar velocity dispersion, $\sigma_{\rm vel}$, the total galaxy $K$-band luminosity, $L_K$, the large-scale halo X-ray luminosity, $L_{\rm X}$, and the hot gas temperature, $T_{\rm X}$ --- for all the galaxies from the list of \citet{gaspari19}, including also radio upper limits. All the values for the optical and X-ray parameters, follows directly from \citeauthor{gaspari19} (Tables 2 and 3 therein). Radio luminosities for ellipticals and lenticular, follow from the flux measurements summarized in Appendix\,\ref{A:data}. Radio luminosities for the nine additional spirals (NGC\,7582, NGC\,3031, NGC\,4151, NGC\,4258, NGC\,7331, NGC\,0224, NGC\,1667, NGC\,1961, and NGC\,4594), follow from the analogous observations \citep{condon1998,condon2002,velzen2012}. In Figure\,\ref{fig:allEdd} we show the same scaling but for the 1.4\,GHz monochromatic luminosities expressed in Eddington units, $L_{\rm r}/L_{\rm Edd}$.

For ellipticals and lenticulars in the sample, we study the level of correlations of radio luminosities with the optical and X-ray parameters of the galaxies and their halos, using standard Pearson's product-moment correlation coefficient. Since the uncertainties in the estimates of those source parameters are in several cases relatively large, however, we resample our dataset for given measurement errors. That is, for each point in our dataset, we simulate data using random number generator implemented in R-studio, where the new data are drown from a normal distribution $\mathcal{N}(\mu_i,\sigma_i)$, with $\mu_i$ and $\sigma_i$ corresponding to the measured value and the uncertainty of that value in our original dataset, respectively. To minimize the possibility of the degeneration for each point in our original dataset, we drown 10,000 observations. As a result, instead of performing the correlation analysis with single points, i.e., observing values from our original data set, we investigate the correlation between the clusters of points, for which the distribution is the densest the closer to the original observed values and is given by the 2-dimensional Gaussian distribution. The Pearson's product-moment correlation values, $\rho$, obtained in such a way for the relations presented in Figures\,\ref{fig:all} and \ref{fig:allEdd}, are listed in Table\,\ref{tab:all-corr}. 

As follows, for ellipticals we do not see any correlation between radio luminosities and black hole masses, while the correlations between $L_{\rm r}$ and the remaining parameters of the galaxies and their halos, are medium at most. In the case of lenticulars, on the other hand, the correlations between $L_{\rm r}$ and $M_{\rm bulge}$, $\sigma_{\rm vel}$, $L_K$, or $L_{\rm X}$, become strong, with the corresponding correlation coefficients $0.5 <\rho < 0.7$. Note however a smaller number of lenticulars when compared with ellipticals in the sample (17 versus 45 radio-detected sources). For the joint sample of ellipticals and lenticulars, the two pronounced ($\rho \gtrsim 0.5$) correlations remaining are those with the optical galaxy parameters, namely $L_{\rm r} - M_{\rm bulge}$ and $L_{\rm r} - L_{\rm K}$. When radio luminosities are expressed in Eddington units, all the medium and strong correlations vanish.

\begin{deluxetable}{cccc}[!th]
%\tabletypesize{\footnotesize}
%\tabletypesize{\scriptsize}
\tablecaption{The Pearson's correlation coefficients $\rho$ for the elliptical (E), lenticulars (L), and all early-type (E+L) galaxies.}
\label{tab:all-corr}
\tablewidth{0pt}
\tablehead{
\colhead{Relation} & \colhead{$\rho_{\rm E}$}  & \colhead{$\rho_{\rm L}$} & \colhead{$\rho_{\rm E+L}$}
}
\startdata
   $L_{\rm r}$\,\,vs.\,\,$M_{\rm BH}$& 0.06 & 0.32 & 0.23\\
    $L_{\rm r}$\,\,vs.\,\,$M_{\rm bulge}$& 0.35 & 0.58 & 0.52\\  
    $L_{\rm r}$\,\,vs.\,\,$\sigma_{\rm vel}$& 0.23 & 0.54 & 0.41\\
    $L_{\rm r}$\,\,vs.\,\,$\,L_{\rm K}$& 0.35 & 0.68 & 0.54\\    
    $L_{\rm r}$\,\,vs.\,\,$\,L_{\rm X}$& 0.27 & 0.50 & 0.43\\    
    $L_{\rm r}$\,\,vs.\,\,$T_{\rm X}$& 0.29& 0.32 & 0.40\\    
    $L_{\rm r}/L_{\rm Edd}$\,\,vs.\,\,$M_{\rm BH}$ & --0.28 & 0.09& --0.11\\
    $L_{\rm r}/L_{\rm Edd}$\,\,vs.\,\,$M_{\rm bulge}$ & 0.10 & 0.37 & 0.25\\
    $L_{\rm r}/L_{\rm Edd}$\,\,vs.\,\,$\sigma_{\rm vel}$ & --0.05 & 0.37 & 0.13\\
    $L_{\rm r}/L_{\rm Edd}$\,\,vs.\,\,$L_{\rm K}$ & 0.12 & 0.50 & 0.29\\
    $L_{\rm r}/L_{\rm Edd}$\,\,vs.\,\,$L_{\rm X}$ & --0.04 & 0.30 & 0.11\\
    $L_{\rm r}/L_{\rm Edd}$\,\,vs.\,\,$T_{\rm X}$ & --0.0 & 0.16 & 0.09\\
\enddata
%\tablecomments{}
\end{deluxetable}

\section{High-resolution Radio Maps.} 
\label{A:VLASS}

\begin{figure}[ht]
    \centering
\includegraphics[width=0.29\textwidth]{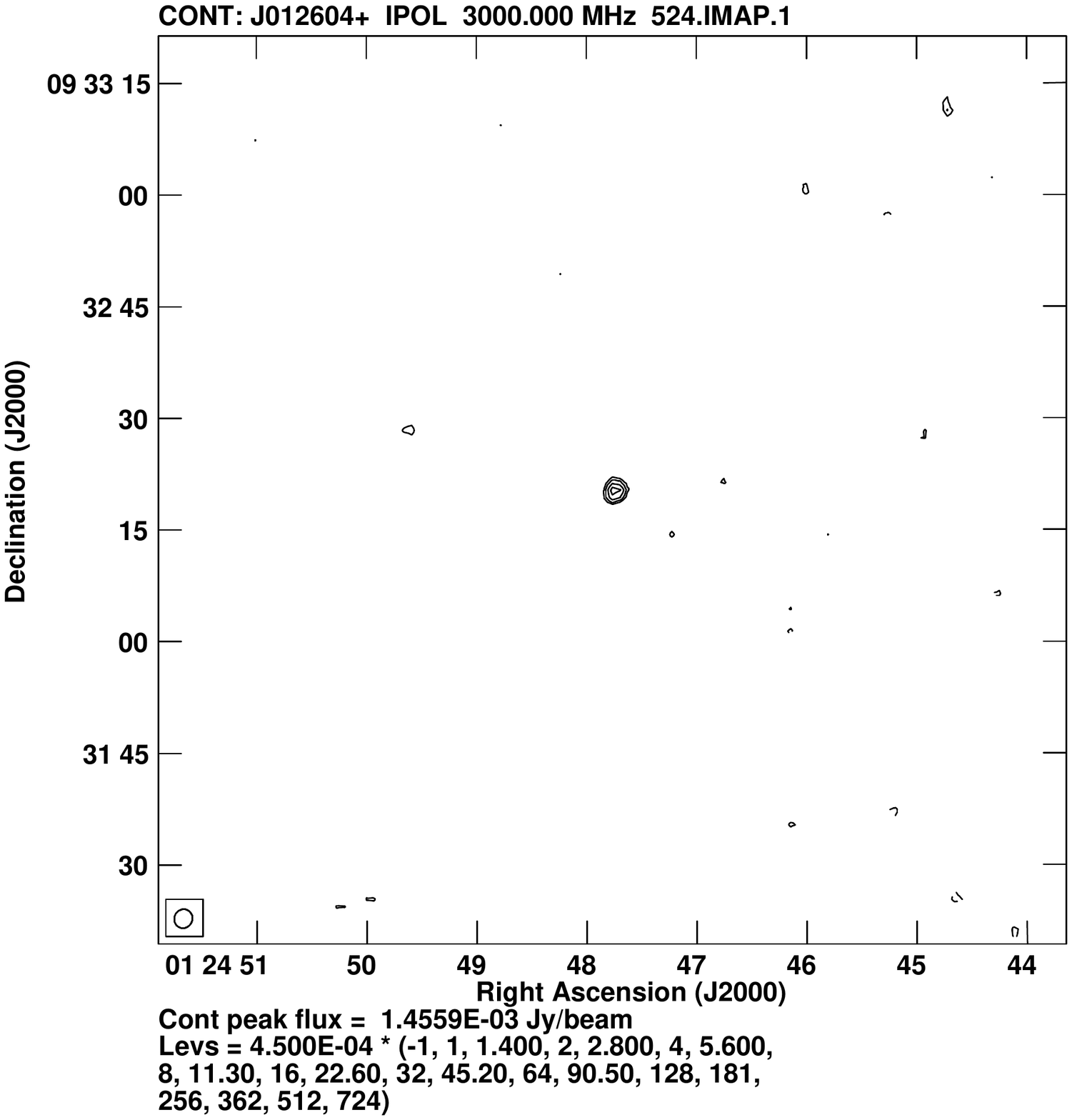}
\includegraphics[width=0.29\textwidth]{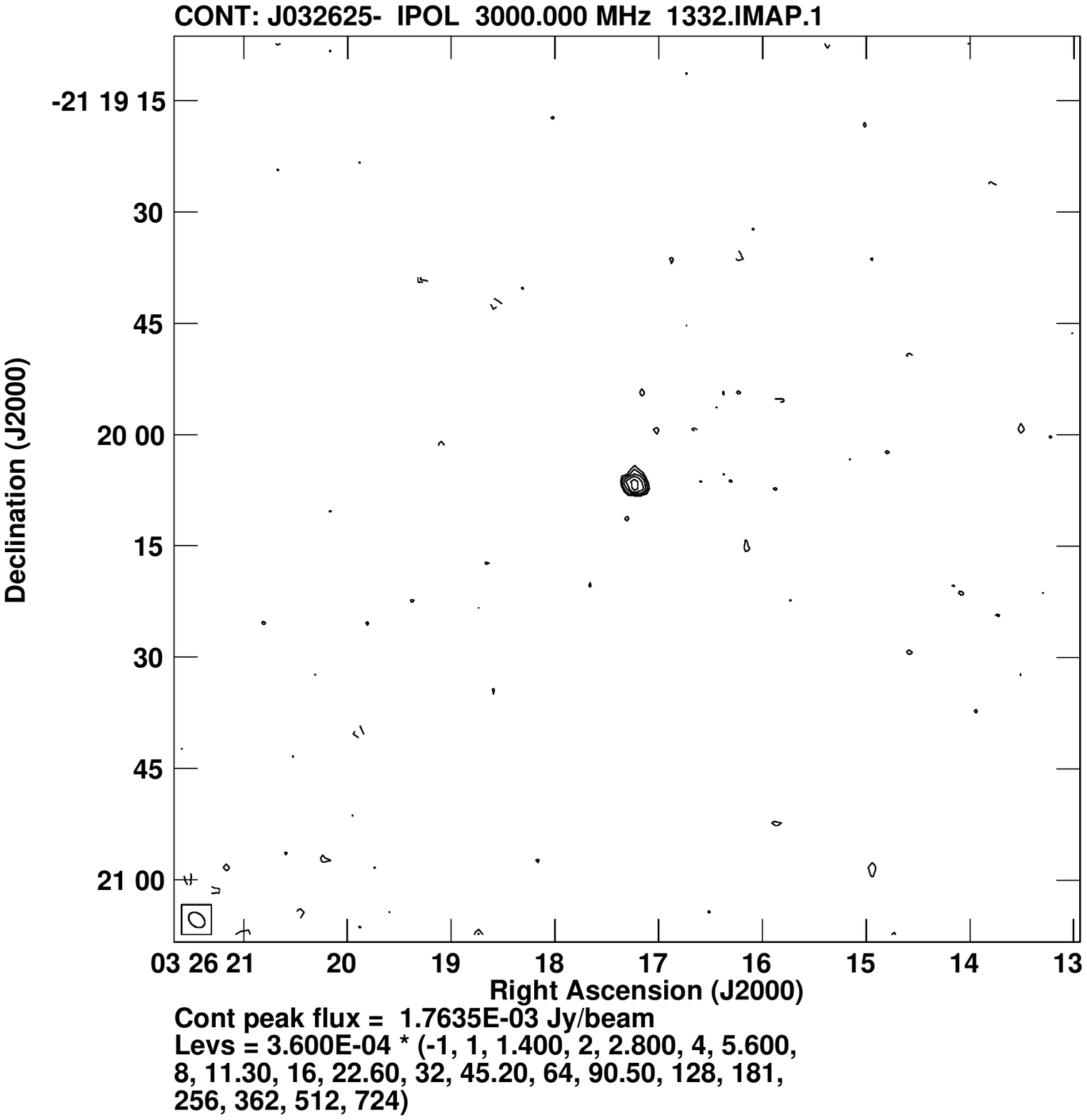}
\includegraphics[width=0.29\textwidth]{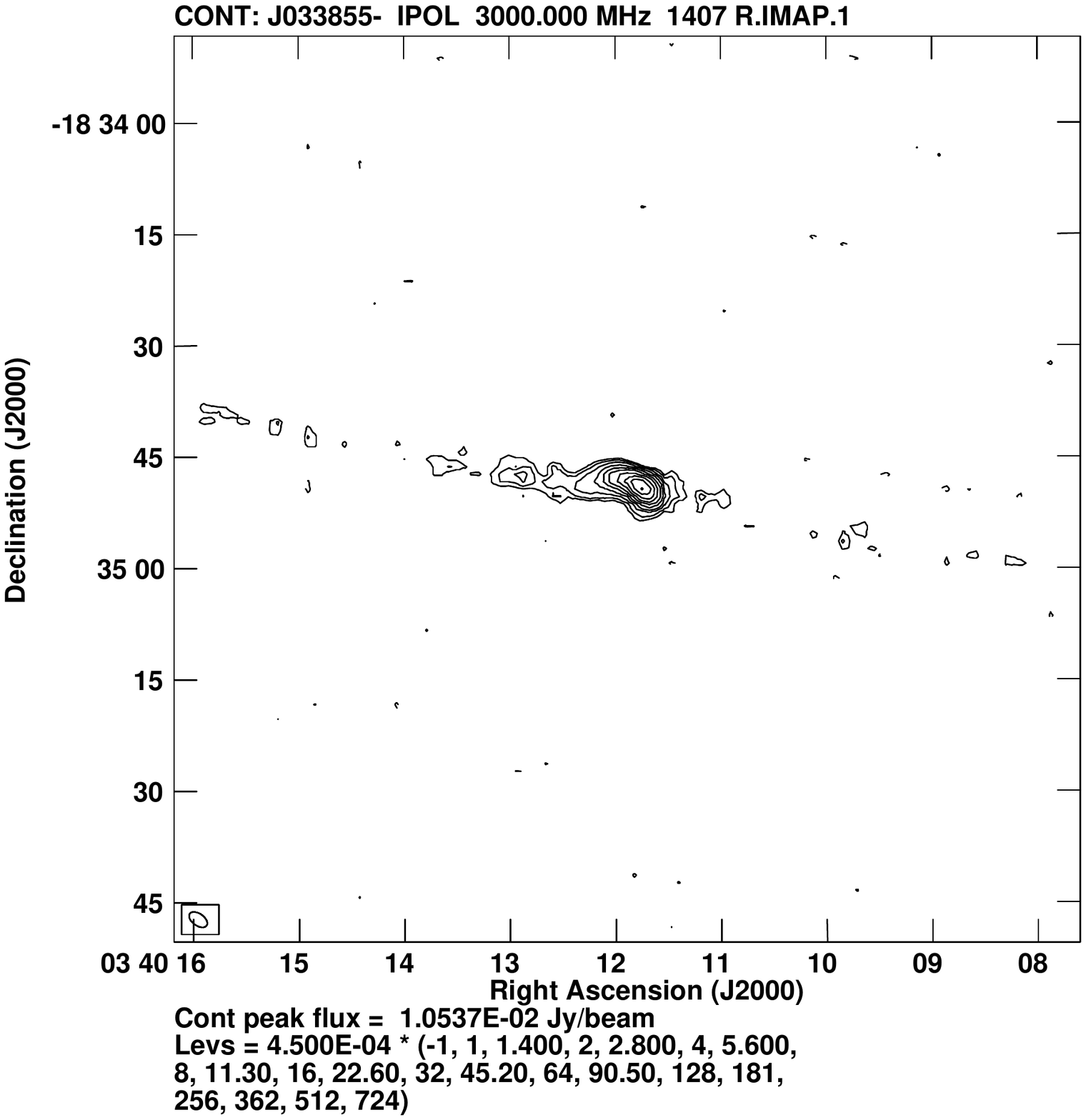}
\includegraphics[width=0.29\textwidth]{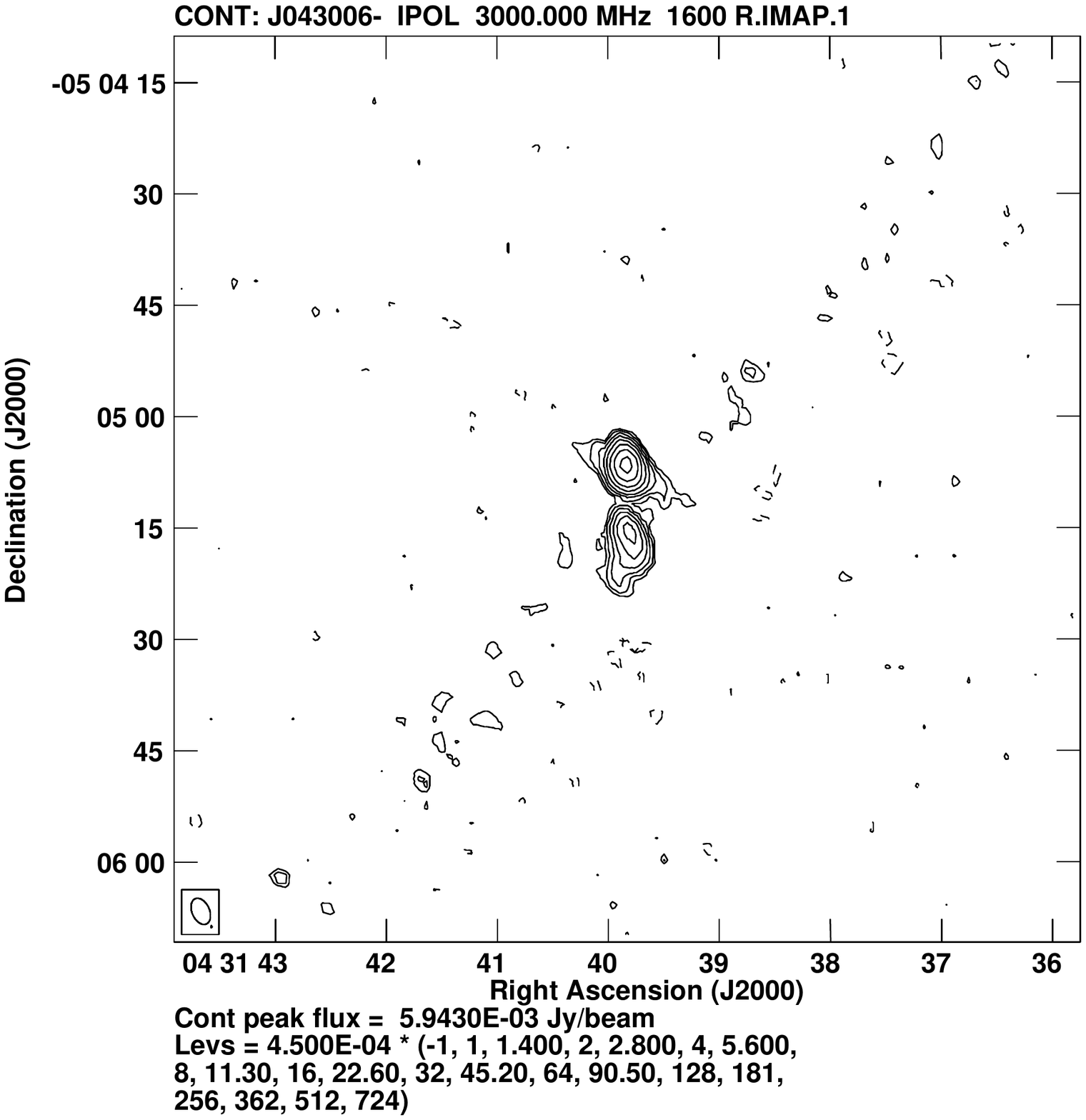}
\includegraphics[width=0.29\textwidth]{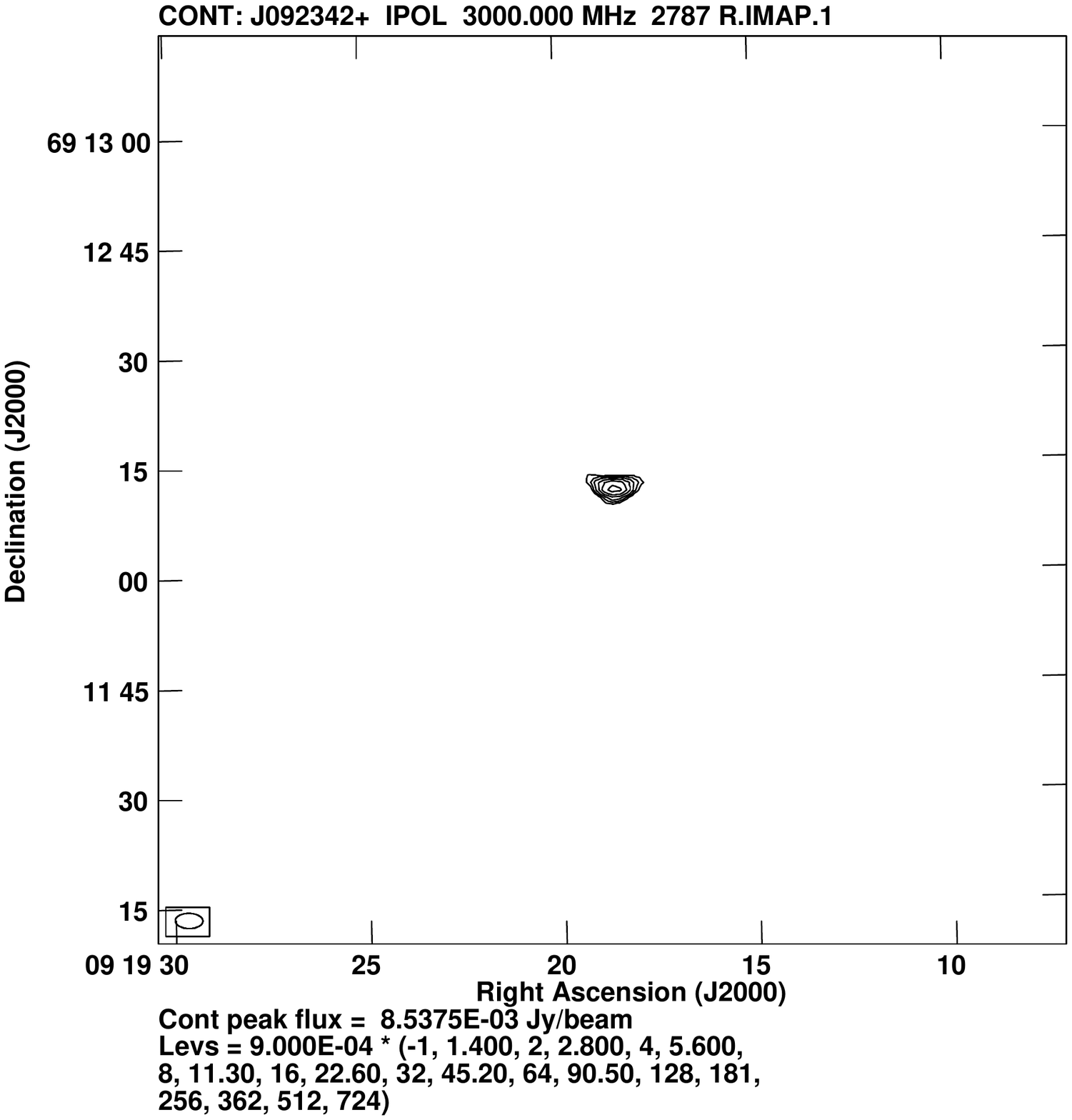}
\includegraphics[width=0.29\textwidth]{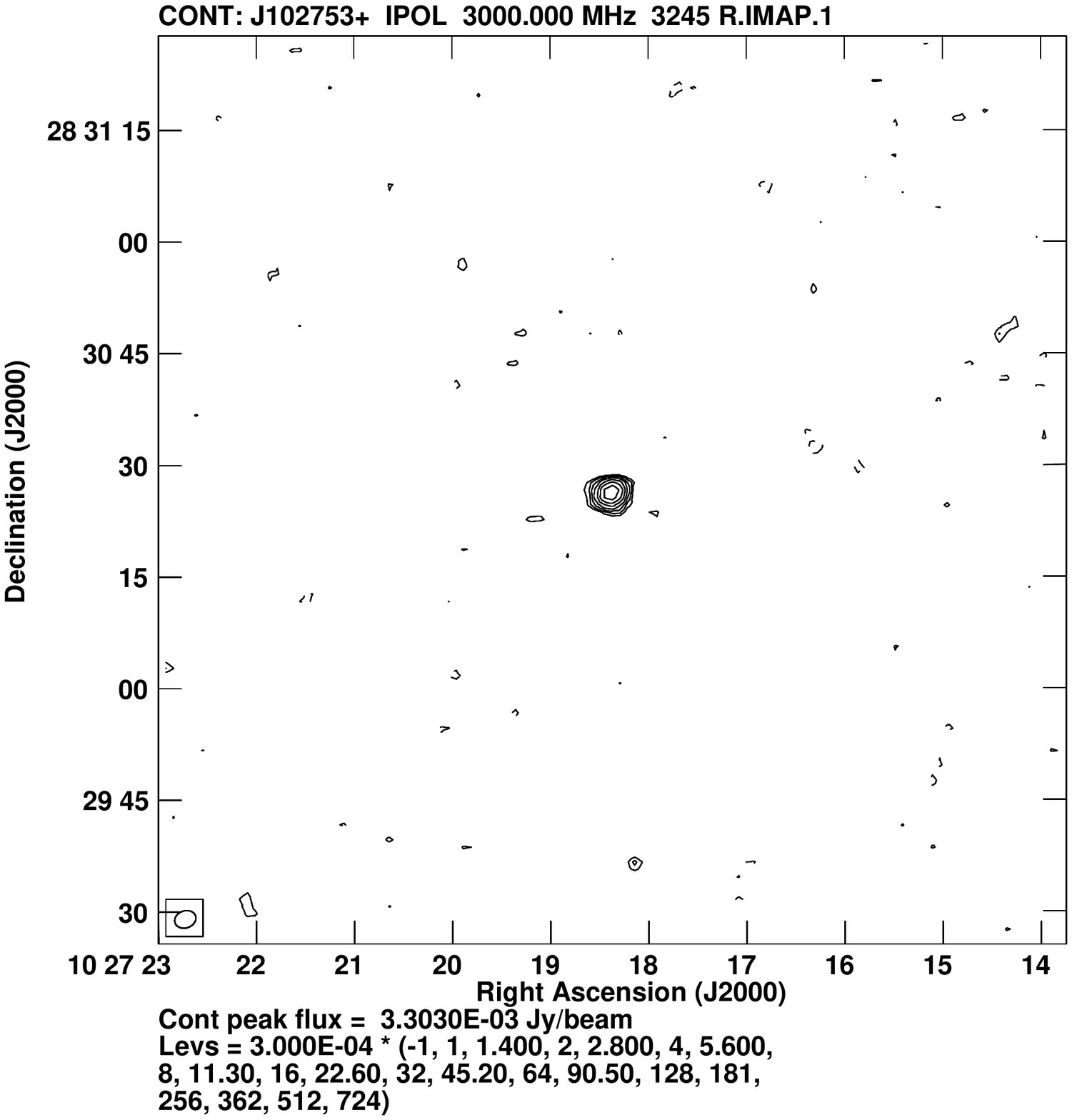}
\includegraphics[width=0.29\textwidth]{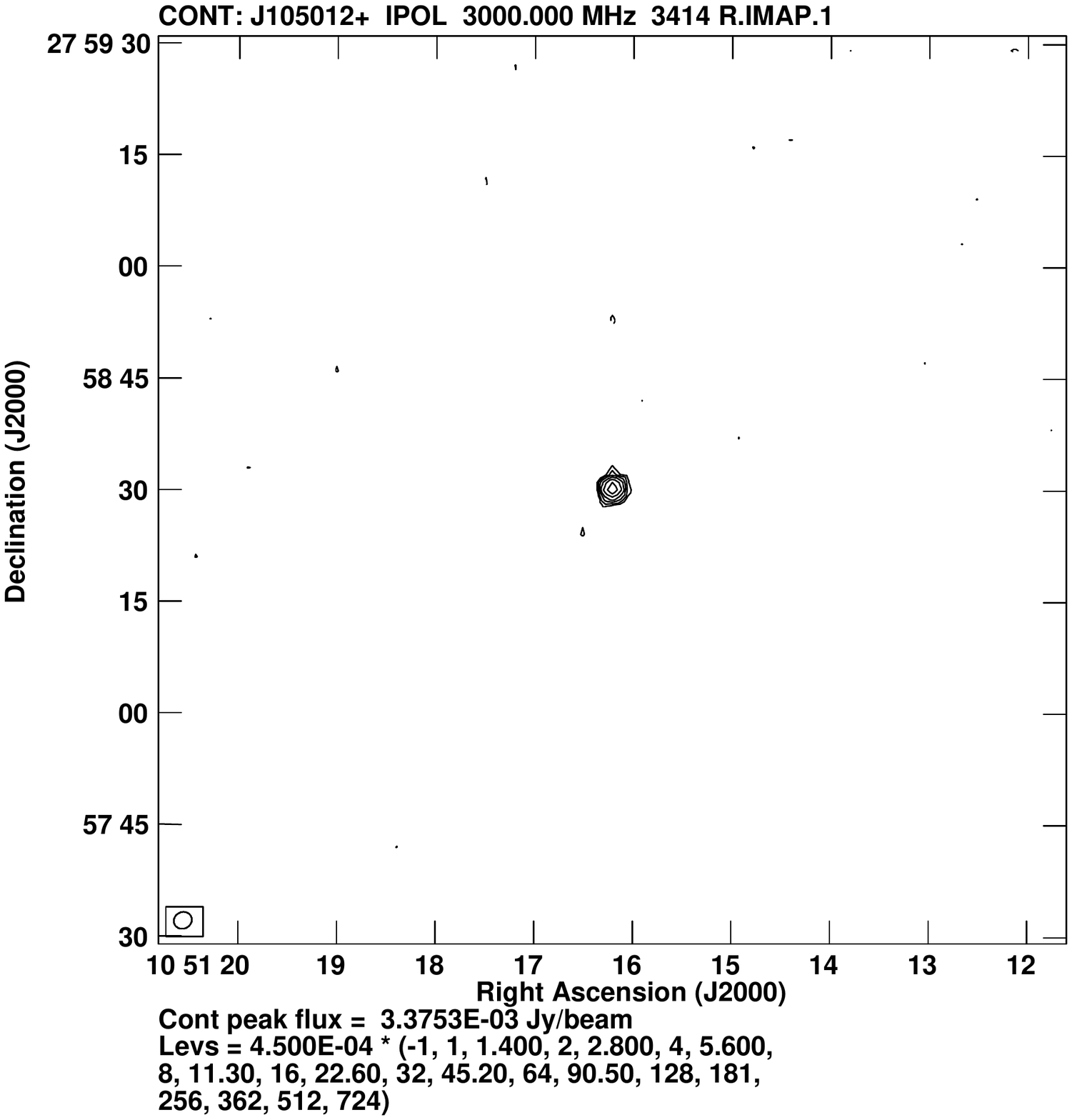}
\includegraphics[width=0.29\textwidth]{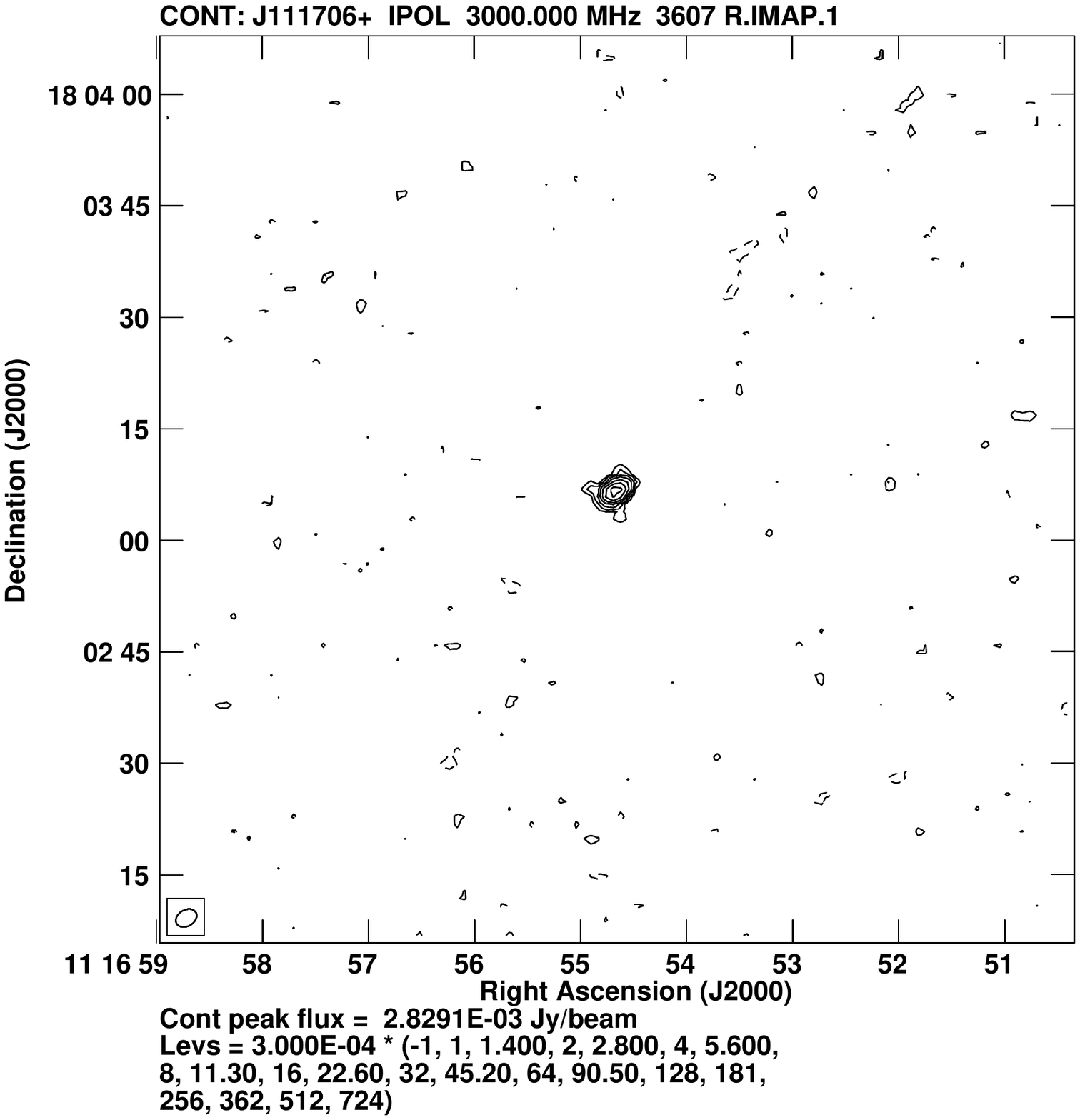}
\includegraphics[width=0.29\textwidth]{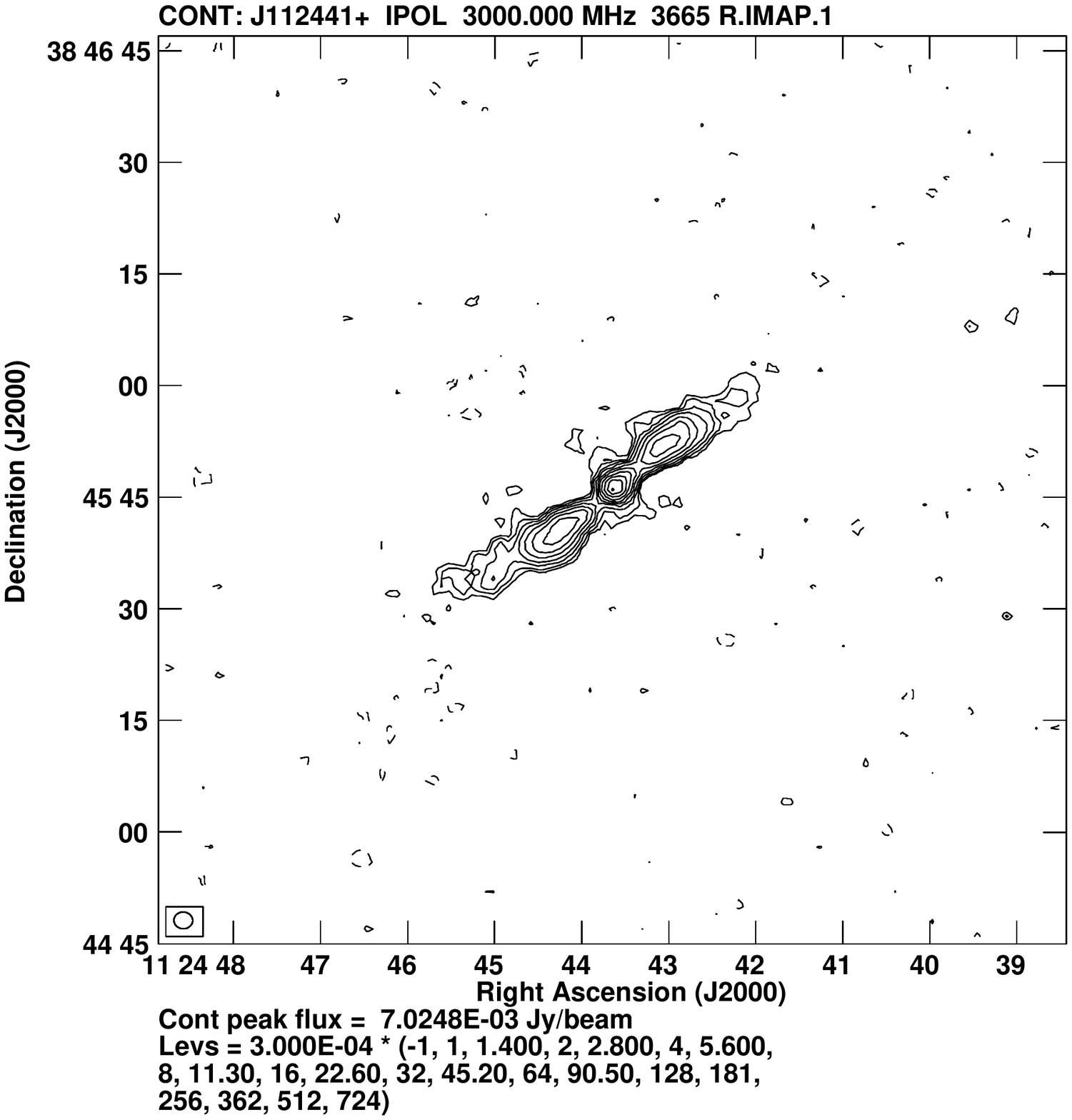}
\caption{VLASS contour maps available for radio-dim galaxies NGC\,0524, NGC\,1332, NGC\,1407 (top panel, left to right), NGC\,1600, NGC\,2787, NGC\,3245 (middle panel), and NGC\,3414, NGC\,3607, NGC\,3665 (bottom panel).}
\label{fig:A:VLASS-RD-1}
\end{figure}

\begin{figure}[ht]
    \centering
\includegraphics[width=0.29\textwidth]{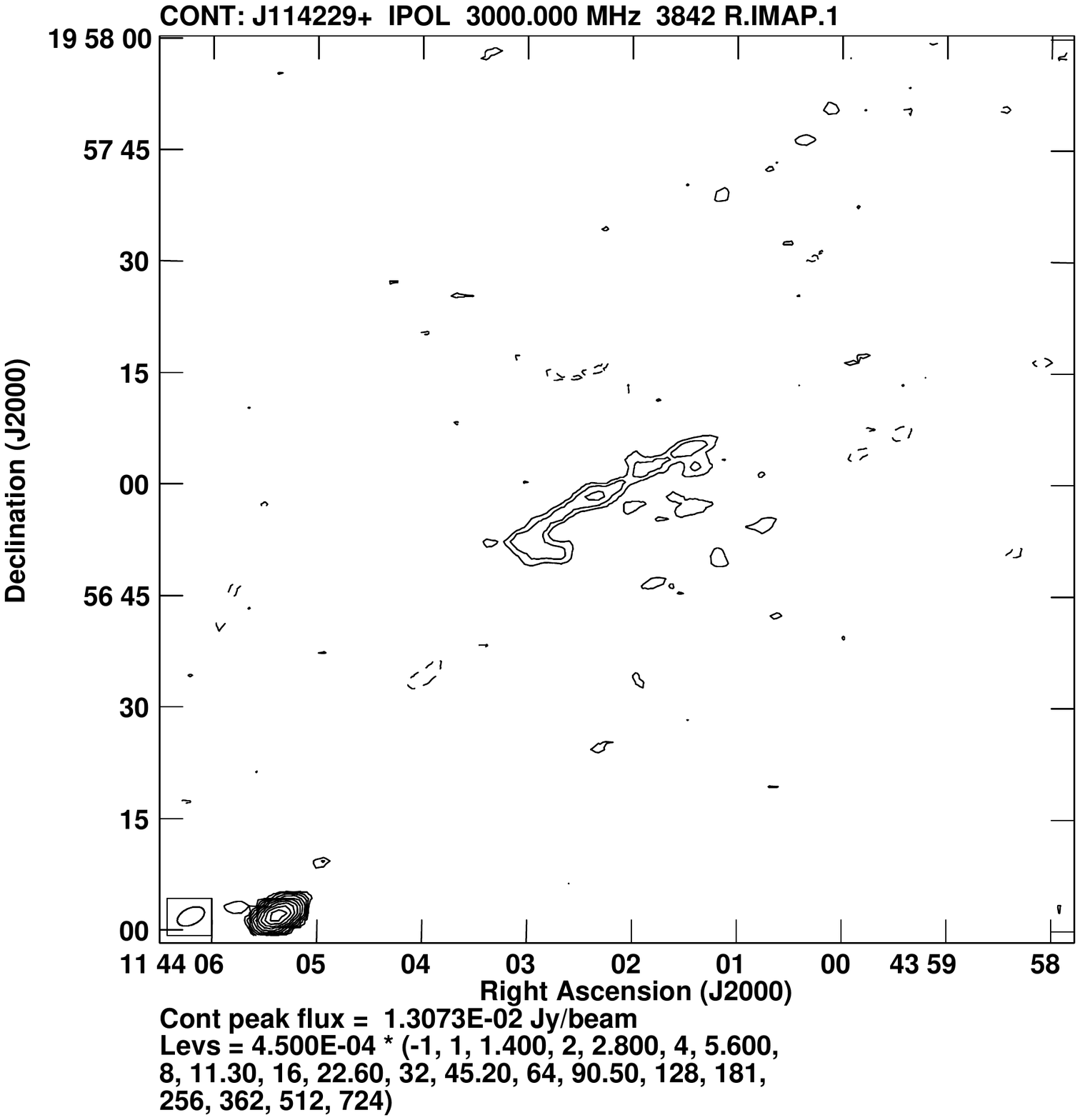}
\includegraphics[width=0.29\textwidth]{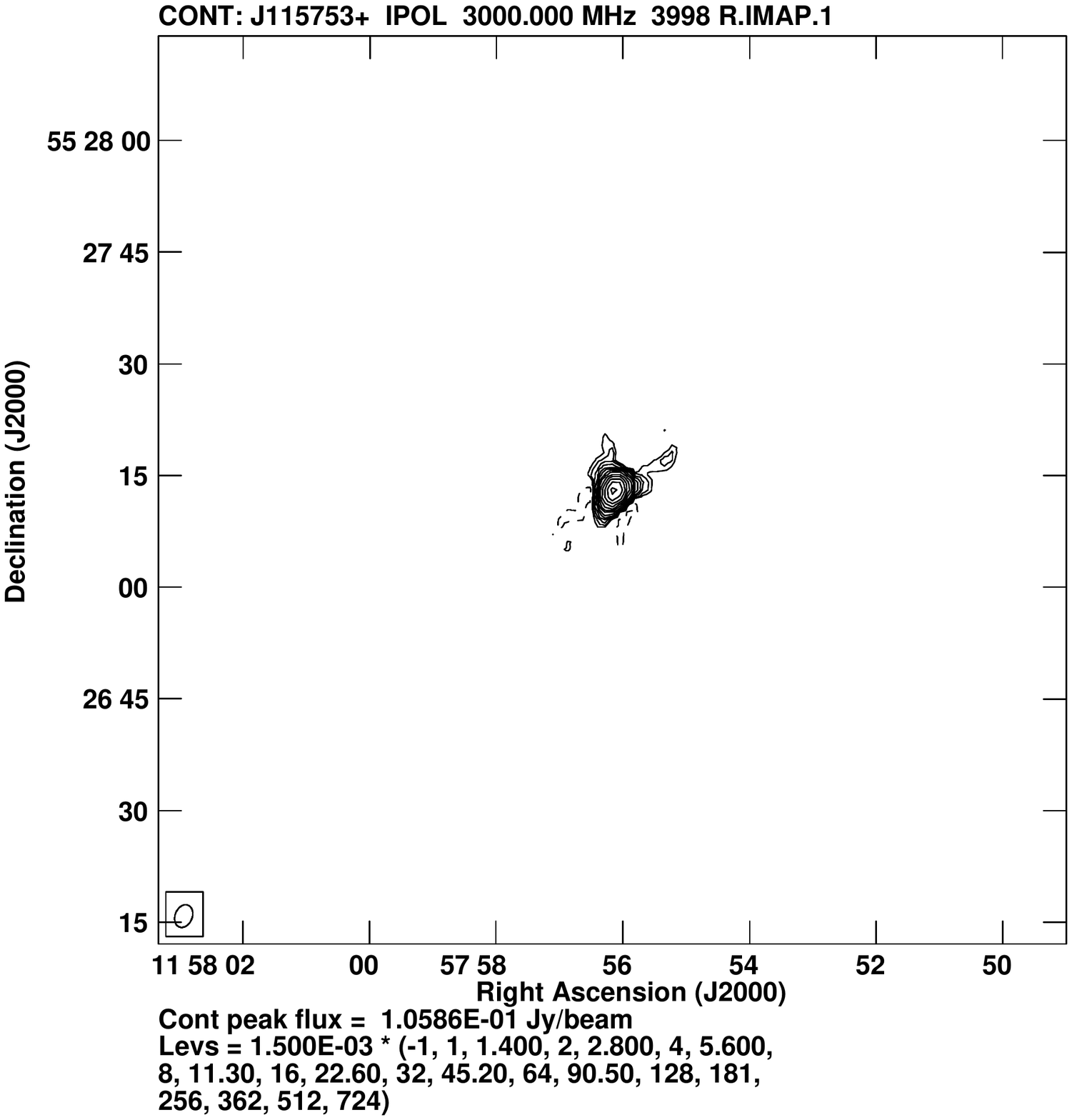}
\includegraphics[width=0.29\textwidth]{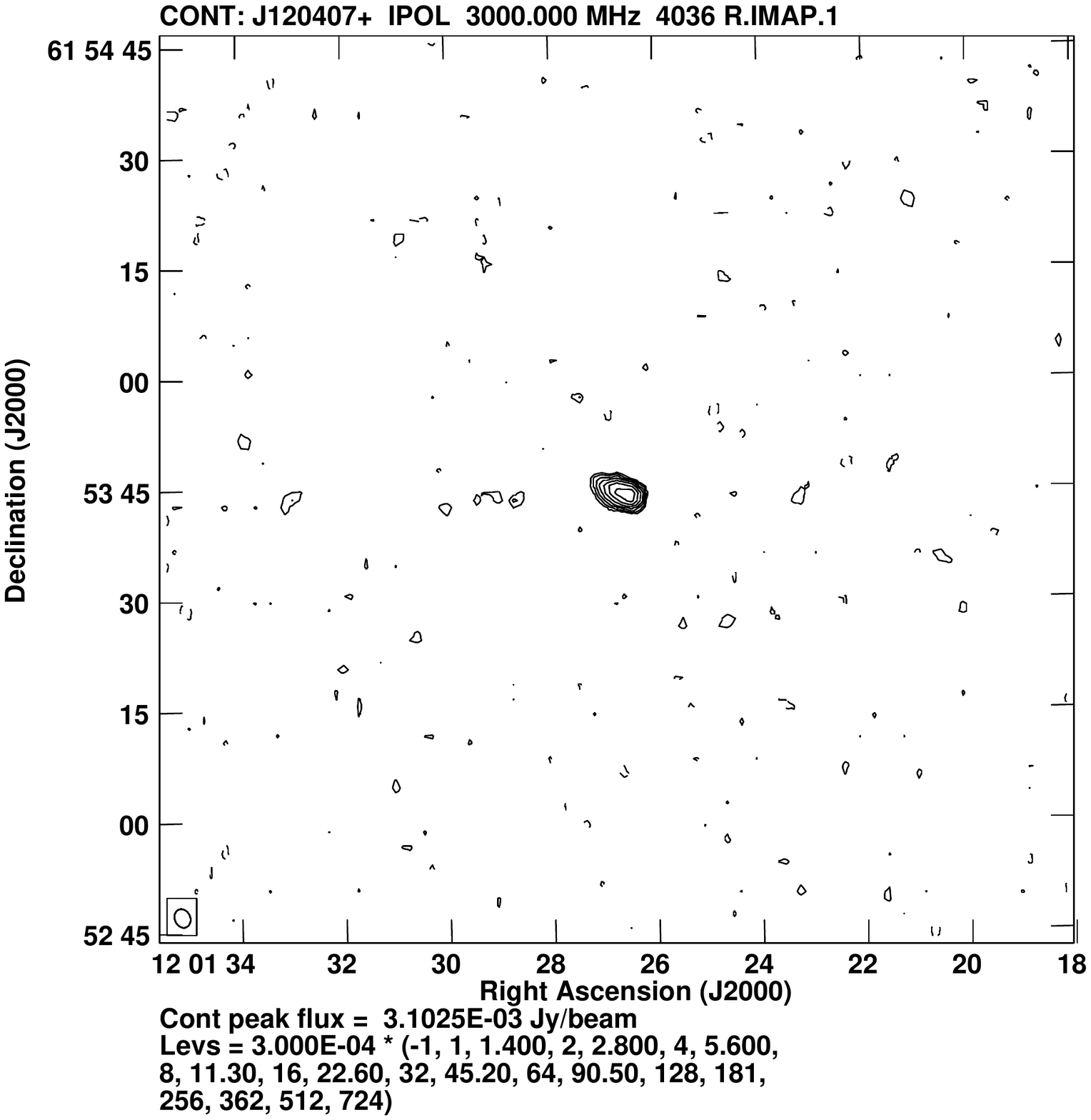}
\includegraphics[width=0.29\textwidth]{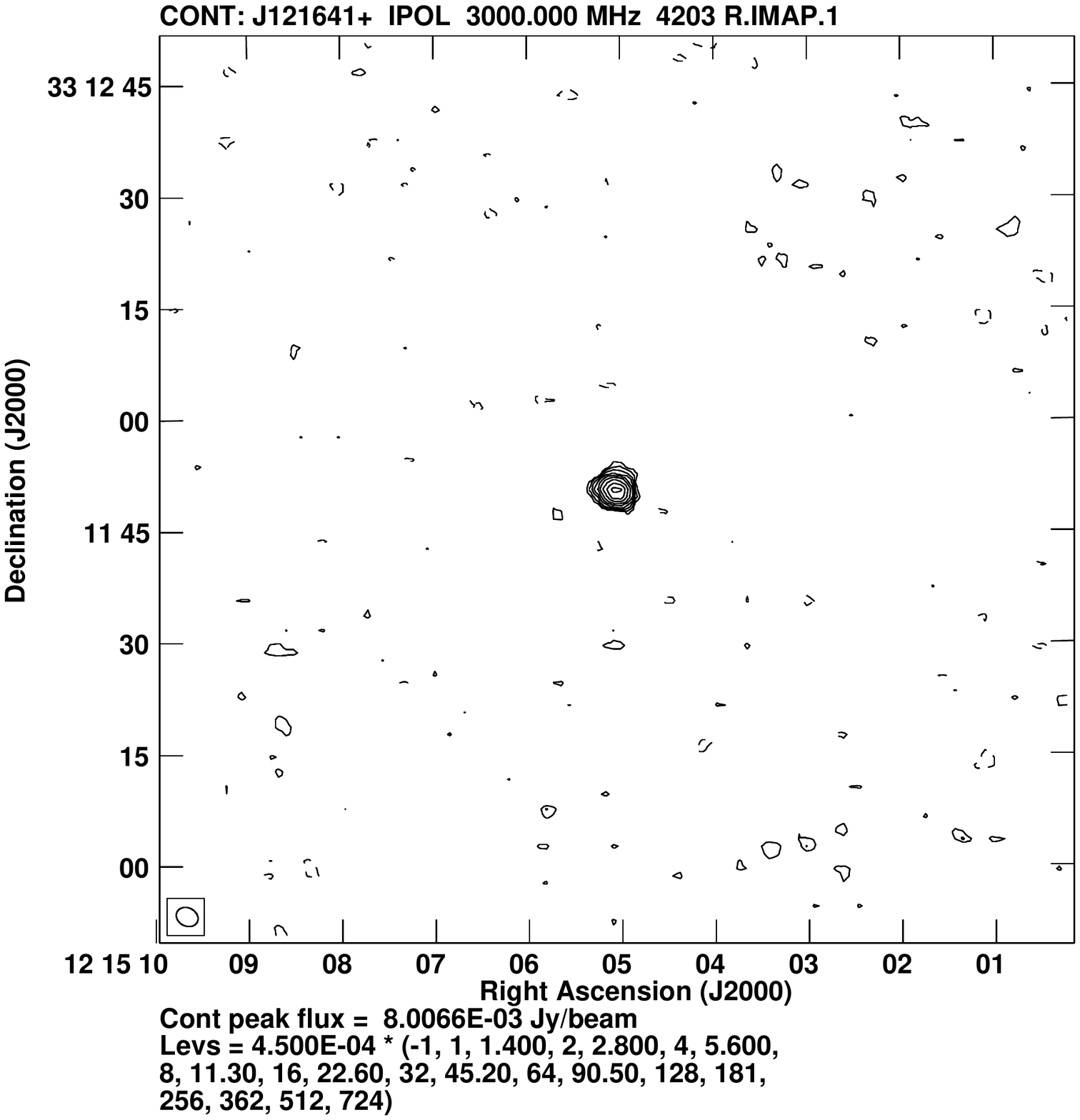}
\includegraphics[width=0.29\textwidth]{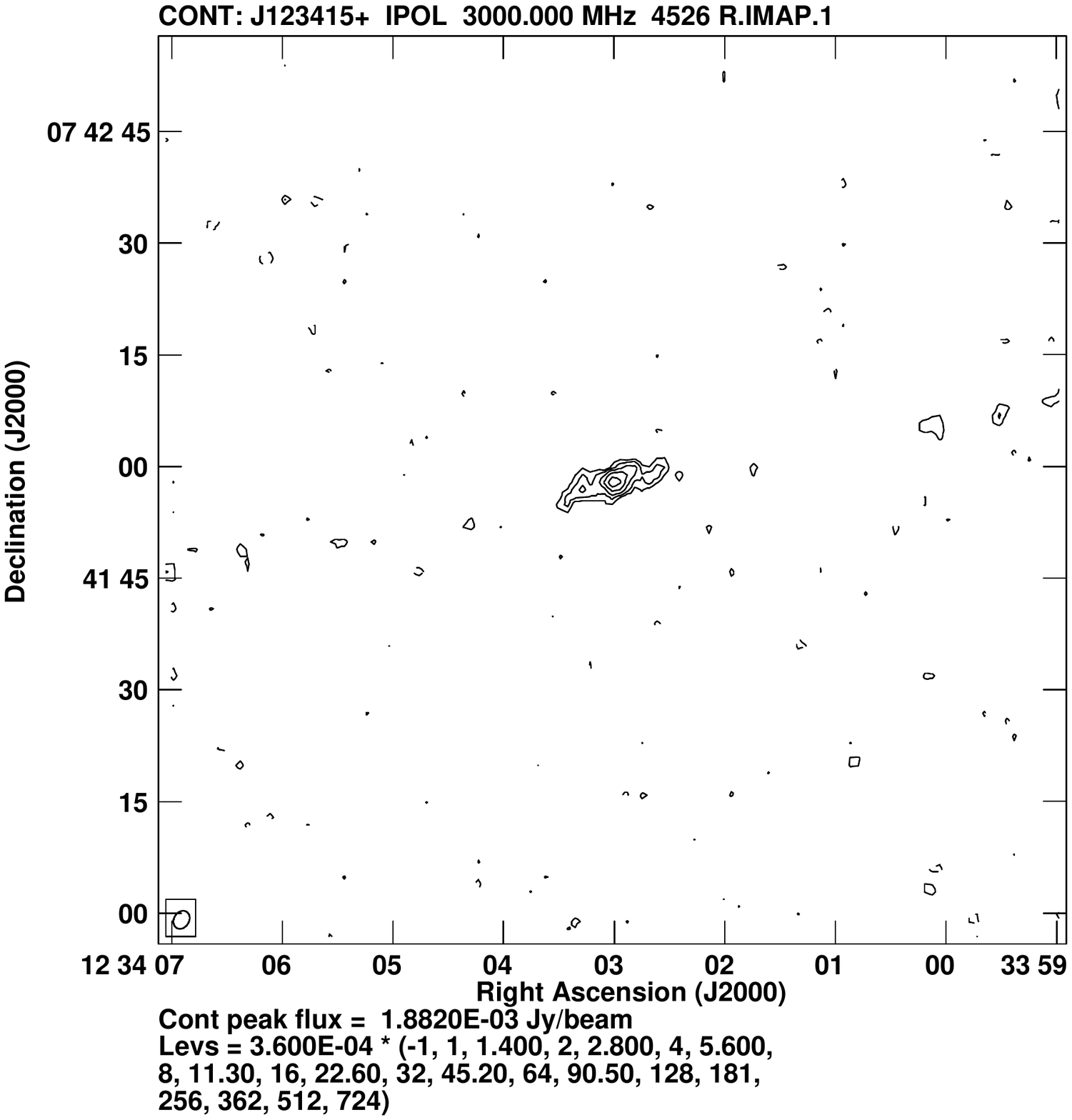}
\includegraphics[width=0.29\textwidth]{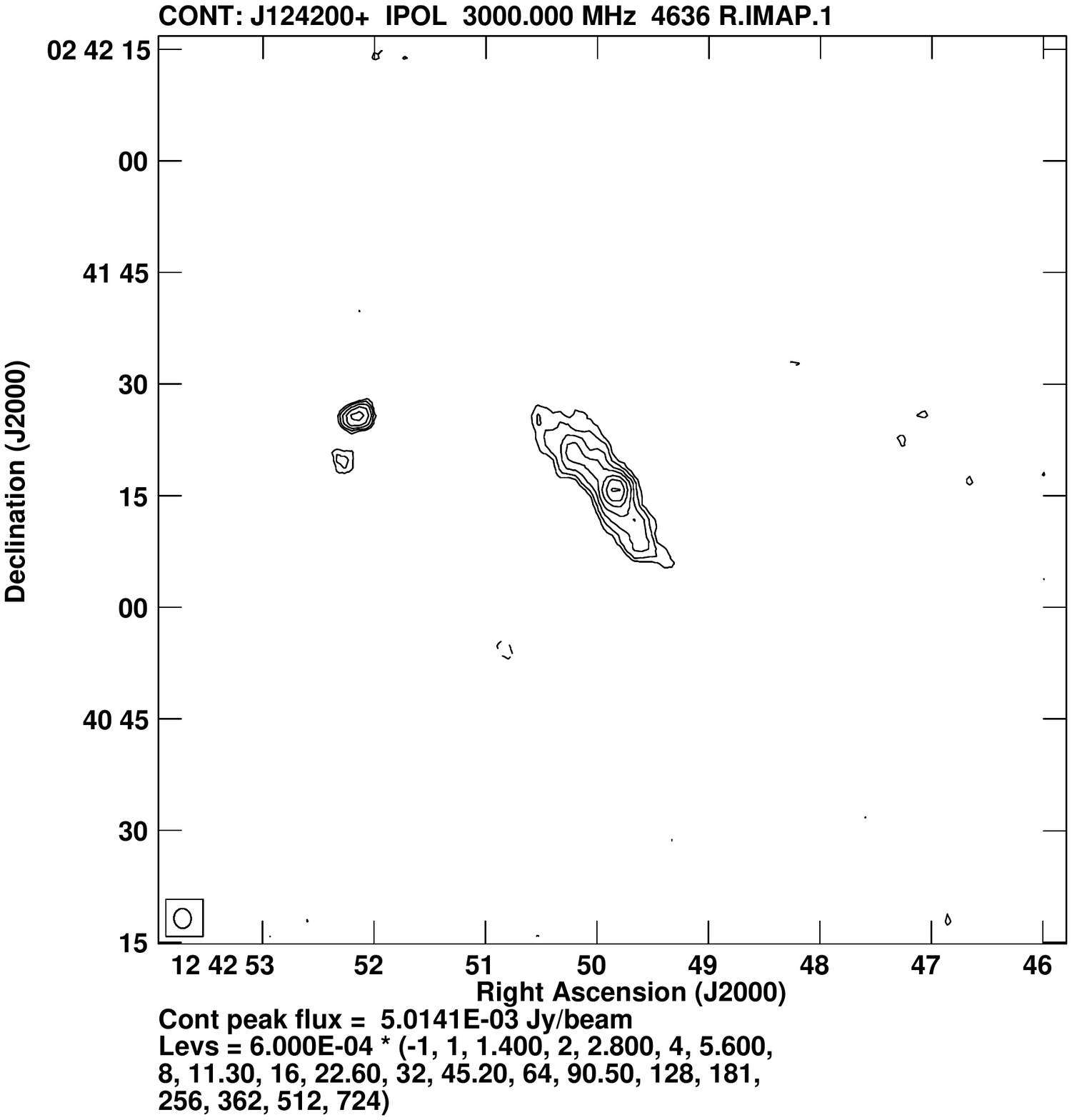}
\includegraphics[width=0.29\textwidth]{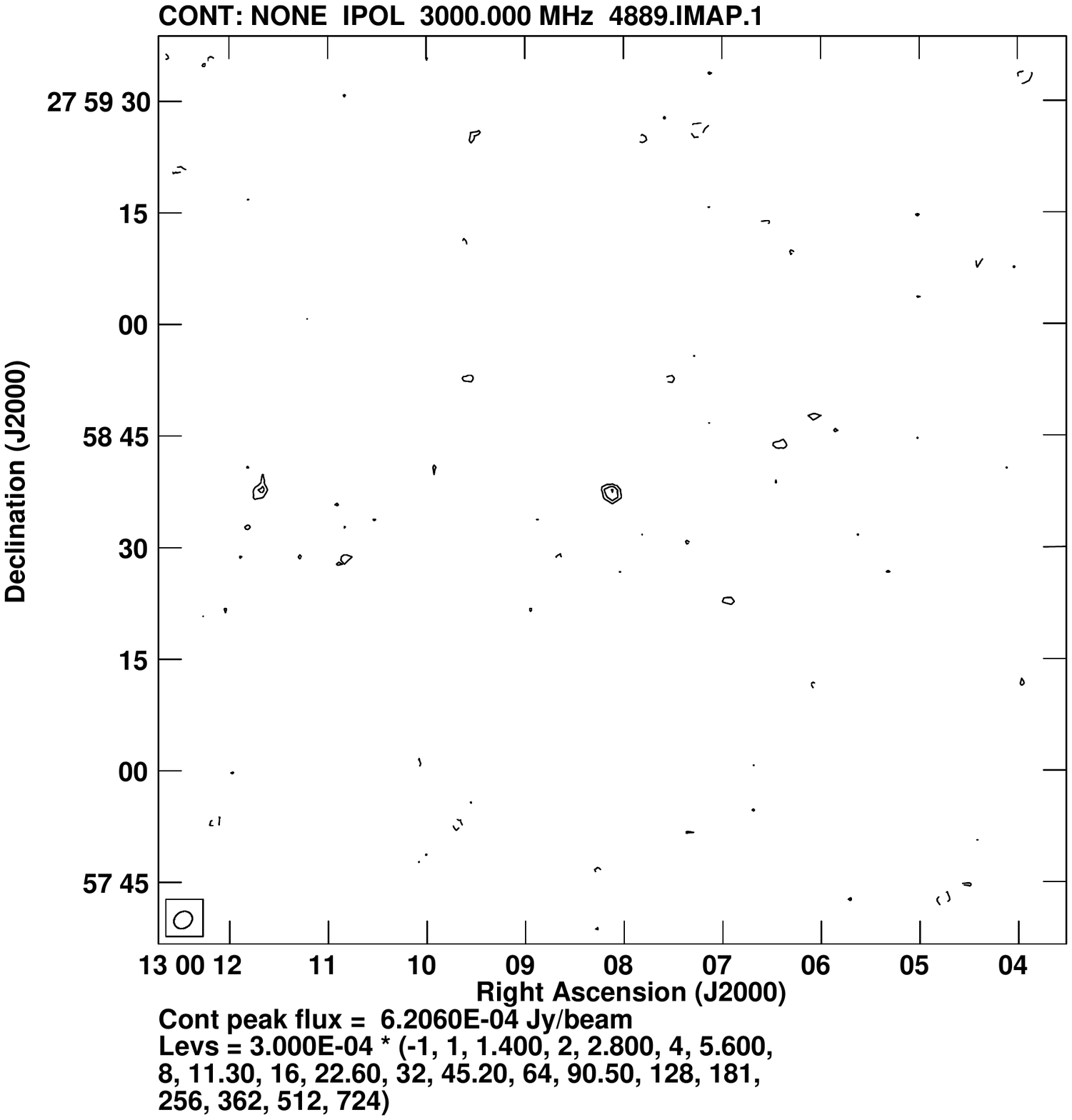}
\includegraphics[width=0.29\textwidth]{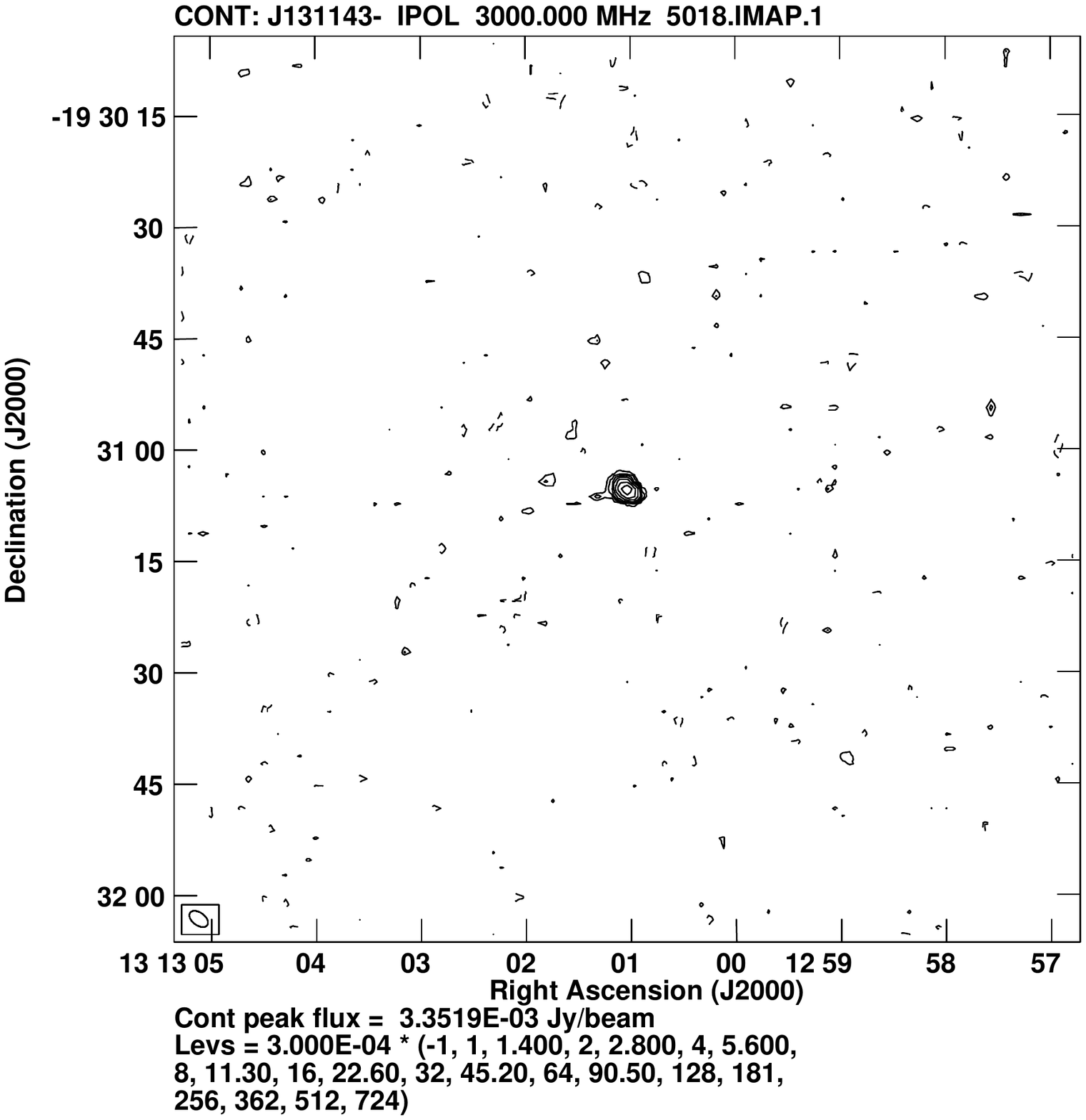}
\includegraphics[width=0.29\textwidth]{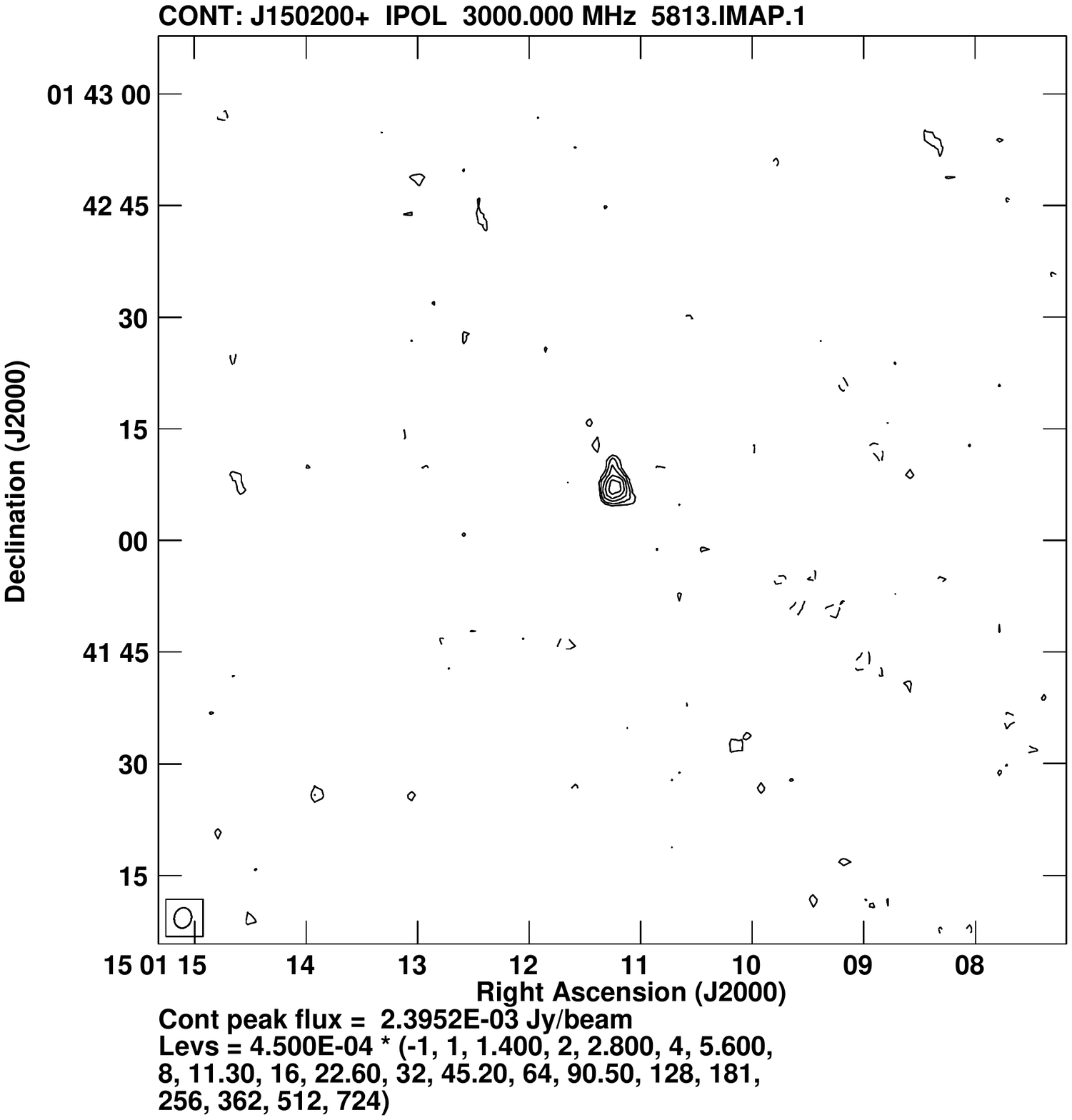}
\caption{VLASS contour maps available for radio-dim galaxies NGC\,3842, NGC\,3998, NGC\,4036 (top panel, left to right), NGC\,4203, NGC\,4526, NGC\,4636 (middle panel), and NGC\,4889, NGC\,5018, NGC\,5813 (bottom panel).}
\label{fig:A:VLASS-RD-2}
\end{figure}

\begin{figure}[ht]
    \centering
\includegraphics[width=0.29\textwidth]{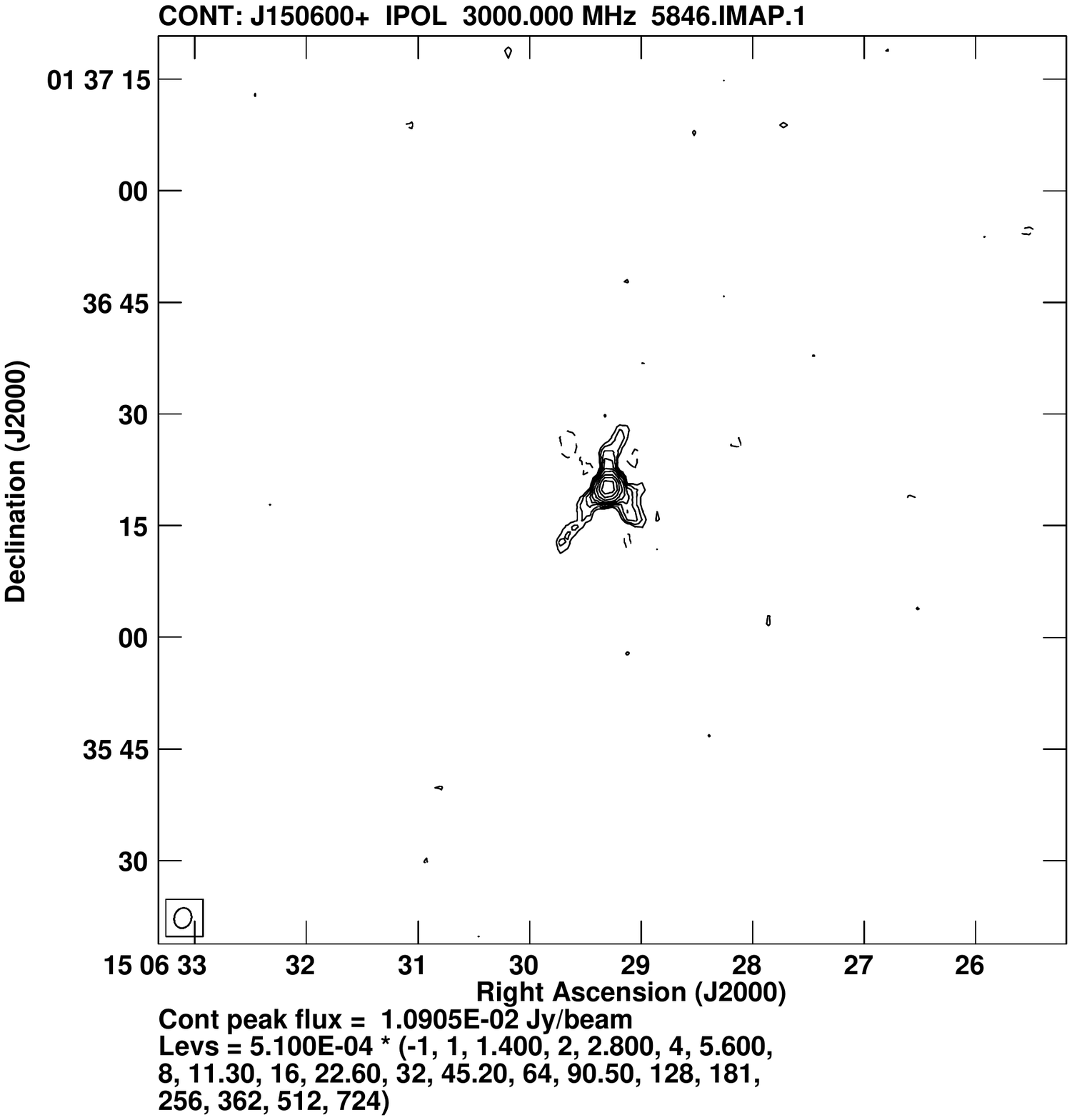}
\includegraphics[width=0.29\textwidth]{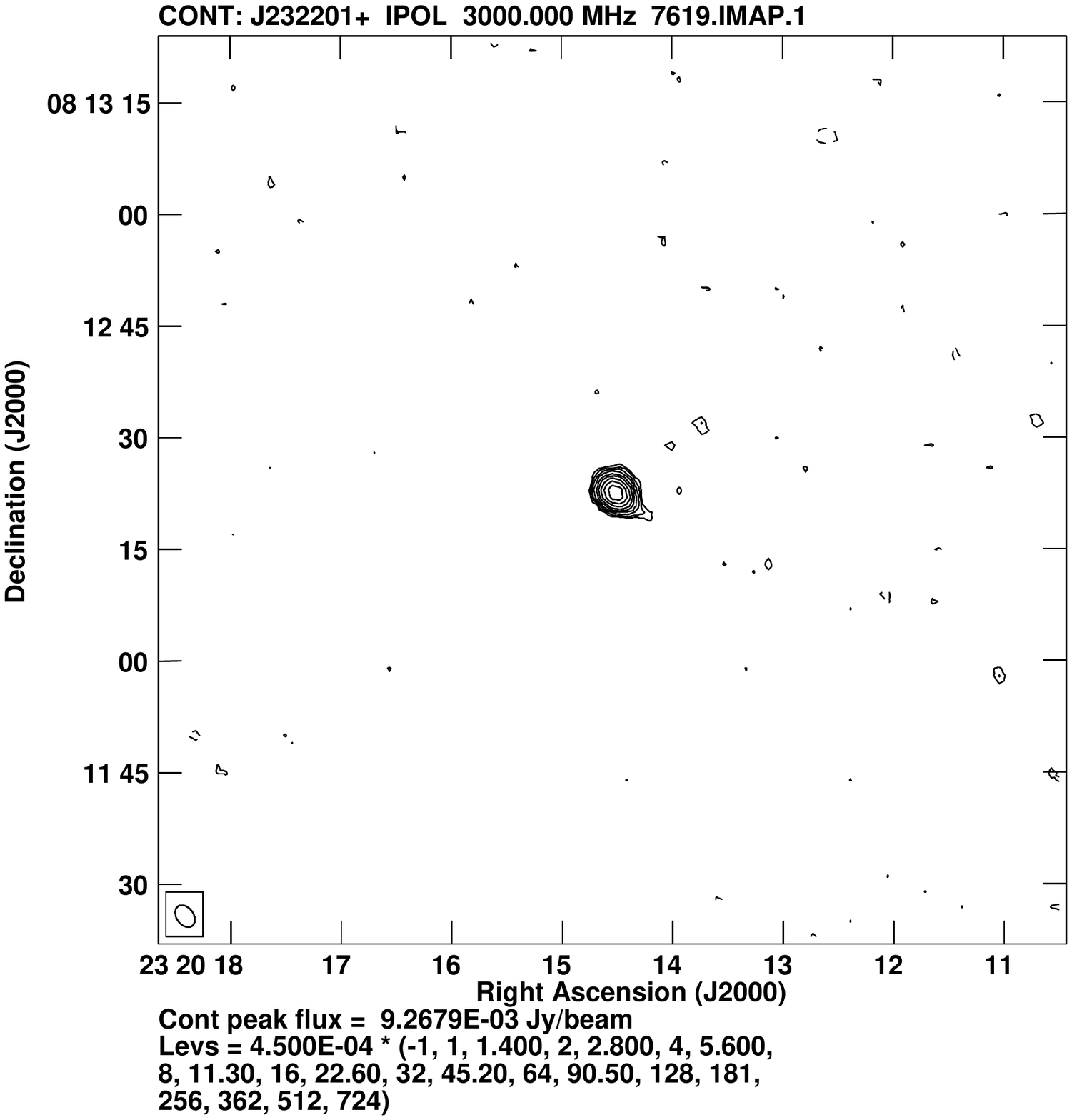}
\includegraphics[width=0.29\textwidth]{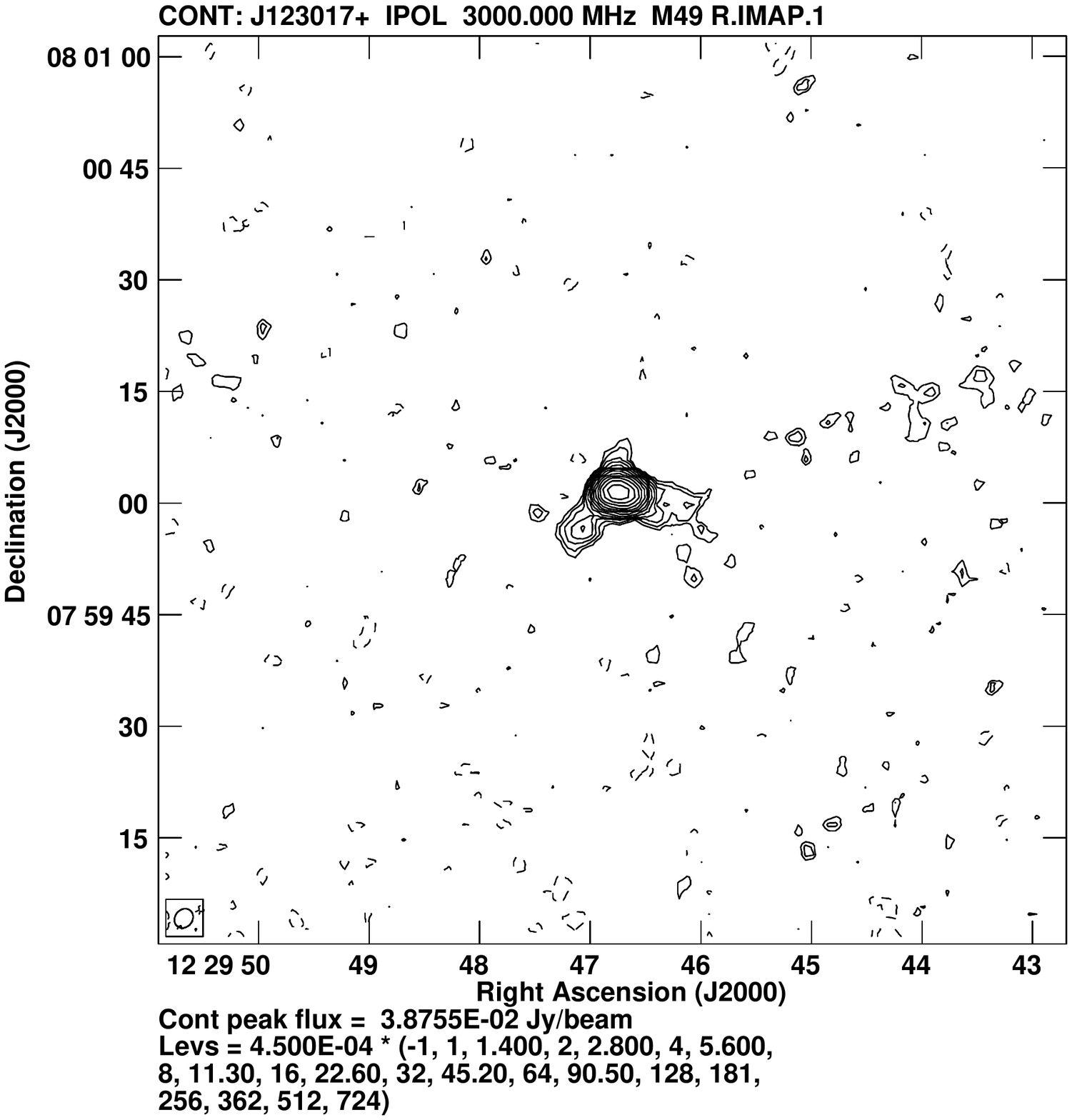}
\includegraphics[width=0.29\textwidth]{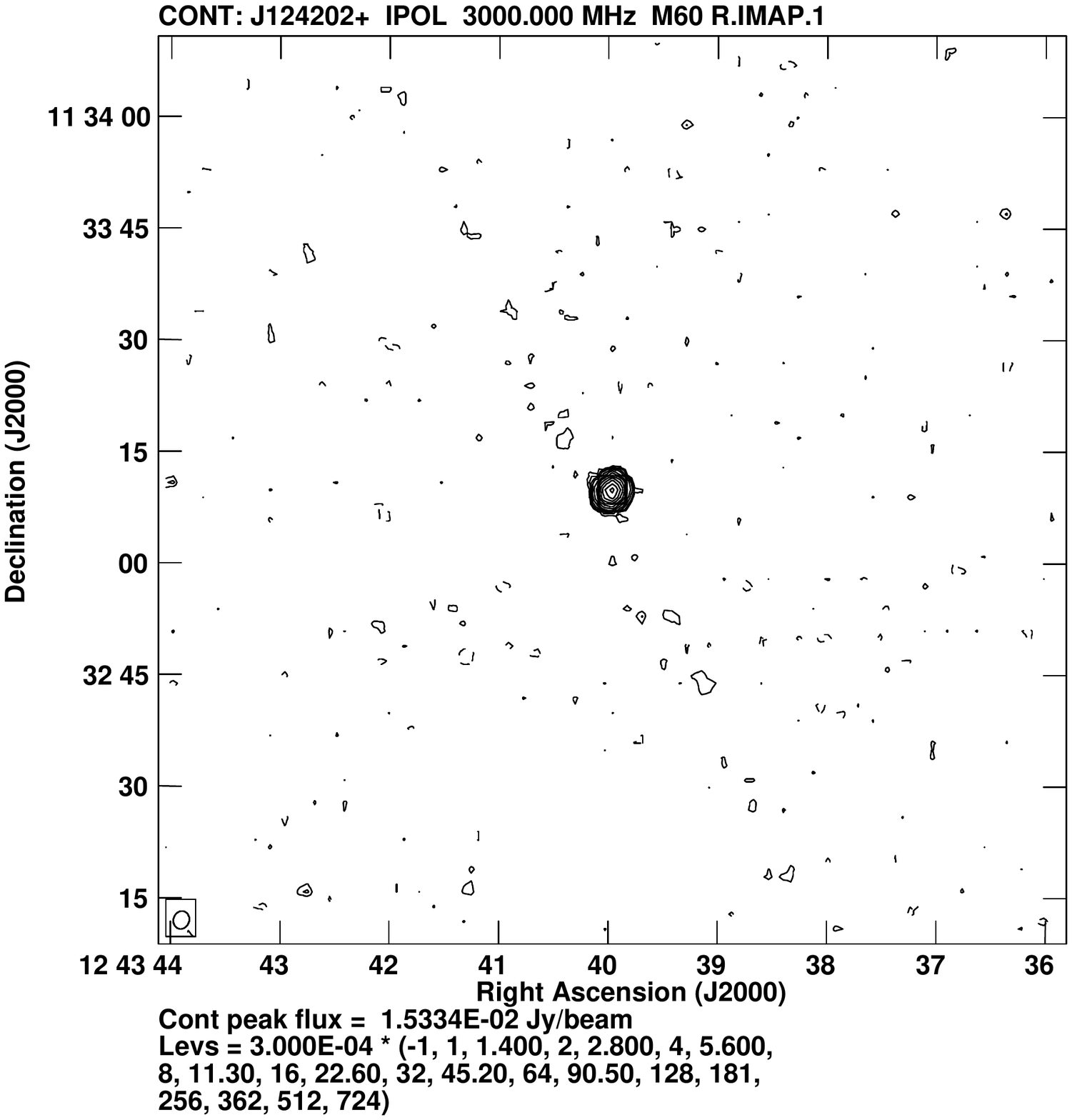}
\includegraphics[width=0.29\textwidth]{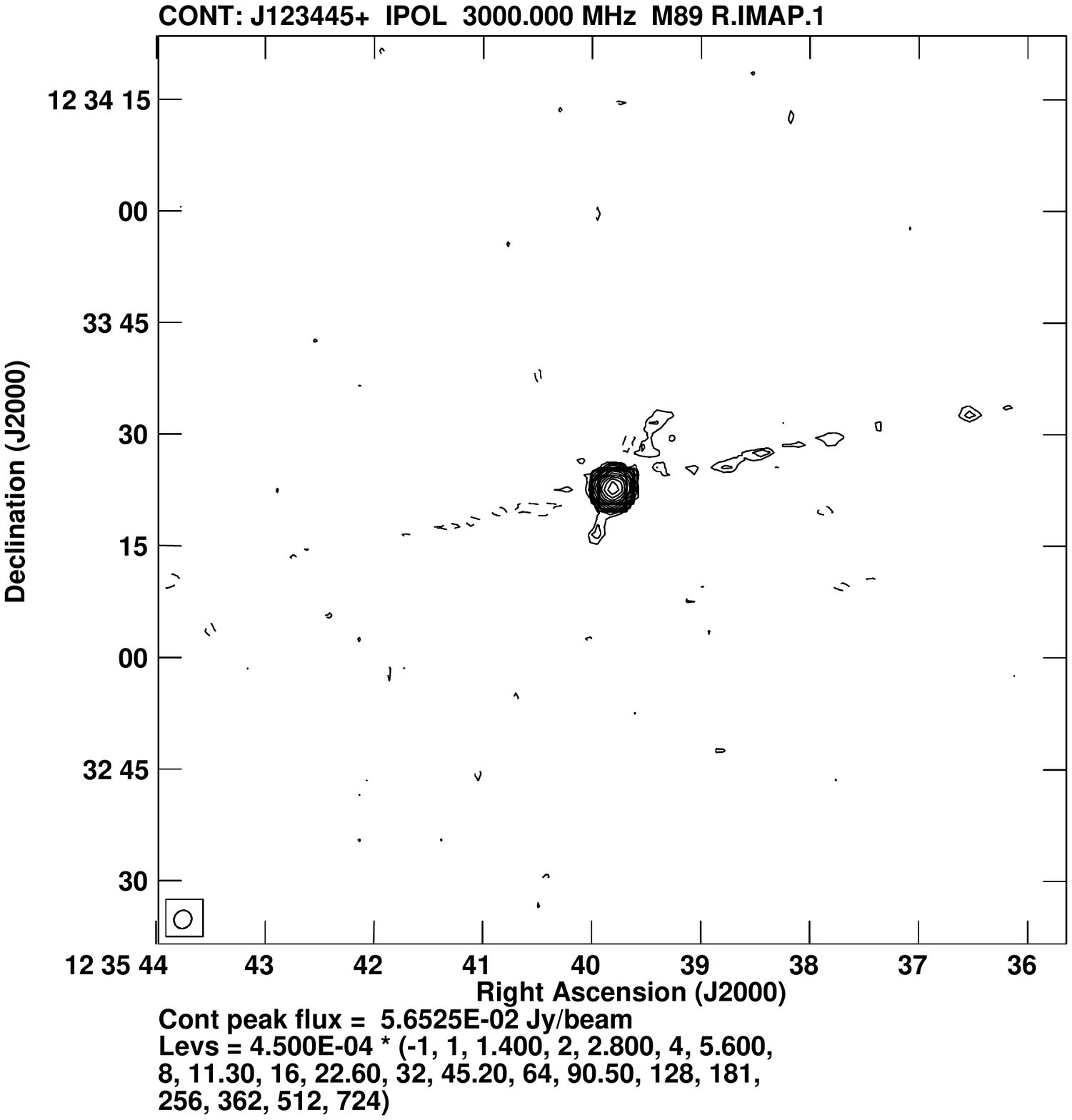}
\includegraphics[width=0.29\textwidth]{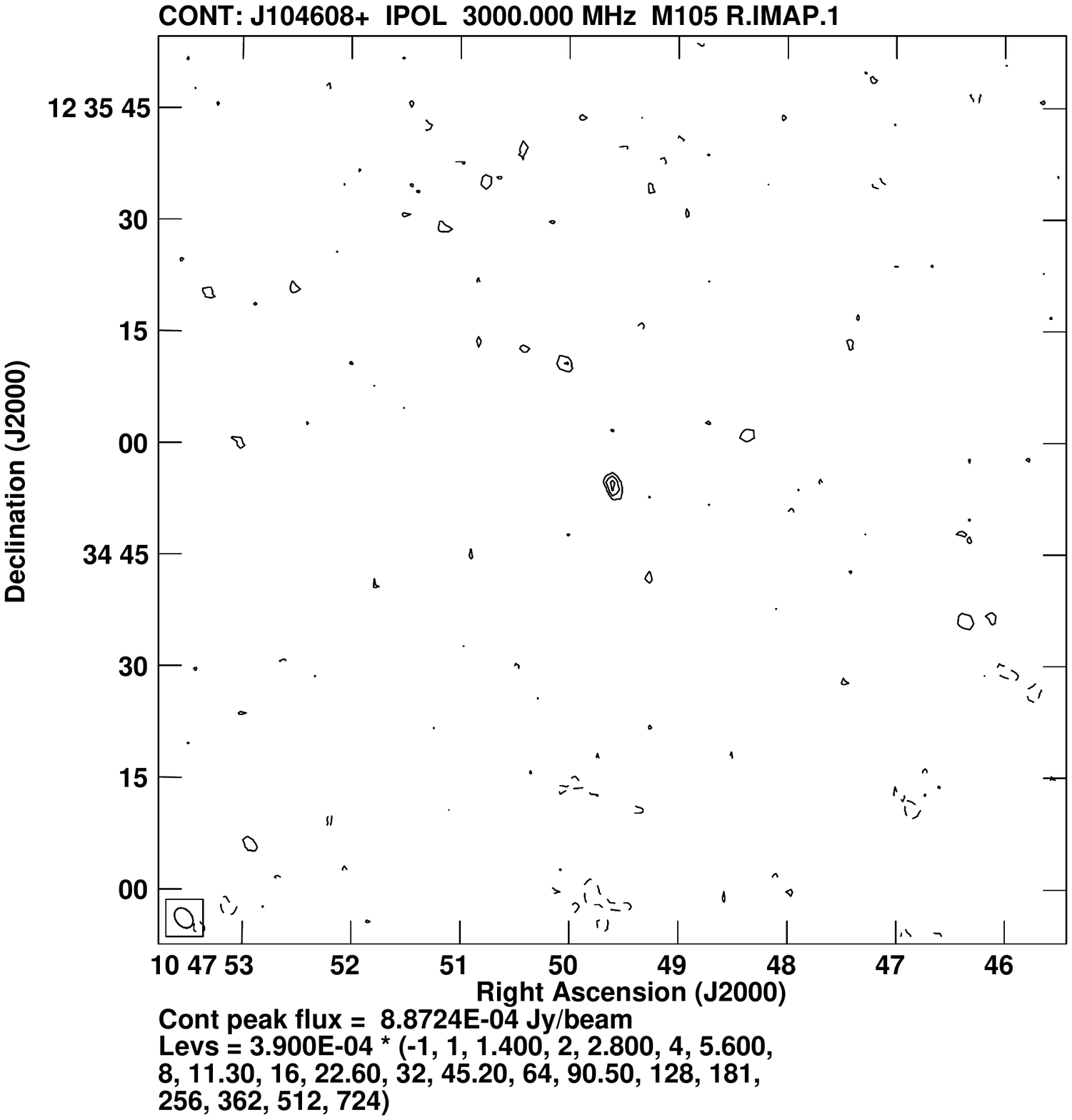}
\includegraphics[width=0.29\textwidth]{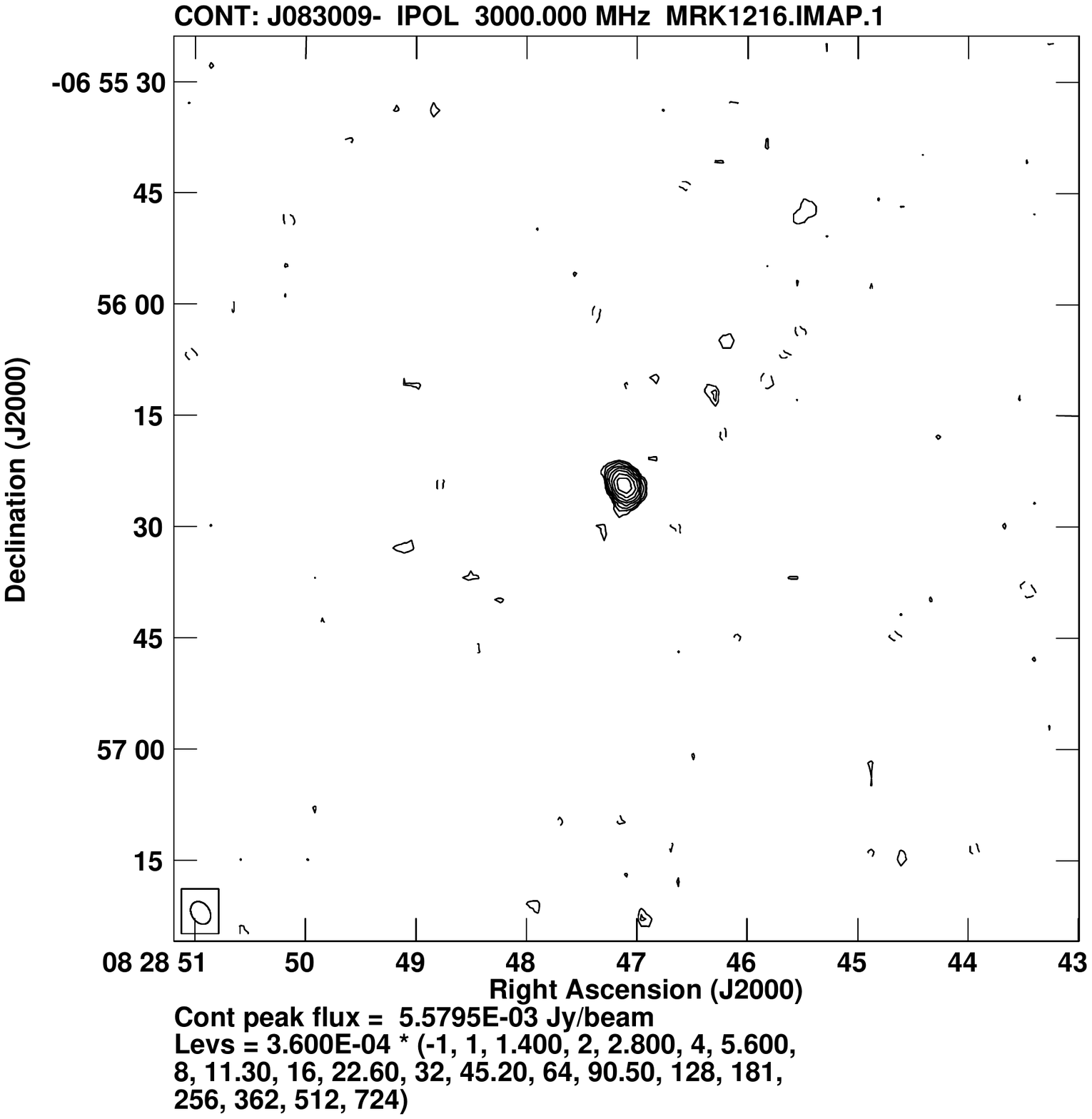}
\caption{VLASS contour maps available for radio-dim galaxies NGC\,5846, NGC\,7619, NGC\,4472/M\,49 (top panel, left to right), NGC\,4649/M\,60, NGC\,4552/M\,89, NGC\,3379/M\,105 (middle panel), and Mrk\,1216 (bottom panel).}
\label{fig:A:VLASS-RD-3}
\end{figure}

\begin{figure}[ht]
    \centering
\includegraphics[width=0.29\textwidth]{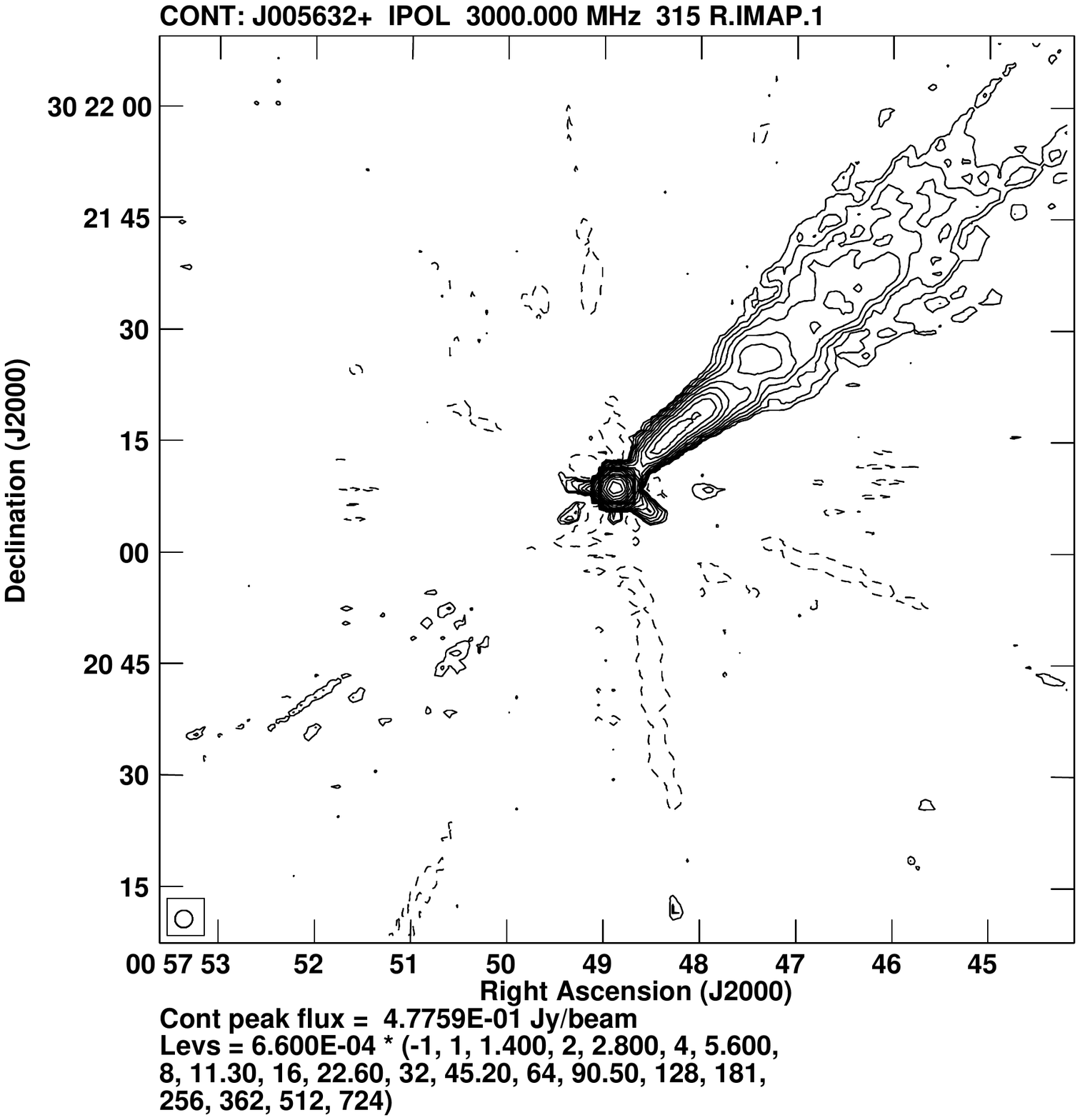}
\includegraphics[width=0.29\textwidth]{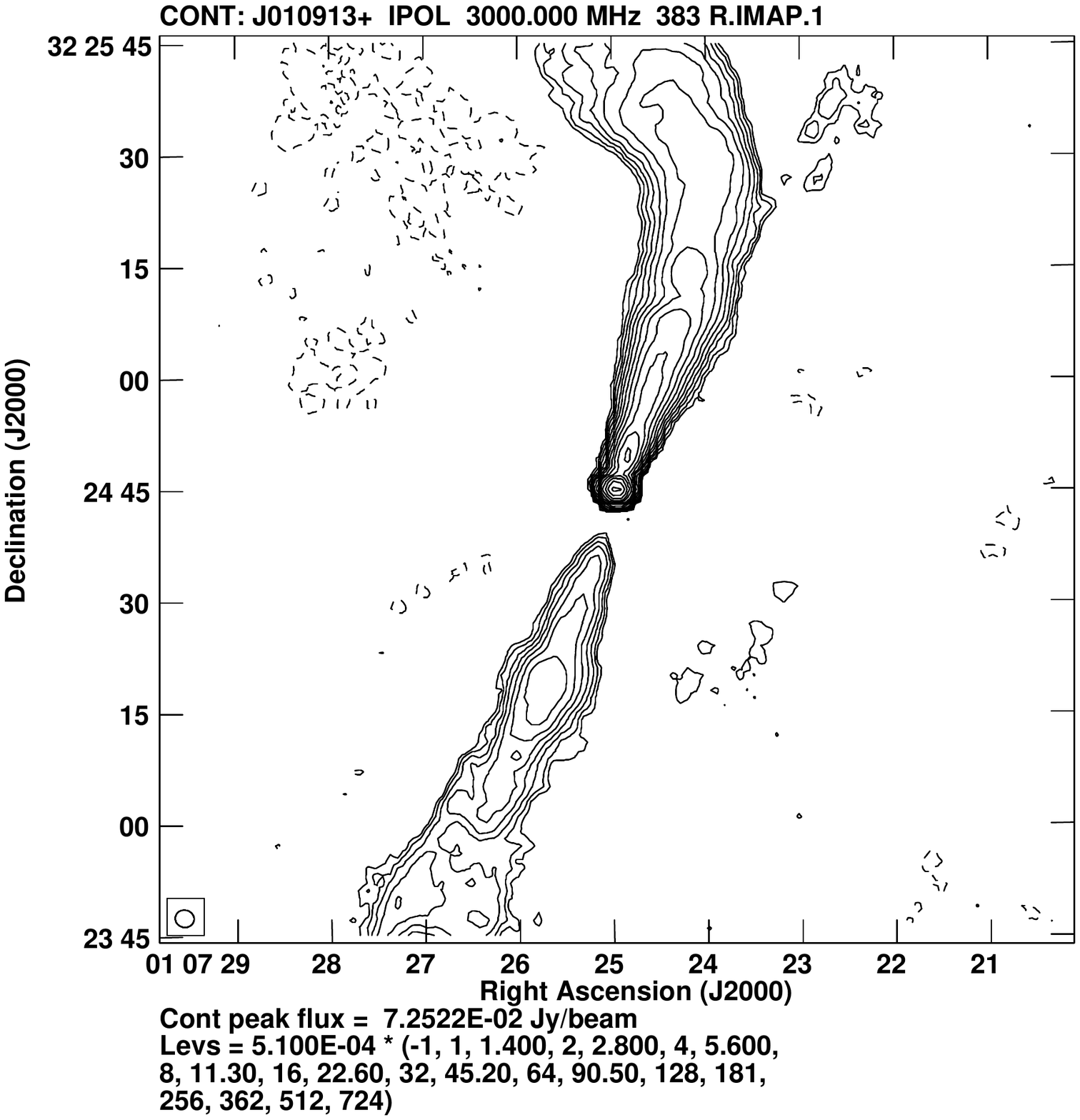}
\includegraphics[width=0.29\textwidth]{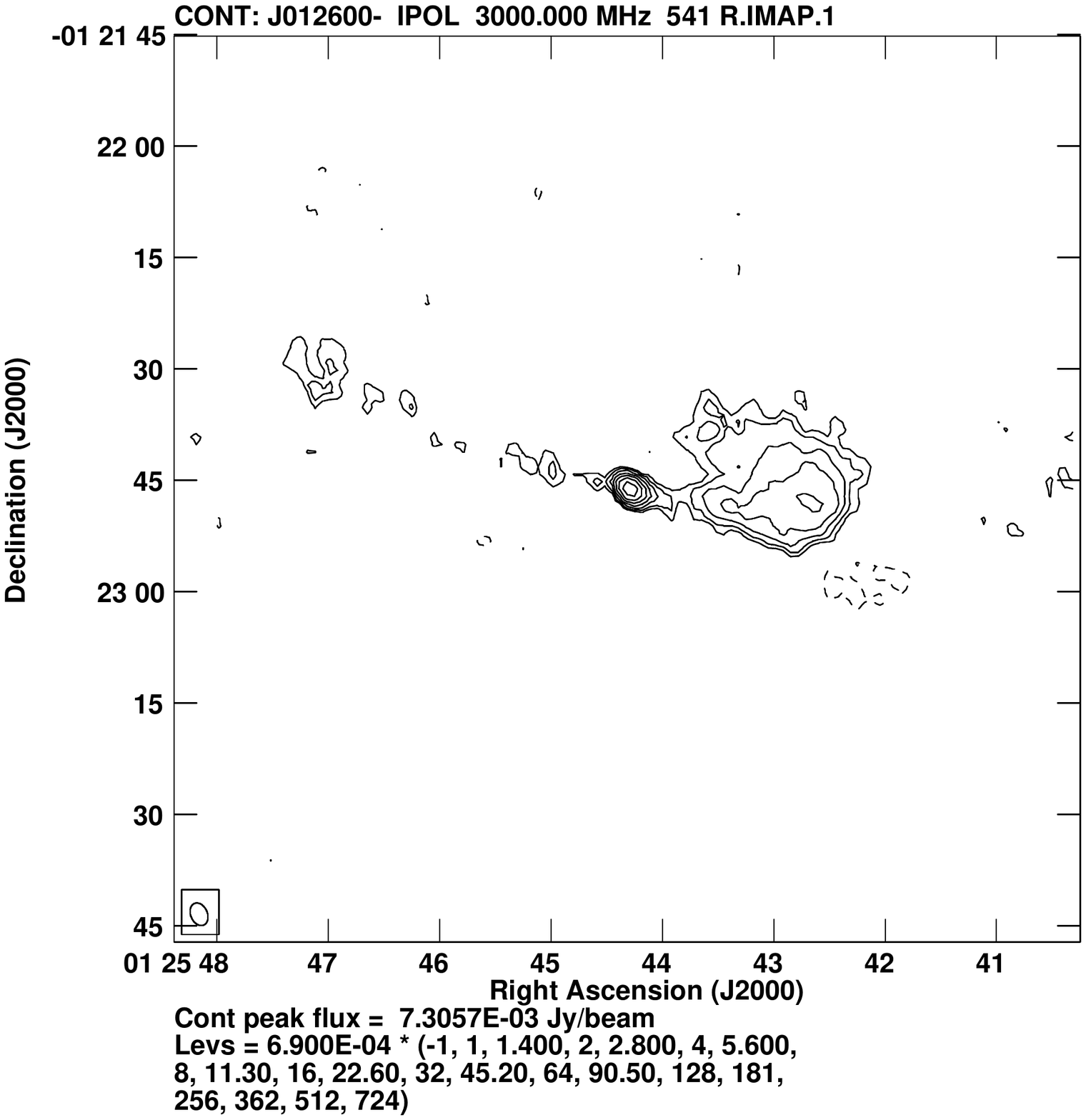}
\includegraphics[width=0.29\textwidth]{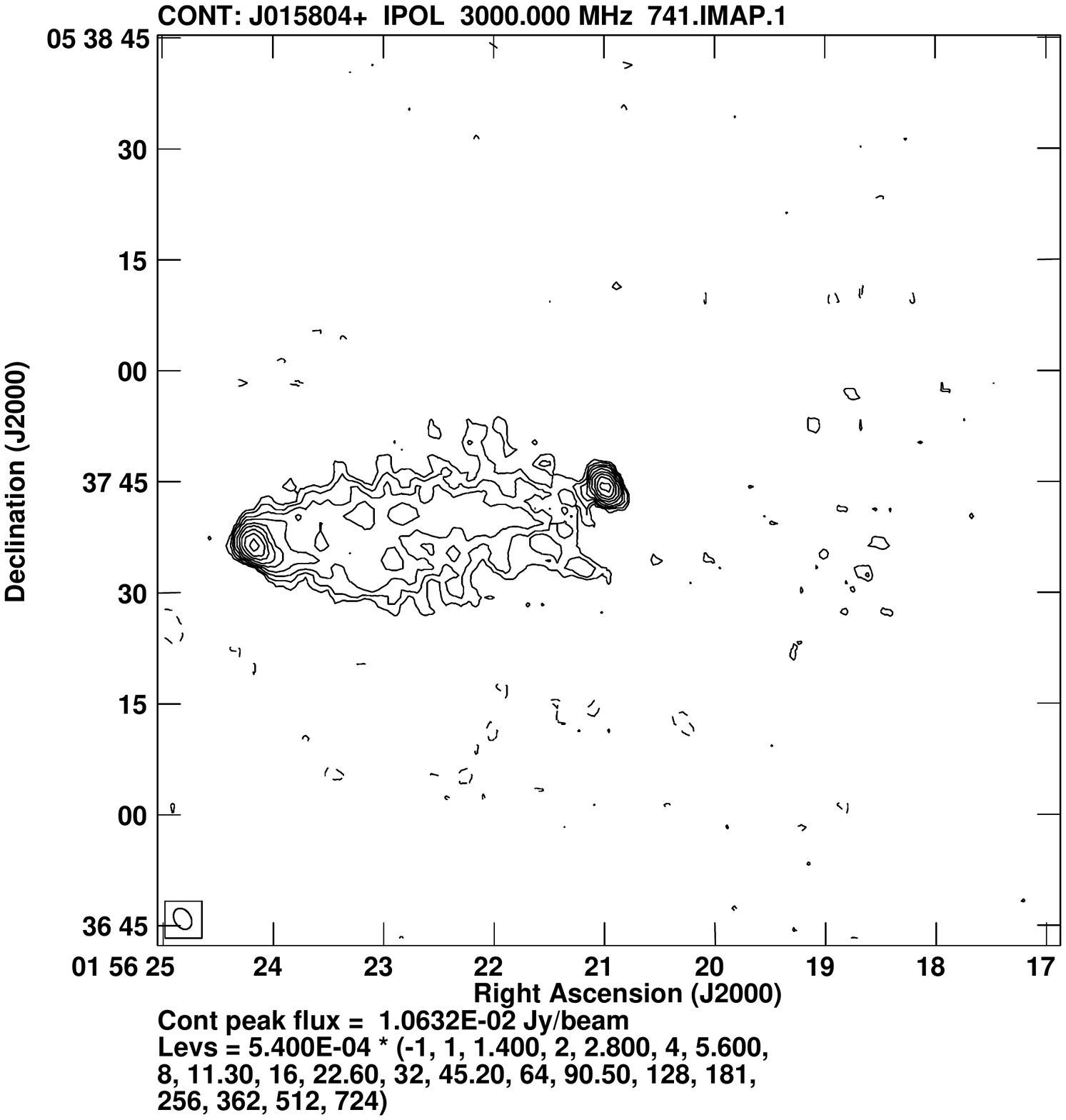}
\includegraphics[width=0.29\textwidth]{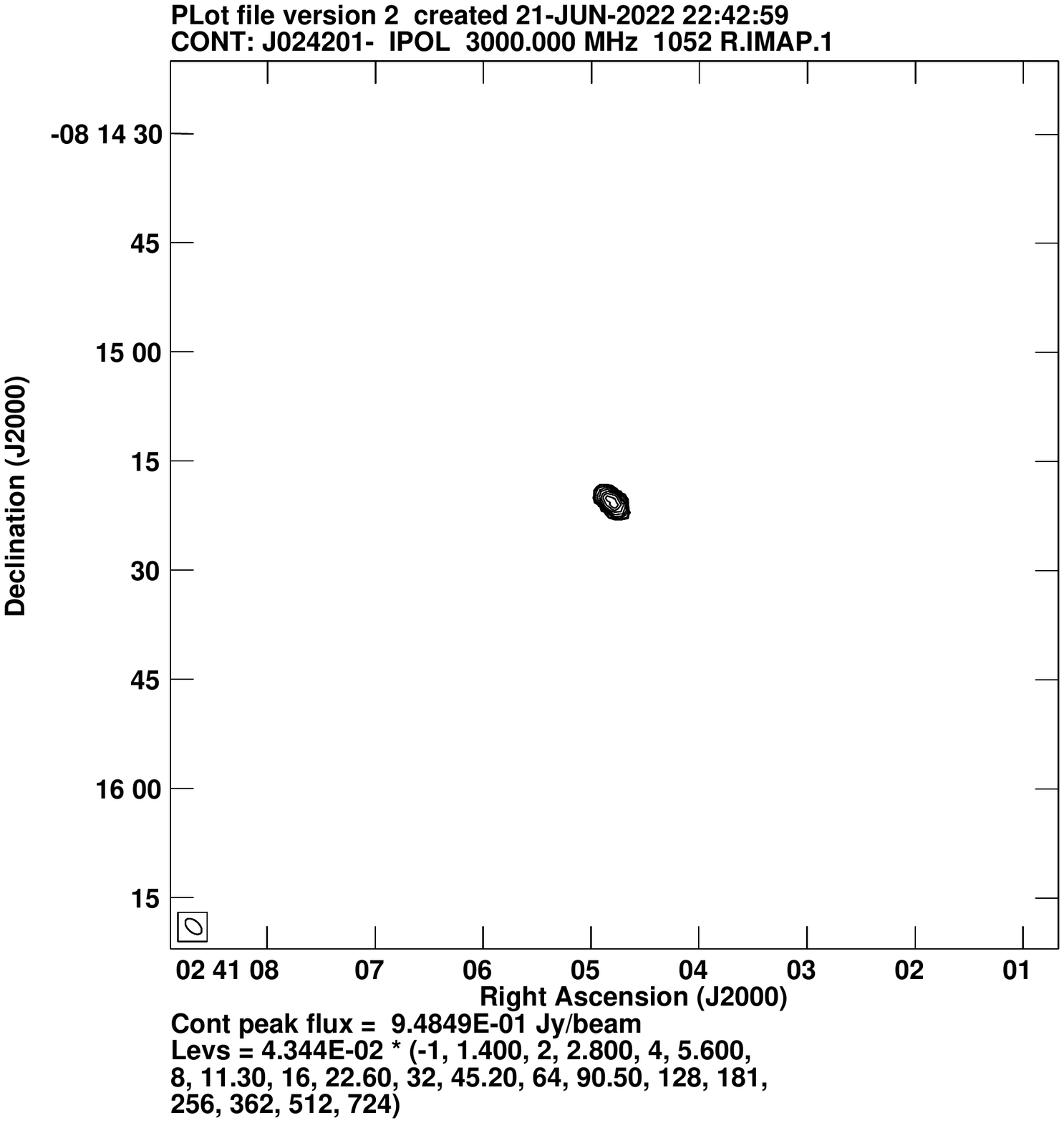}
\includegraphics[width=0.29\textwidth]{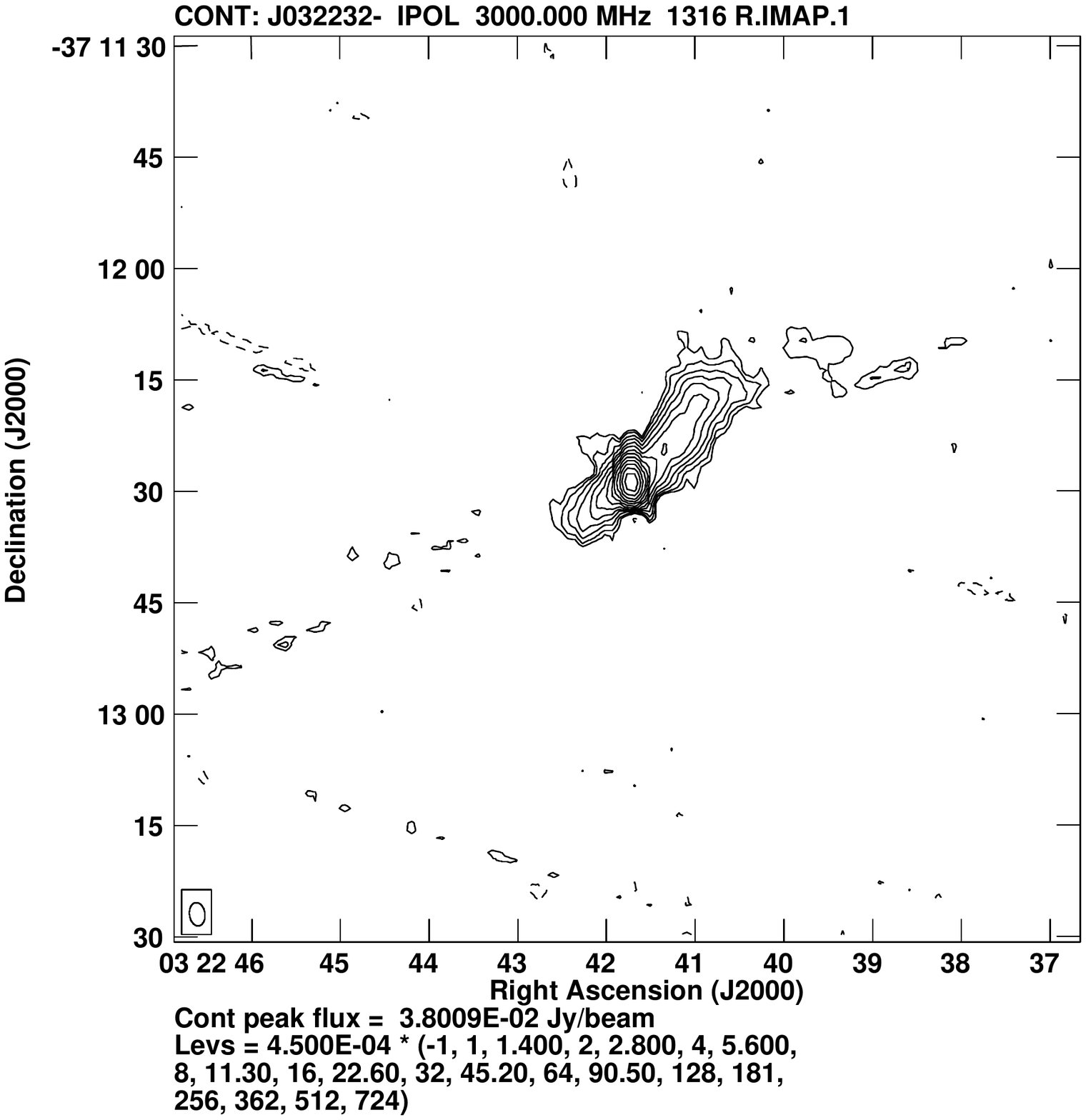}
\includegraphics[width=0.29\textwidth]{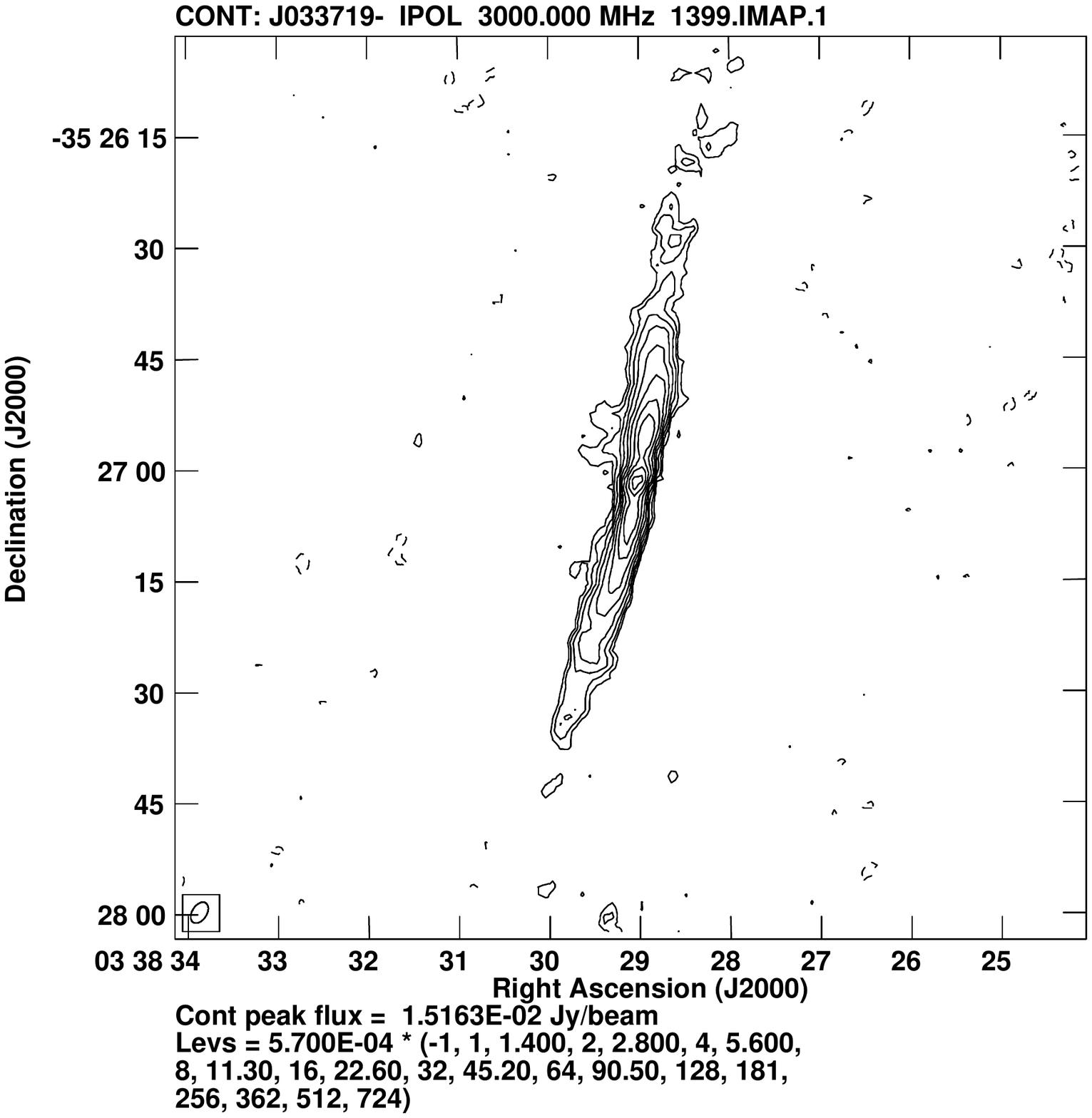}
\includegraphics[width=0.29\textwidth]{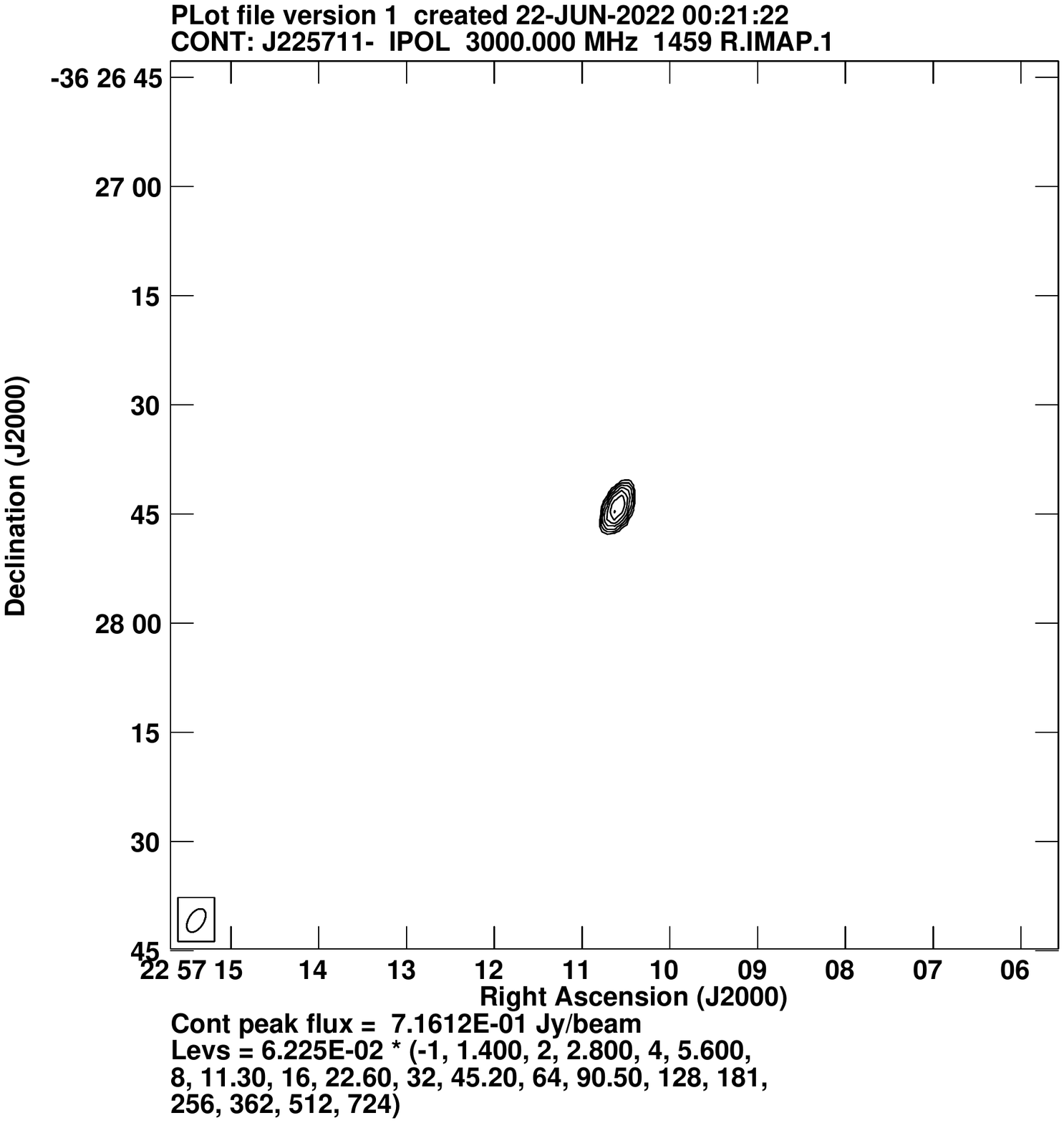}
\includegraphics[width=0.29\textwidth]{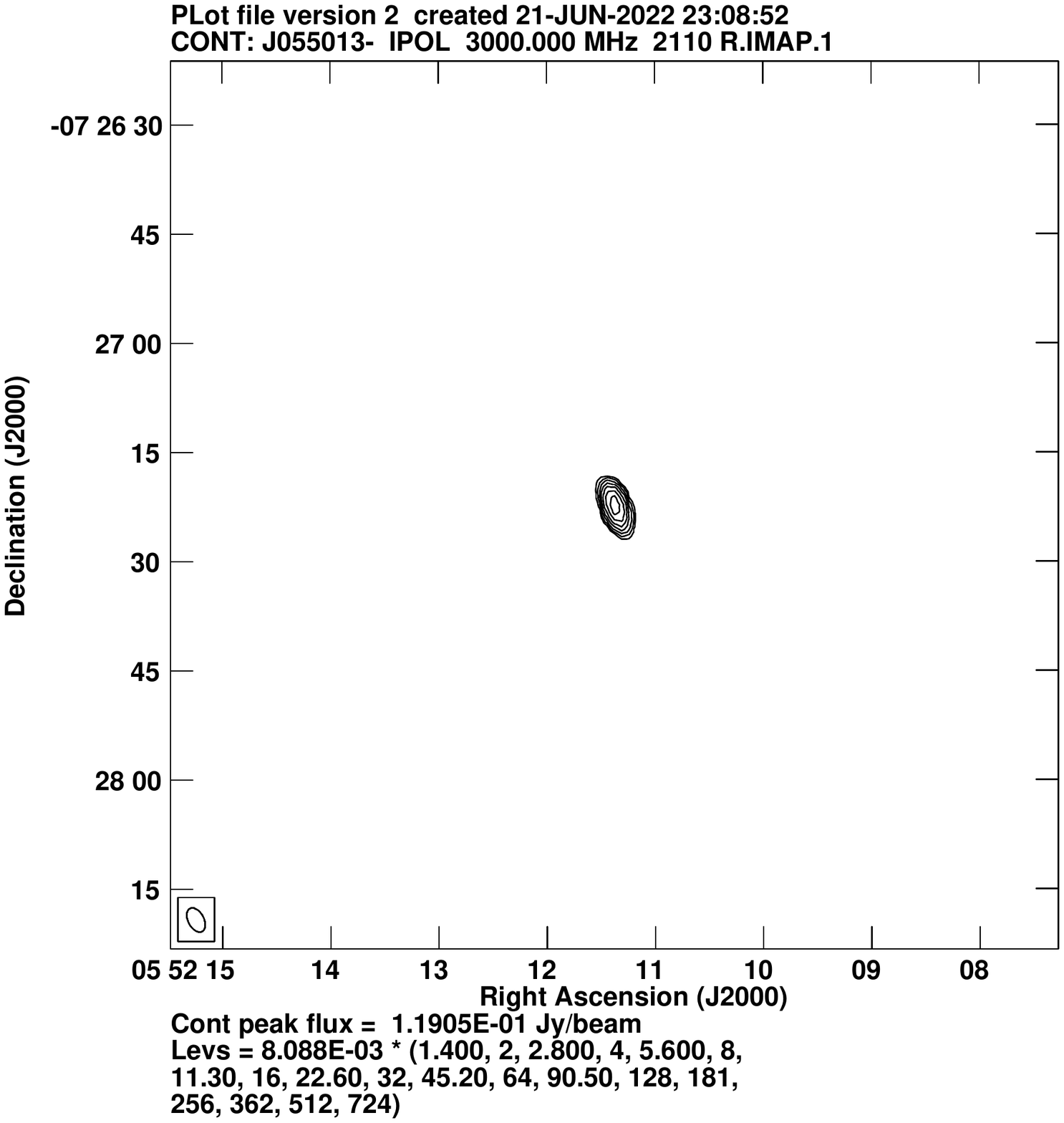}
\caption{VLASS contour maps available for radio-bright galaxies NGC\,0315, NGC\,0383, NGC\,0541 (top panel, left to right), NGC\,0741, NGC\,1052, NGC\,1316 (middle panel), and NGC\,1399, IC\,1459, NGC\,2110 (bottom panel).}
\label{fig:A:VLASS-RB-1}
\end{figure}

\begin{figure}[ht]
    \centering
\includegraphics[width=0.29\textwidth]{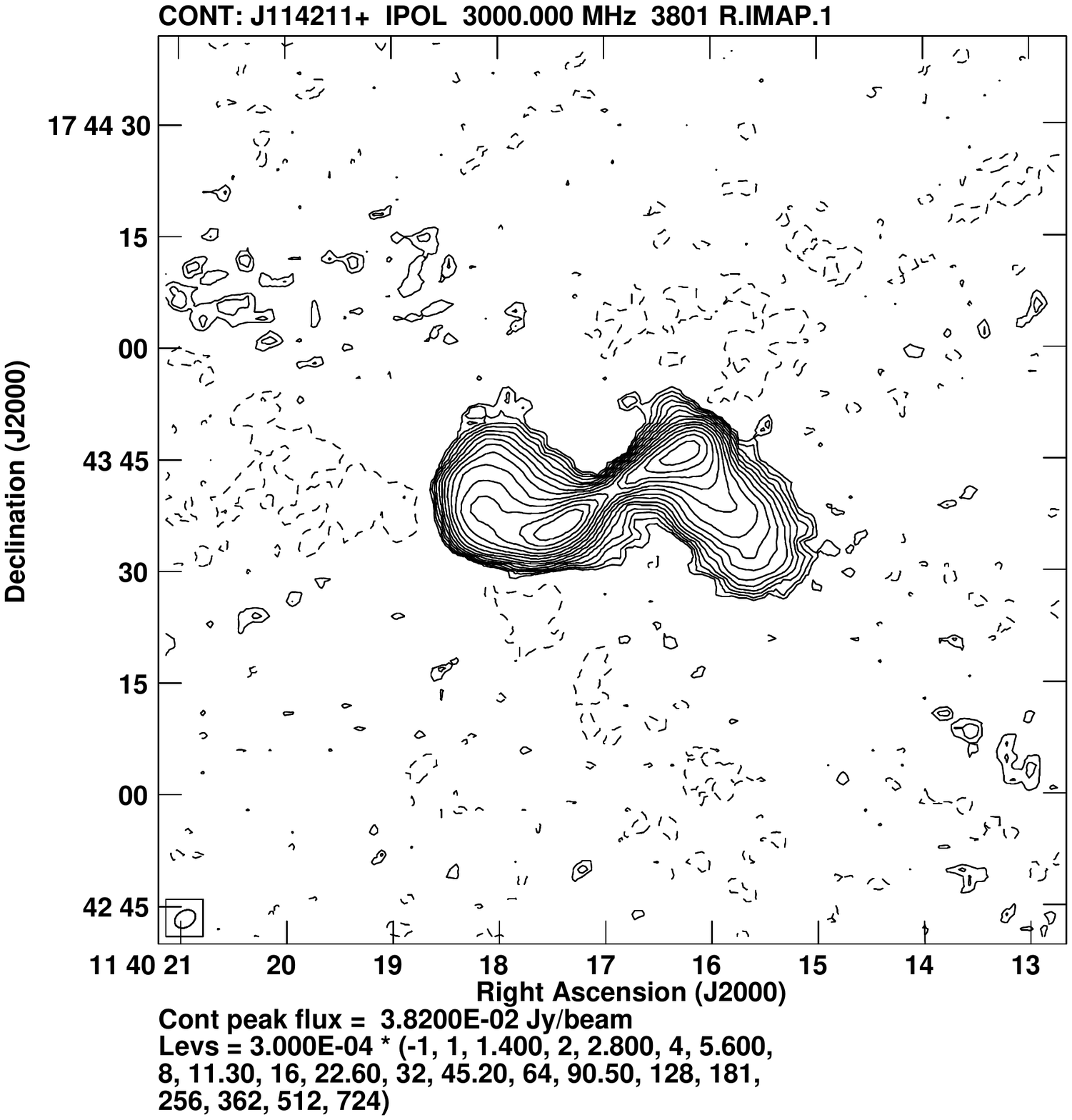}
\includegraphics[width=0.29\textwidth]{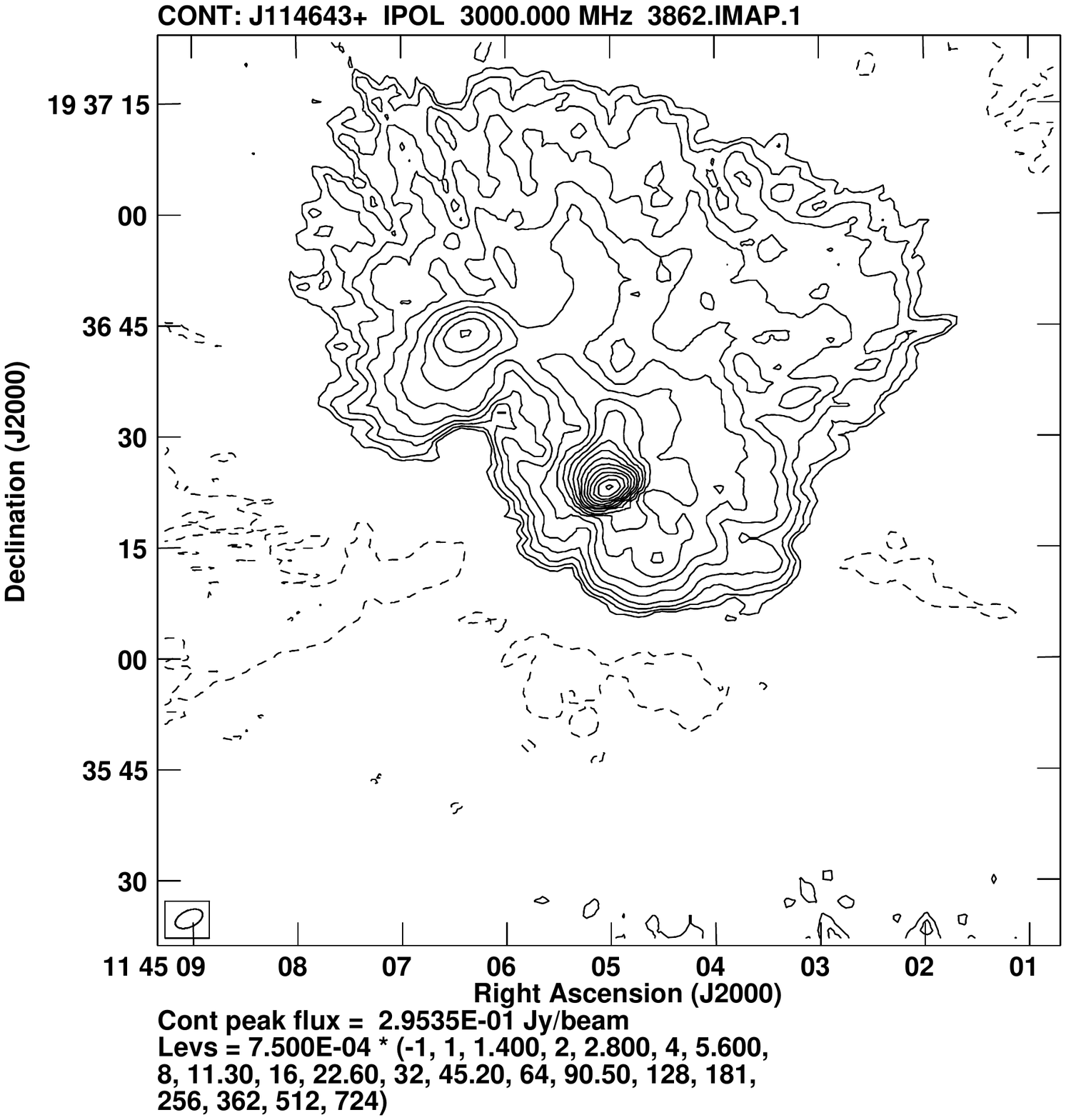}
\includegraphics[width=0.29\textwidth]{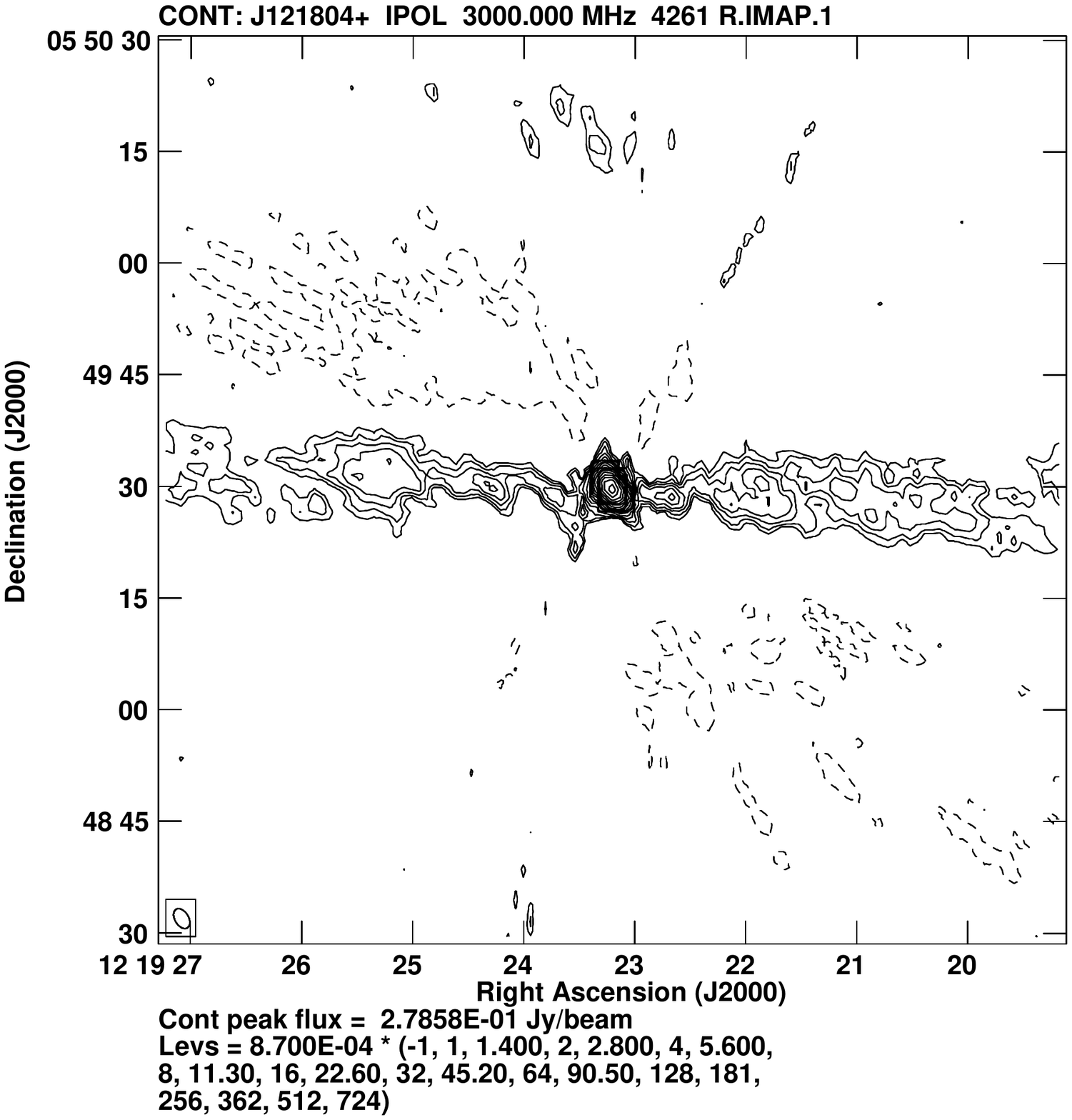}
\includegraphics[width=0.29\textwidth]{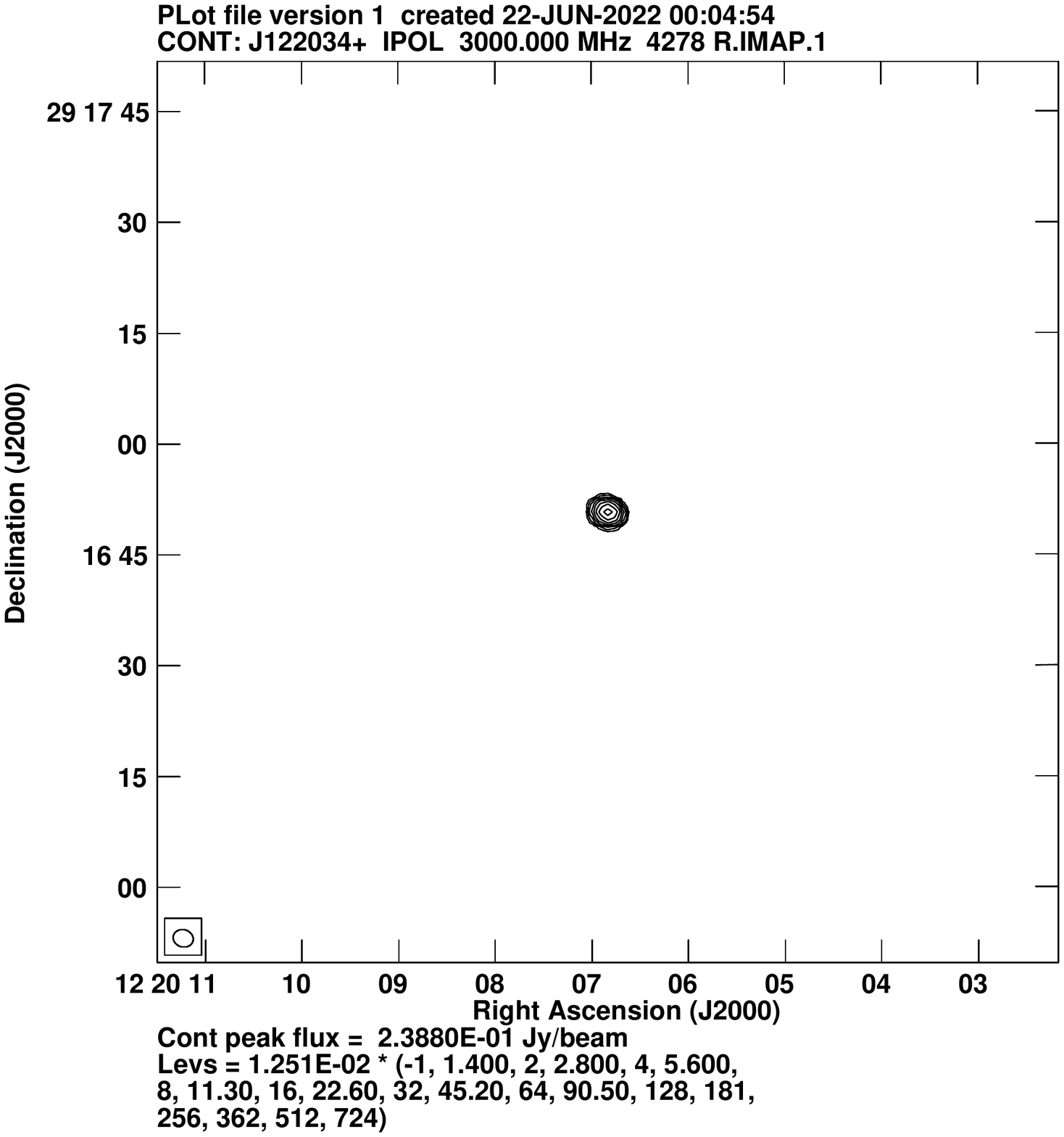}
\includegraphics[width=0.29\textwidth]{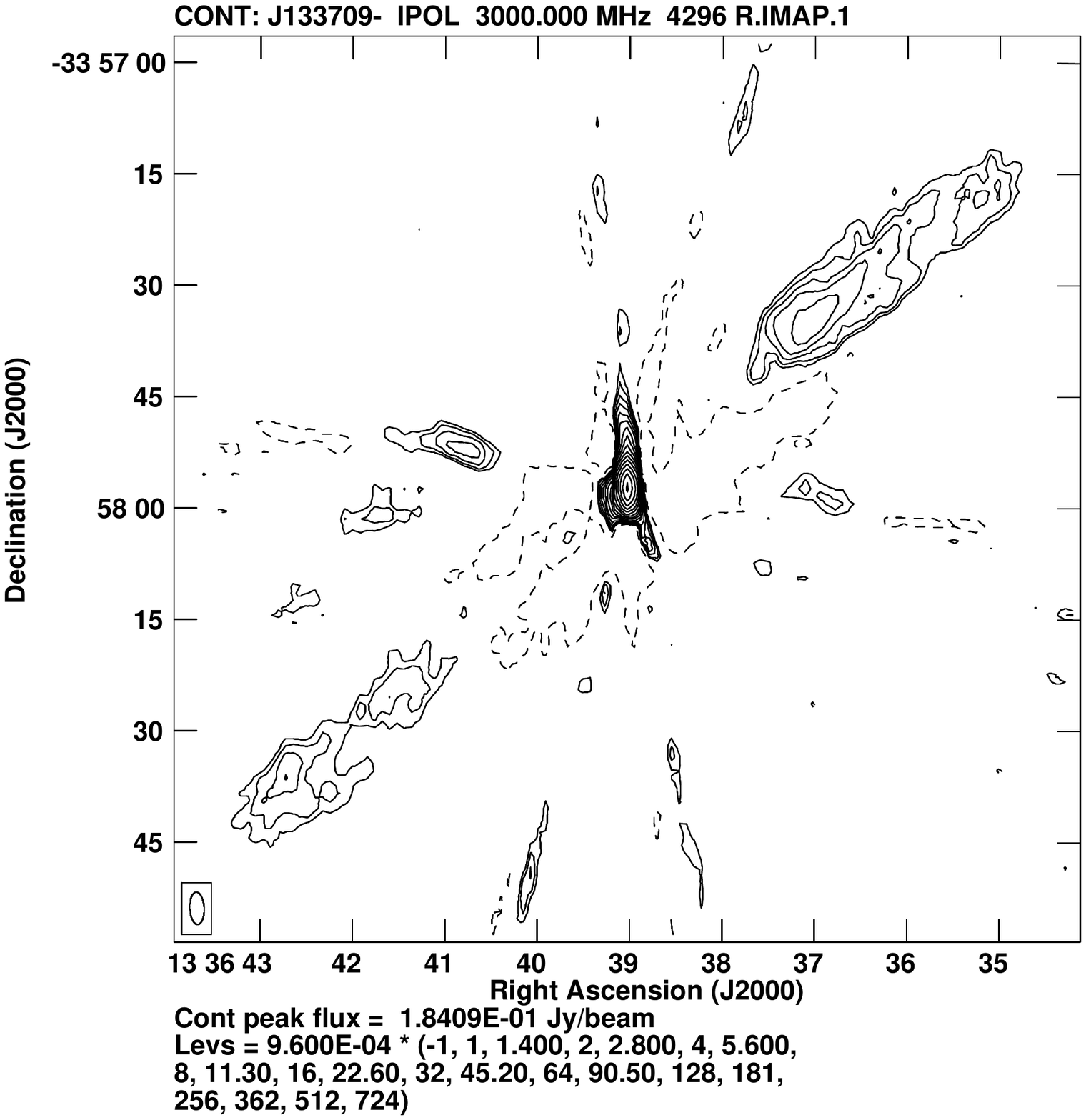}
\includegraphics[width=0.29\textwidth]{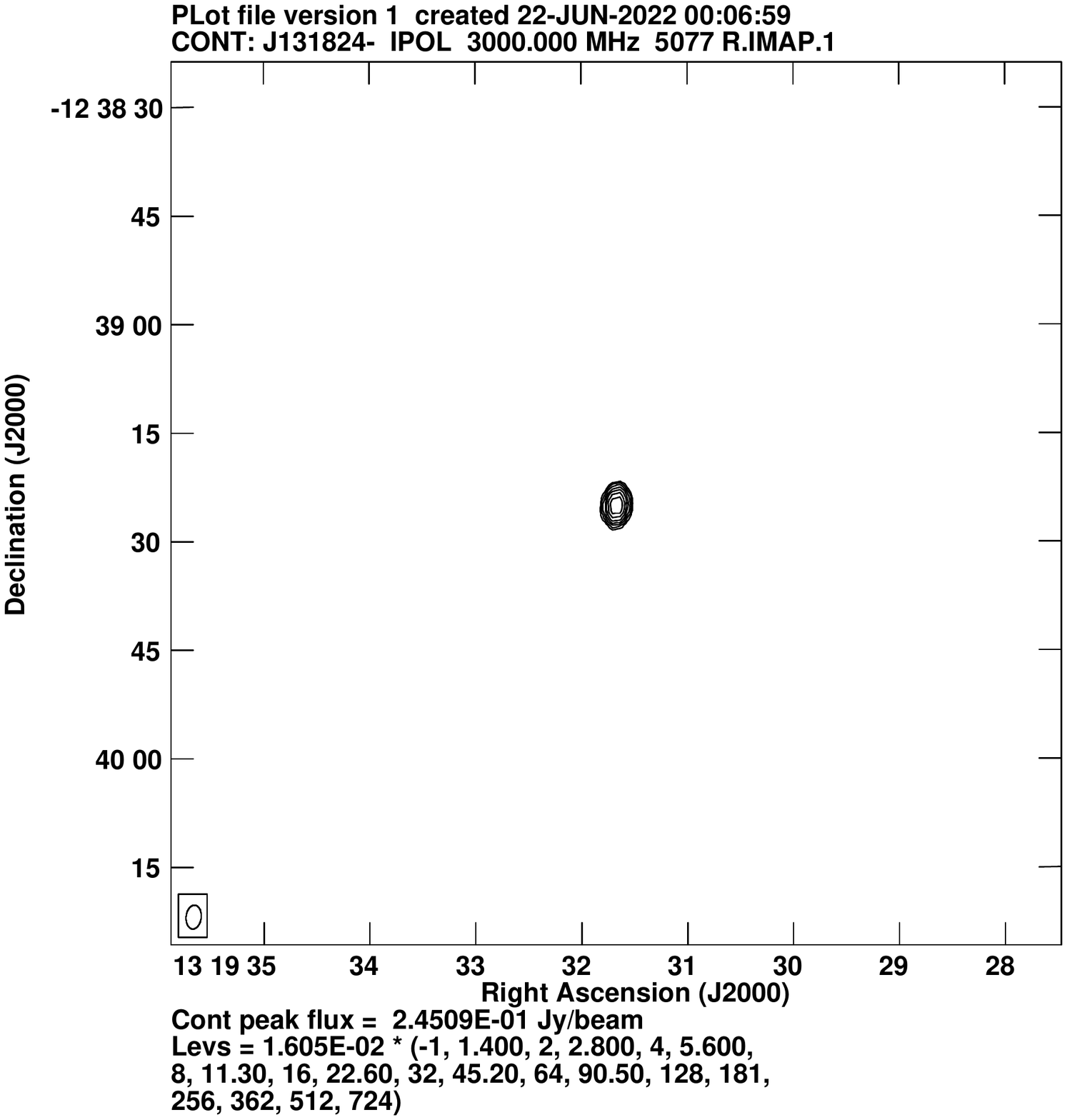}
\includegraphics[width=0.29\textwidth]{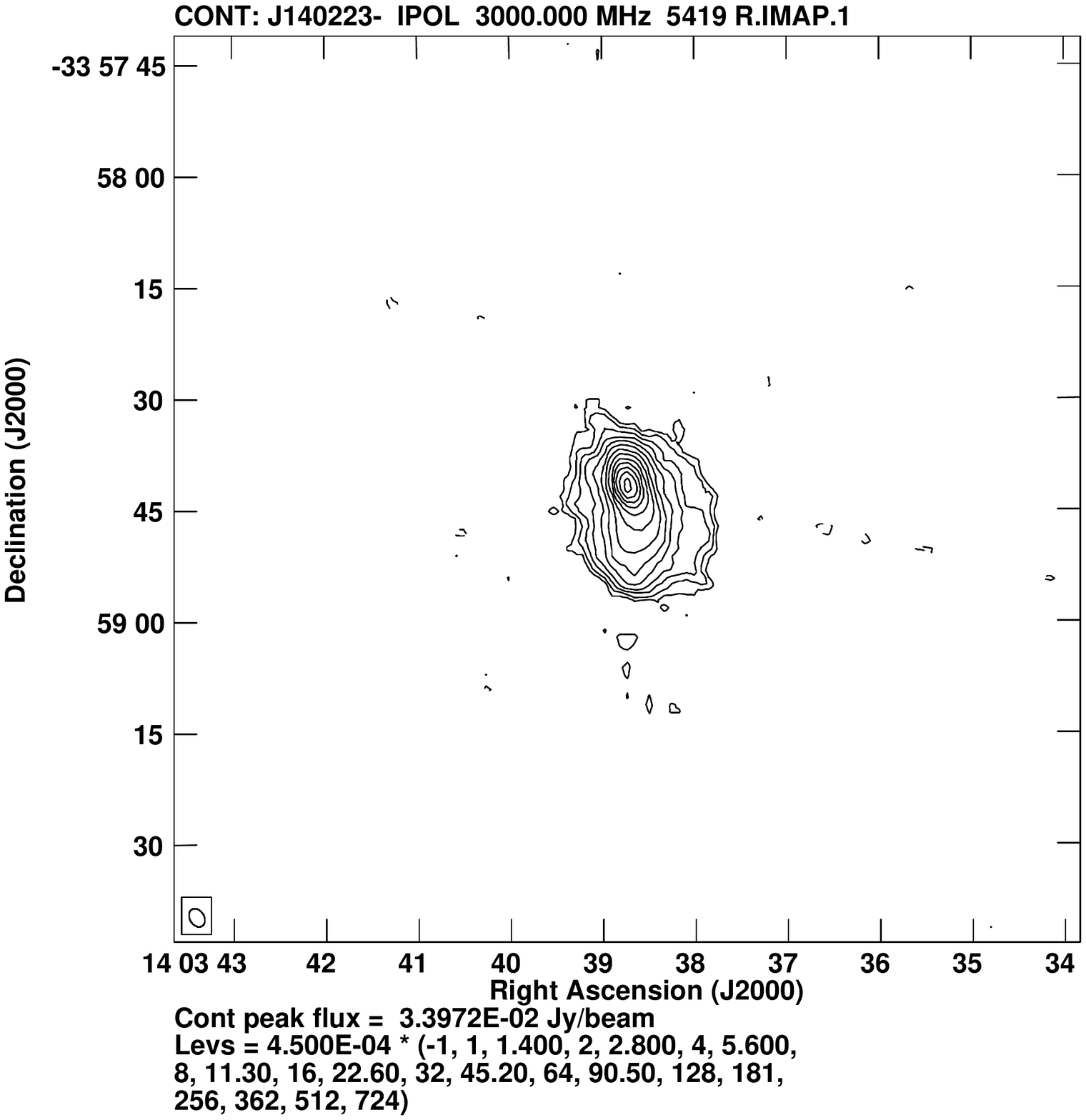}
\includegraphics[width=0.29\textwidth]{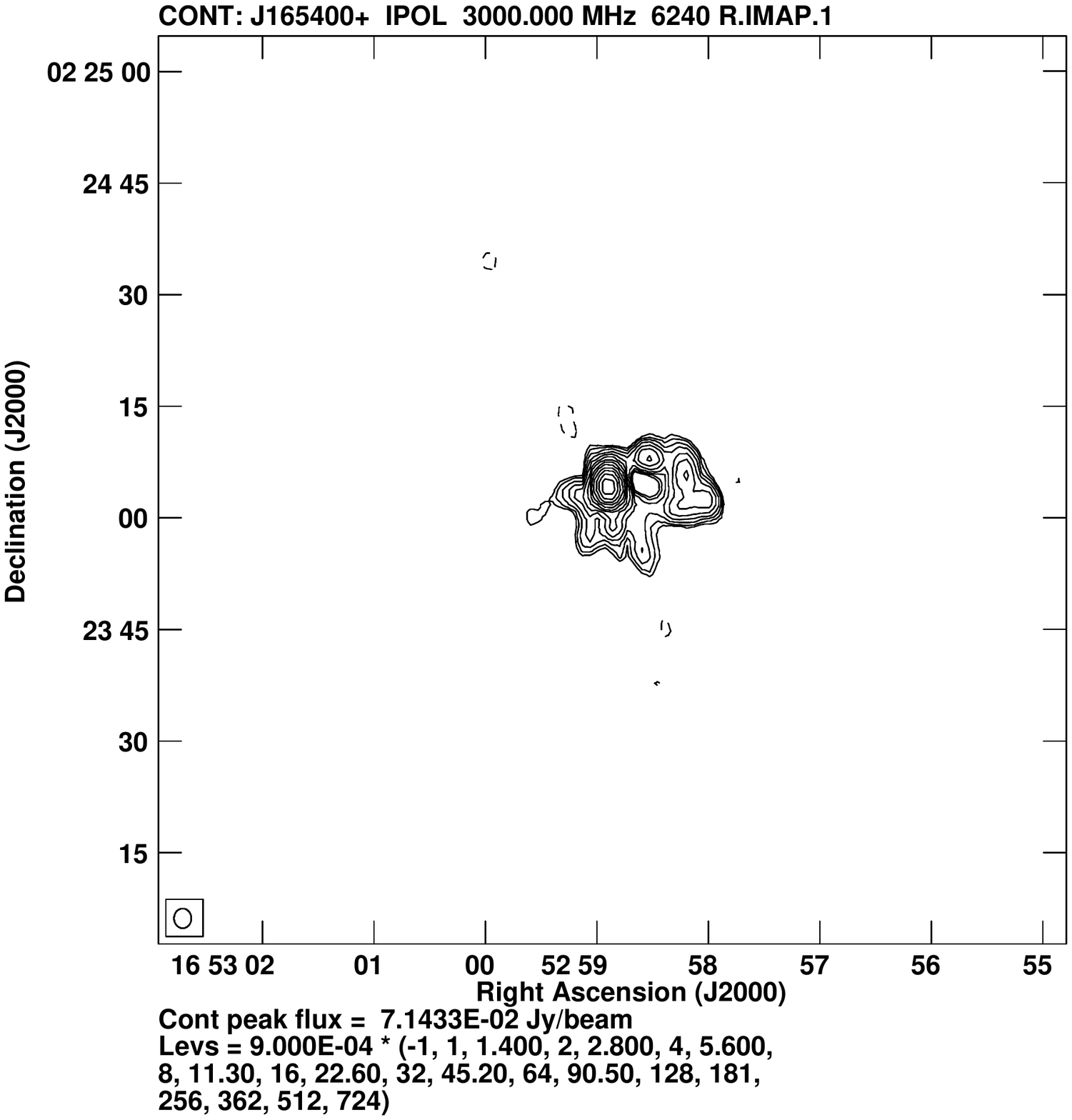}
\includegraphics[width=0.29\textwidth]{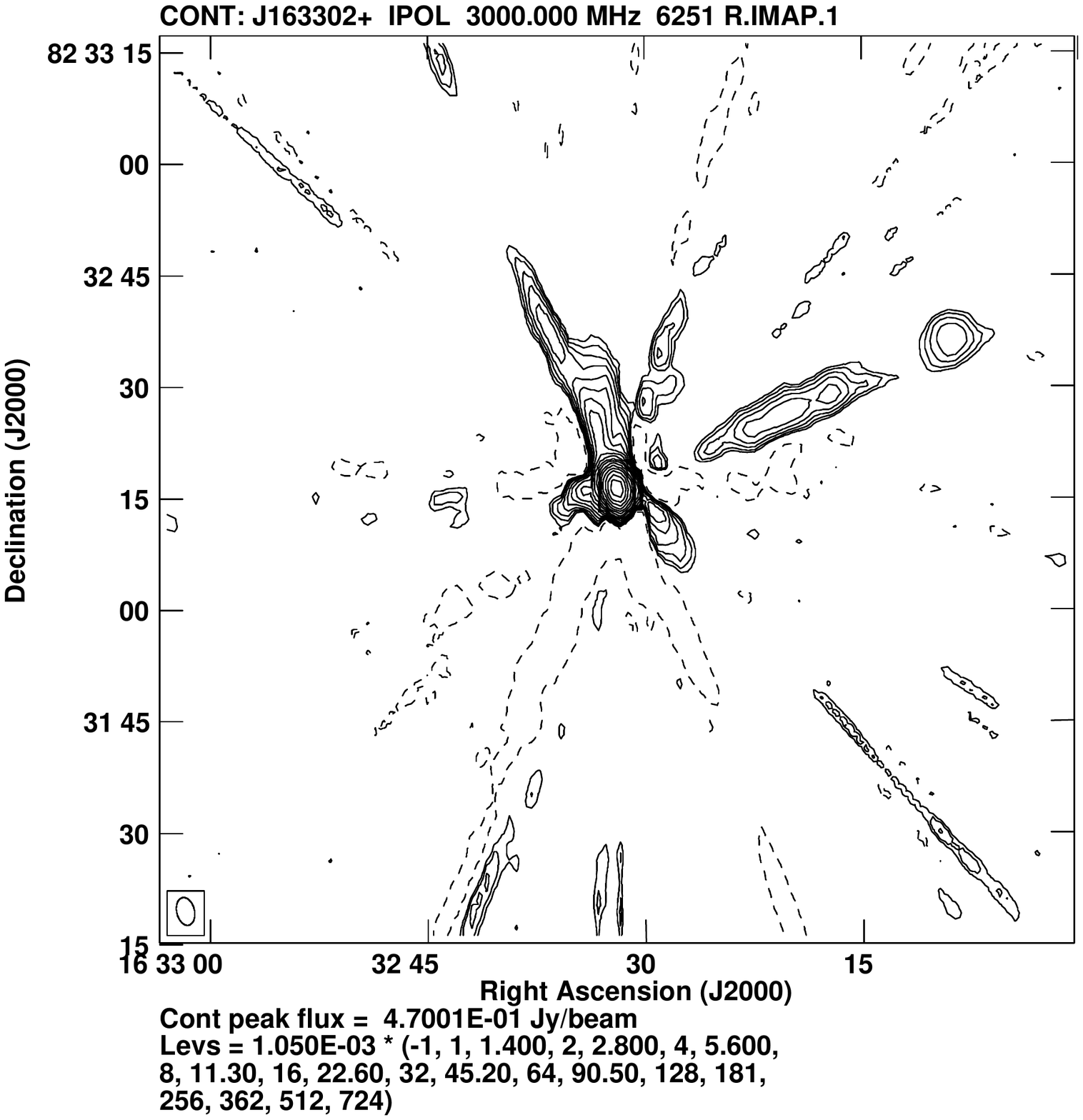}
\caption{VLASS contour maps available for radio-bright galaxies NGC\,3801, NGC\,3862, NGC\,4261 (top panel, left to right), NGC\,4278, IC\,4296, NGC\,5077 (middle panel), and NGC\,5419, NGC\,6240S, NGC\,6251 (bottom panel).}
\label{fig:A:VLASS-RB-2}
\end{figure}

\begin{figure}[ht]
    \centering
\includegraphics[width=0.29\textwidth]{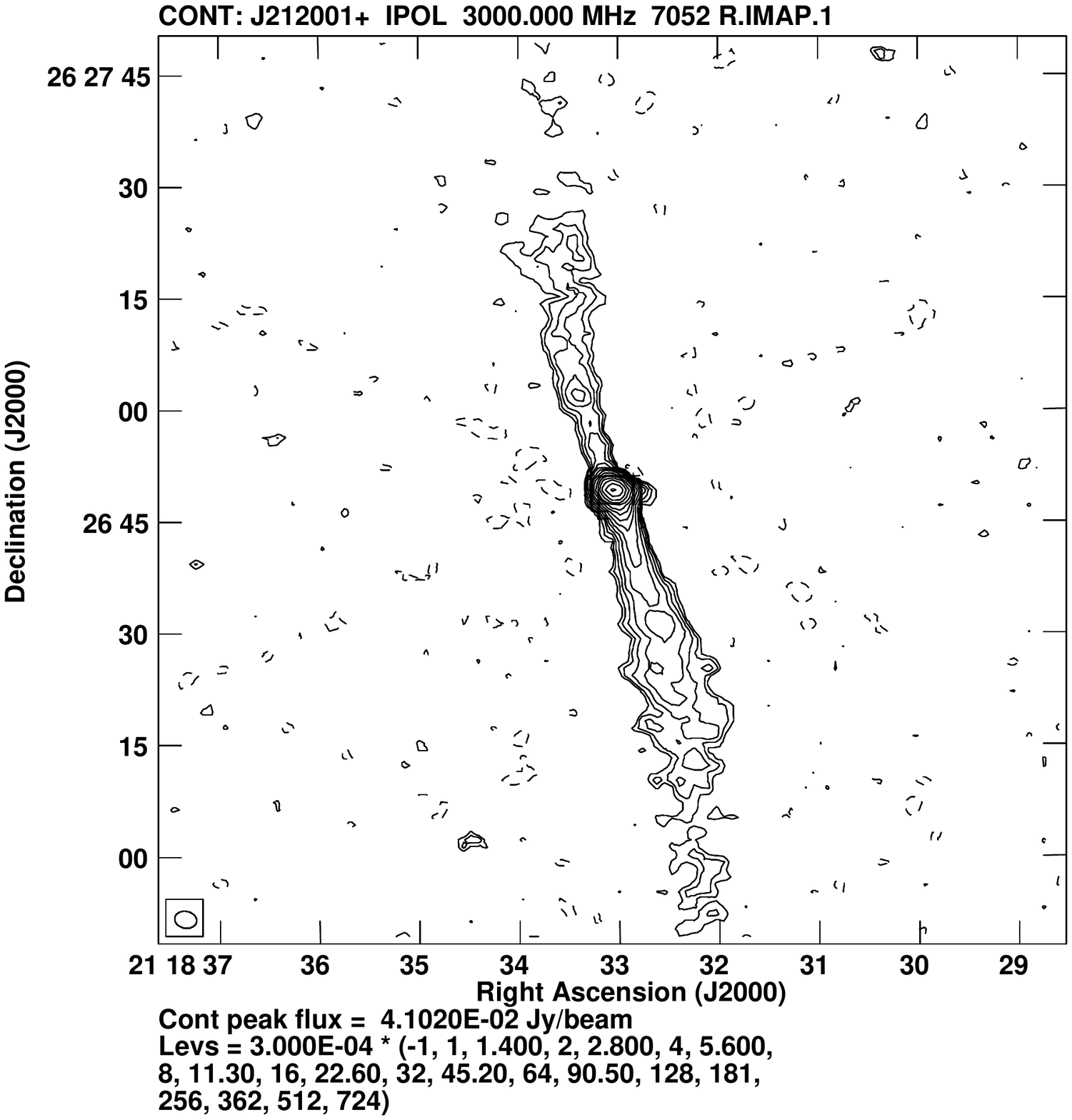}
\includegraphics[width=0.29\textwidth]{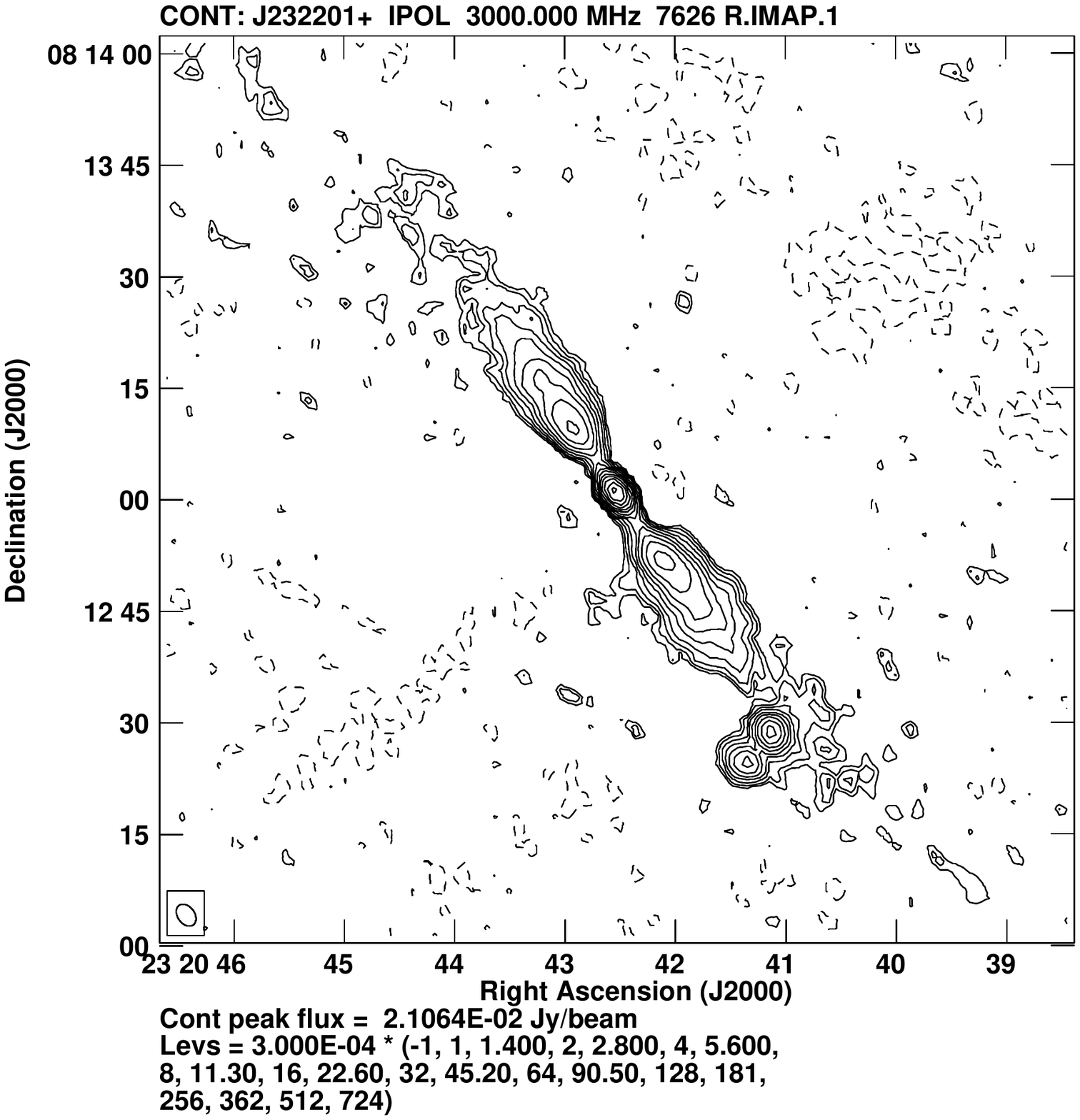}
\includegraphics[width=0.29\textwidth]{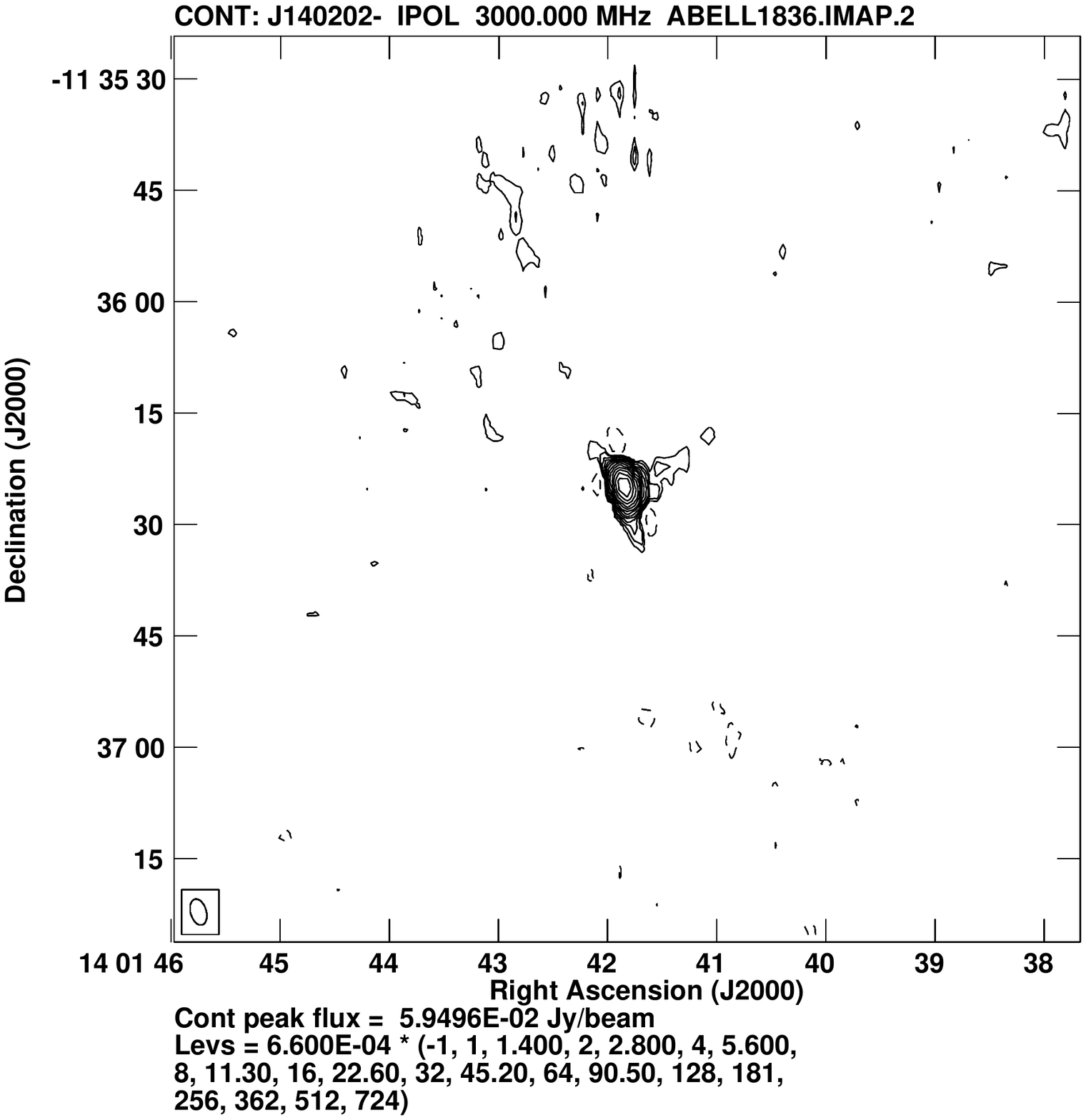}
\includegraphics[width=0.29\textwidth]{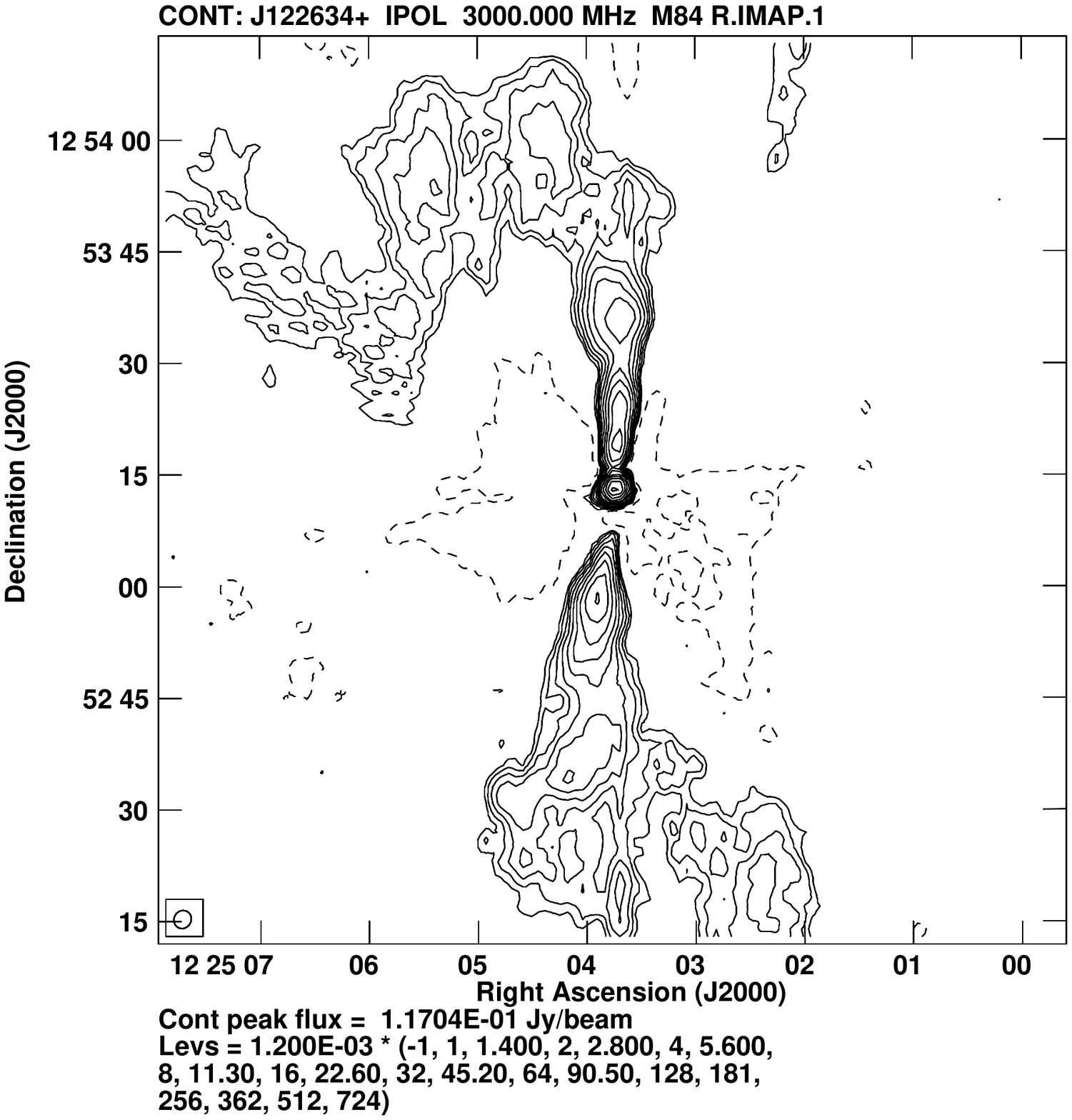}
\includegraphics[width=0.29\textwidth]{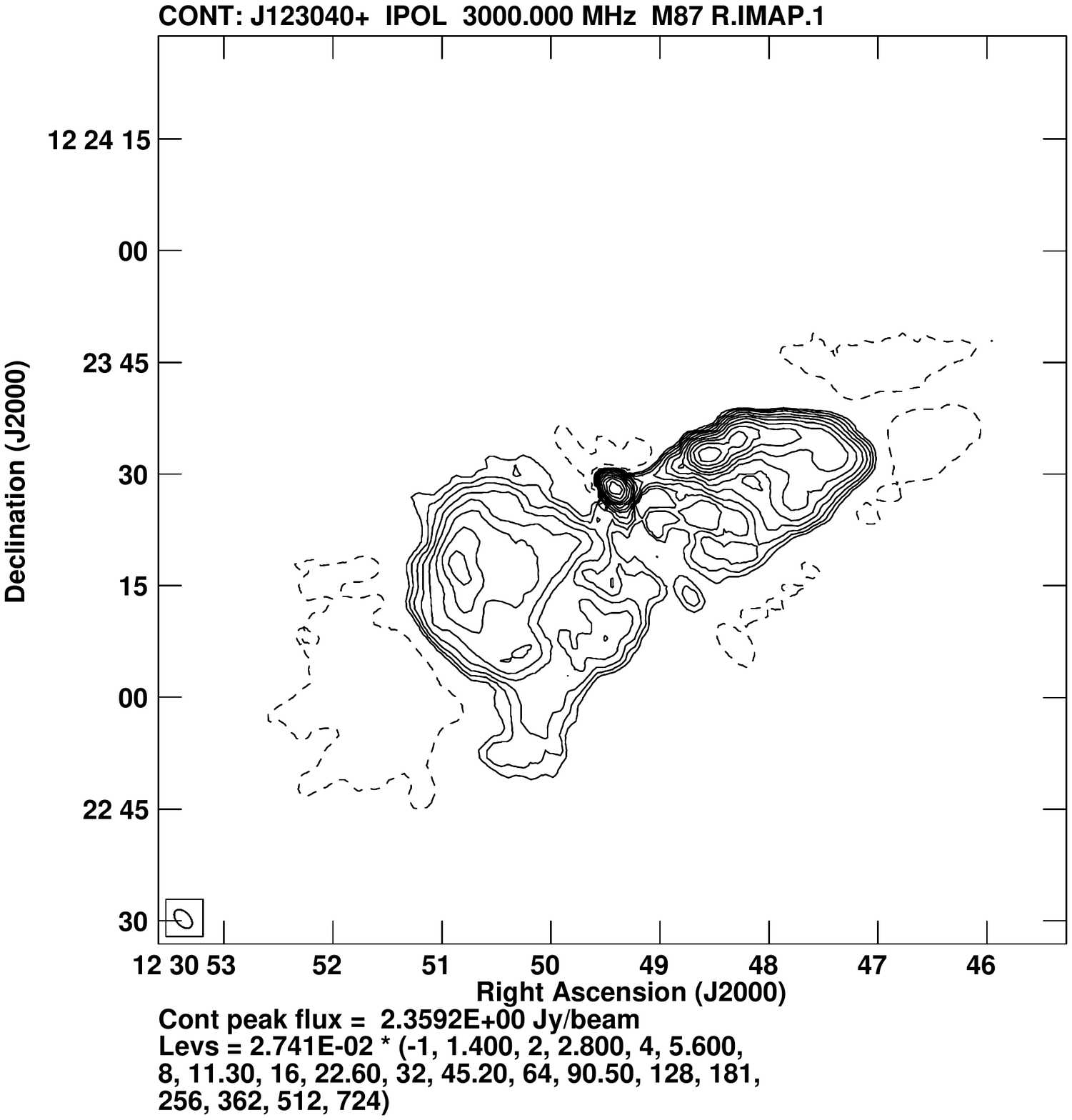}
\includegraphics[width=0.29\textwidth]{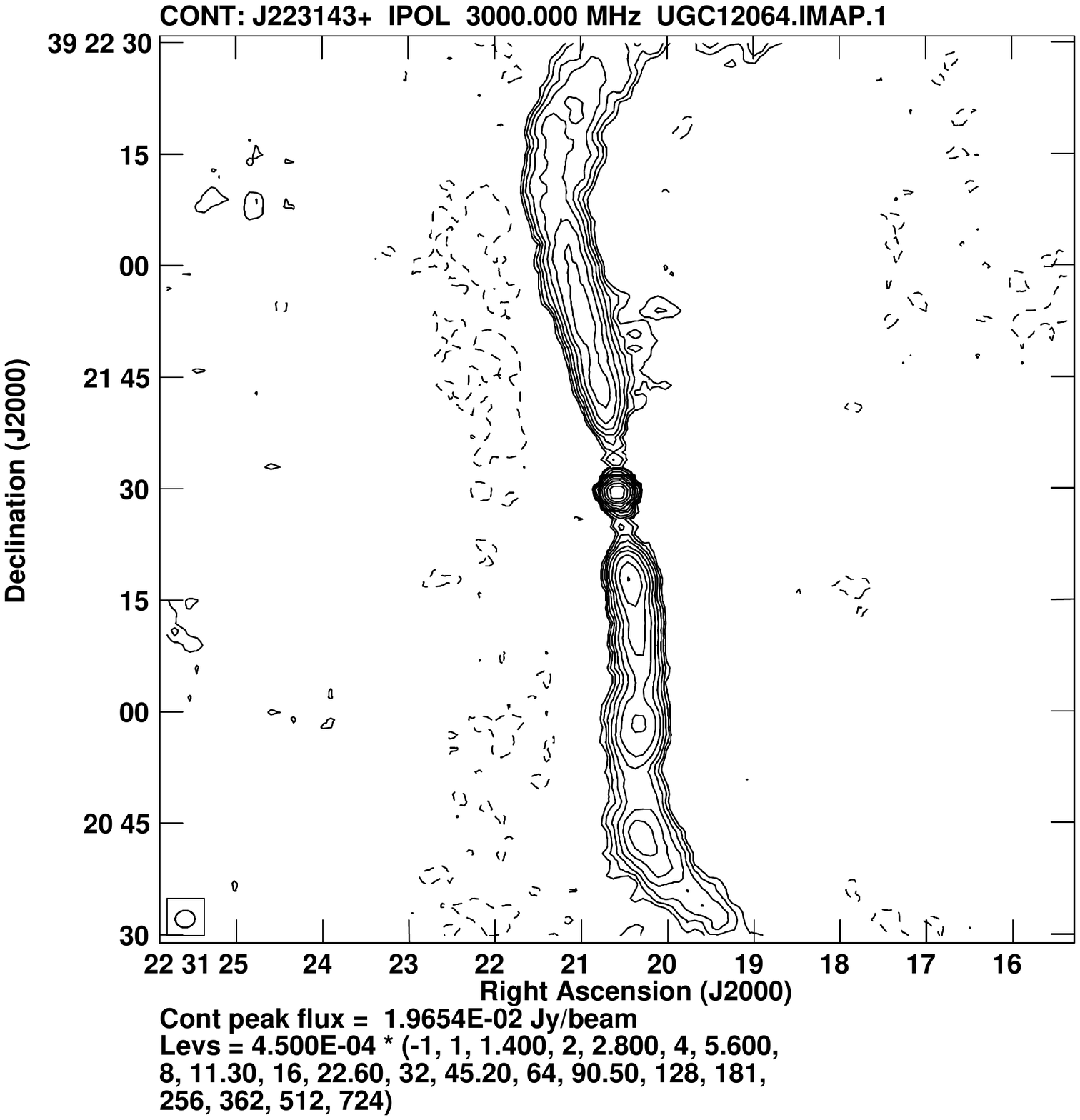}
\caption{VLASS contour maps available for radio-bright galaxies NGC\,7052, NGC\,7626, Abell\,1836/PKS\,B1358--113 (top panel, left to right), NGC\,4374/M\,84, NGC\,4486/M\,87, UGC\,12064 (bottom panel).}
\label{fig:A:VLASS-RB-3}
\end{figure}

We have inspected the $2.5^{\prime\prime}$-resolution VLASS maps at 3\,GHz for the radio-detected early-type galaxies from \citet{gaspari19}. Among 36 radio-dim sources, 11 maps display only featureless noise (NGC\,4697, NGC\,3608, NGC\,3585, NGC\,3923, NGC\,3091, NGC\,1550, NGC\,5328, NGC\,4596, NGC\,4564, NGC\,4026, and NGC\,3115), and so in Figures\,\ref{fig:A:VLASS-RD-1}, \ref{fig:A:VLASS-RD-2}, and \ref{fig:A:VLASS-RD-3} we show the VLASS contour maps for the central $2^{\prime}$ regions of the remaining 25 radio-dim targets exclusively. In the case of 26 radio-bright sources from the sample, the southernmost object NGC\,5128/Centaurus\,A is not covered by the survey, while NGC\,6086 is undetected by VLASS. In Figures\,\ref{fig:A:VLASS-RB-1}, \ref{fig:A:VLASS-RB-2}, and \ref{fig:A:VLASS-RB-3}, we therefore present the VLASS contour maps for the central $2^{\prime}$ regions of the remaining 24 radio-bright targets.



\end{document}